%% file: aanda.tex
\documentclass[longauth]{aa}  

\usepackage{graphicx}
\usepackage{txfonts}
\usepackage{hyperref}
\hypersetup
{
    colorlinks= true,
    linkcolor = blue,
    filecolor = blue,      
    urlcolor  = blue,
    citecolor = blue
}
\usepackage{cleveref}
\usepackage{outlines}
\usepackage{xcolor}
\usepackage{booktabs}
\usepackage{subfigure}
\usepackage{rotating} 
\usepackage{colortbl} 
\usepackage{xcolor}   
\usepackage{multirow} 
\usepackage{lipsum}
\usepackage{lastpage}
\begin{document}

   \title{Chemical templates of the Central Molecular Zone}
   \subtitle{Shock and protostellar object signatures under Galactic Center conditions}

\input{author}

   \date{Received [Month] xx, xxxx; accepted [Month] xx, xxxx}

 
  \abstract
   {The Central Molecular Zone (CMZ) of the Milky Way exhibits extreme conditions, including high gas densities, elevated temperatures, enhanced cosmic-ray ionization rates, and large-scale dynamics. This makes it a perfect laboratory for astrochemical studies. With large-scale molecular surveys revealing increasing chemical and physical complexity in the CMZ, it is essential to develop robust methods to decode the chemical information embedded in this extreme region.}
   {A key step to interpreting the molecular richness found in the CMZ is by building chemical templates tailored to its diverse conditions. In particular, understanding how CMZ environments affect shock and protostellar chemistry is crucial. The combined impact of high ionization, elevated temperatures, and dense gas remains insufficiently explored for observable tracers.}
   {In this study, we utilized \texttt{UCLCHEM}, a gas-grain time-dependent chemical model, to link physical conditions with their corresponding molecular signatures and identify key tracers of temperature, density, ionization, and shock activity. To achieve this, we ran a grid of models of shocks and protostellar objects representative of typical CMZ conditions, focusing on twenty-four species, including complex organic molecules.}
   {Shocked and protostellar environments show distinct evolutionary timescales ($\lesssim 10^4$ vs. $\gtrsim 10^4$ years), with 300~K emerging as a key temperature threshold for chemical differentiation. We find that cosmic-ray ionization and temperature are the main drivers of chemical trends. HCO$^+$, H$_2$CO, and CH$_3$SH trace ionization, while HCO, HCO$^+$, CH$_3$SH, CH$_3$NCO, and HCOOCH$_3$ show consistent abundance contrasts between shocks and protostellar regions over similar temperature ranges.}
   {We characterized the behavior of twenty-four species in protostellar and shock-related environments. While our models underpredict some complex organics in shocks, they reproduce observed trends for most species, supporting scenarios involving need for recurring shocks in Galactic Center clouds and enhanced ionization towards Sgr B2(N2). Future work should assess the role of shock recurrence and metallicity in shaping chemistry.}


   \maketitle
%

\section{Introduction}\label{Intro}

The Central Molecular Zone (CMZ) encompassing the inner $\simeq$~500~pc of the Milky Way \citep{morris1996} is the most extreme environment in the Galaxy. It hosts the central supermassive black hole, Sgr A$^*$, and the highest number of supernovae per unit volume \citep{henshaw2023}. The CMZ also holds up to 10\% of the Galaxy’s molecular gas, despite accounting only for 0.1\% of its surface area \citep{dahmen1998, ferriere2007, roman-duval2016}.

The bulk physical properties found in the CMZ also stand out compared to typical Galactic environments. The temperatures of both gas and dust \citep[$T_\mathrm{dust}=20-50$~K and $T_\mathrm{gas}=50-100$~K; e.g.,][]{morris1983,guesten1985,marsh2016,ginsburg2016,krieger2017,tang2021}, as well as the average H$_2$ volume densities of the dense gas component \citep[a few 10$^4$ cm$^{-3}$; e.g.,][]{guesten1983,longmore2013,mills2018,battersby2025}, are considerably higher than those observed in the Galactic disk  \citep[$T_\mathrm{dust\;\&\;gas}=10-30$~K and $n_\mathrm{H_2}$ of a few 10$^2$~cm$^{-3}$; e.g.,][]{chen2016,friesen2017,spilker2021}.  Additionally, the Galactic Center is where we observe cosmic-ray ionization rates (CRIRs, $\zeta$) 10 -- 100 times greater \citep[e.g.,][]{vandertak2000,lepetit2016,oka2019} than those measured in the Galactic disk, where its values are usually of the order of $10^{-17}\;\mathrm{s}^{-1}$ \citep[e.g.,][]{indriolo2015,padovani2020,padovani2022}.

Both the extreme conditions and proximity make the CMZ a unique laboratory for studying the interplay between large-scale dynamics \citep[e.g., cloud-cloud collisions;][]{tsuboi2015,zeng2020}, star formation \citep[see][for a recent review]{henshaw2023}, and chemistry \citep[][and references therein]{jimenez-serra2025} in an environment far from typical Galactic disk conditions. To such an extent, the CMZ is the best Galactic analog of high-redshift star-forming gas in extragalactic nuclei \citep{kruijssen2013}. In fact, its proximity \citep[$\sim$8.2 kpc;][]{GRAVITY2019} allows for high-resolution studies, enabling detailed investigations of processes that are unresolved in external galaxies \citep[e.g.][]{martin2021}.

In recent years, we have entered an era of large-scale molecular surveys that enable us to map the CMZ with unprecedented detail. Observations, such as the Atacama Large Millimeter/submillimeter Array (ALMA) CMZ Exploration Survey (ACES), can capture the full molecular content of the CMZ \citep{longmore2025}. Several case studies achieve deeper imaging of molecular line emission in selected regions, e.g., Sgr B2, Sgr C, G0.253+0.016, and the 20 km$\,$s$^{-1}$ cloud \citep[e.g.,][]{rathborne2015,sanchezmonge2017,barnes2019,belloche2019,walker2021,lu2021,moeller2025,yang2025,xu2025}. These datasets reveal a wealth of information on the physical and chemical properties of the CMZ and its extreme complexity, presenting a new challenge: interpreting these data in terms of the underlying physical conditions and chemical processes.

To fully leverage these rich datasets, it is crucial to deepen our understanding of the physical and chemical interplay under the extreme conditions of the CMZ. This requires a robust framework to connect physical structures with state-of-the-art chemical models. The chemical aspect plays a central role here, as molecular species serve as key diagnostics of temperature, density, irradiation, and shock activity \citep[e.g.,][]{herbst2009,bayet2011,jorgensen2020}.

Over the years, several modeling studies have sought to address the chemistry of different environments within the CMZ \citep{requena-torres2006,requena-torres2008,martin2008,goicoechea2013,goicoechea2018,harada2015,garrod2017,zeng2018,bonfand2019,santa-maria2021}. With this paper, we aim to provide a broad set of chemical templates specifically tailored to the unique conditions of the CMZ, which can be applied to a range of environments. These templates will serve as benchmarks for interpreting observations from large-scale molecular surveys and help bridge the gap between observed chemistry and underlying physical conditions. The data presented in this work is publicly available\footnote{All chemical templates are available at \url{https://doi.org/10.5281/zenodo.15674940}}. Therefore, this paper acts as an overview of the results, focusing on the key patterns rather than a detailed chemical analysis of specific molecules.

This paper is organized as follows. Section \ref{sect:methodology} introduces the methods used to model the physical and chemical conditions and the species considered. In Section \ref{sect:results}, we present the results, focusing on protostellar objects and regions influenced by shocks. Subsequently, Section \ref{sect:discussion} discusses the most effective molecular tracers of CMZ conditions and compares our findings with previous studies. Finally, Section \ref{sect:conclusions} summarizes our conclusions and outlines future directions for extending this work. 

\section{Methodology}\label{sect:methodology}

We used \texttt{UCLCHEM}\footnote{https://uclchem.github.io/} \citep{uclchem3.0}, an open source, time-dependent, gas-grain chemical code, to determine the chemical signatures of 24 species (see Tab. \ref{tab:molecules}) in varying physical conditions. Our species selection is based on the molecules observed within the ACES survey \citep{longmore2025}, which covers 6 spectral windows in ALMA Band 3 and spans frequencies from 85.96 to 101.44 GHz. We used the most recent version (v3.4) of the code together with the most recent version (Rate22) of the UMIST Database for Astrochemistry \citep[UDfA; ][]{umist2022}. For the purpose of this work, we modeled two types of physical structures, i.e., protostellar objects and C-type shocks. In the case of protostellar objects, we designed them to resemble cores found in Galactic Center clouds, e.g., the Sgr~B2 cores \citep[e.g.,][]{schmiedeke2016, schworer2019}, while modeled C-type shocks were heavily influenced by the large-scale processes observed throughout the CMZ, e.g., shocks induced by cloud-cloud collisions \citep[e.g., G+0.693-0.027; ][]{zeng2018,zeng2020,colzi2024,jimenez-serra2025}. As such, the models are relevant to star formation and large-scale structures observed in both the CMZ and extragalactic regime \citep[e.g.,][and references therein]{martin2021, henshaw2023}. 

\subsection{Modeling gas-grain chemistry}

\begin{table}[t!]
\caption{Molecules considered in this study}
\label{tab:molecules}
\centering
\setlength{\extrarowheight}{1pt}
\begin{tabular}{l|l|l}
\hline
\hline
\addlinespace[0.04cm] 
&Formula         & Name                 \\
\addlinespace[0.04cm] 
\hline
\addlinespace[0.04cm] 
\multirow{13}{*}{\rotatebox[origin=c]{90}{Simple molecules}} & CS              & Carbon monosulfide   \\
&SO              & Sulfur monoxide      \\
&SiO             & Silicone monoxide    \\
&NS$^+$          & Thionitrosyl ion     \\
&C$_2$S$^\ast$   & Thioxoethenylidene   \\
&HCN             & Hydrogen cyanide     \\
&HNC             & Hydrogen isocyanide  \\
&HCO             & Formyl radical       \\
&HCO$^+$         & Oxomethylium         \\
&H$_2$CO         & Formaldehyde         \\
&HNCO            & Isocyanic acid       \\
&HC$_3$N         & Cyanoacetylene       \\
&CH$_2$CO        & Ketene               \\
\addlinespace[0.04cm] 
\hline
\addlinespace[0.04cm] 
\multirow{11}{*}{\rotatebox[origin=c]{90}{Complex organic molecules}} & CH$_3$CN               & Acetonitrile          \\
&CH$_3$OH               & Methanol              \\
&CH$_3$SH$^\ast$        & Methanethiol          \\
&NH$_2$CHO              & Formamide             \\
&CH$_3$CCH              & Propyne               \\
&CH$_3$CHO              & Acetaldehyde          \\
&CH$_3$NCO$^\ast$      & Methyl isocyanate      \\
&HCOOCH$_3^\ast$        & Methyl formate        \\
&C$_2$H$_5$CN$^\ast$    & Propanenitrile        \\
&C$_2$H$_5$OH$^\ast$    & Ethanol               \\
&CH$_3$OCH$_3^\ast$   & Dimethyl ether        \\
\hline
\end{tabular}
\tablefoot{$^\ast$Species added to the network specifically for the purpose of this study, as they are not a part of the default chemical network. For details, we refer the reader to Sect. \ref{sect:simple-mol} and Sect. \ref{sect:COMs}.}
\end{table}

By being time-dependent and accounting for three phases -- gas, surface, and bulk -- \texttt{UCLCHEM} provides a detailed understanding of the chemical evolution of selected species and their behavior in response to different physical models and varying physical parameters. Two of these phases correspond to dust grains, which are known to play a crucial role in interstellar chemistry by acting as catalysts in the astrophysical sense, e.g., enabling otherwise slow gas-phase reactions \citep{vandishoeck2014,cuppen2017}. 

The outermost layer of dust grains is referred to as the surface, where species can adsorb from or desorb into the gas phase. The layer beneath the surface is called the bulk. Together, these layers form the grain’s ice composition. Species from the bulk can diffuse to the surface and may either be released into the gas phase or destroyed in fast shocks. Reactions on the surface include grain-surface diffusion, chemical reactive desorption, and reaction-diffusion competition. The implementation of these grain-surface mechanisms was thoroughly described in \citet{quenard2018}.

The modeling process with \texttt{UCLCHEM} typically consists of two stages. In the first stage, the model follows the isothermal collapse of a diffuse cloud until it reaches a predefined density. The final chemical composition of this stage is then passed on to the second-stage models. In the second stage, the model follows more advanced physical structures, which currently include protostellar objects \citep{viti2004, awad2010} and shocks \citep{jimenez-serra2008, james2020}.

The default chemical network in \texttt{UCLCHEM} includes 249 gas and grain surface species, with the degree of chemical complexity reaching eight constituent atoms. However, not all the species required by this study are a part of this default network. Thus, we expanded it by adding additional 65 gas-phase and grain surface species, along with grain surface reactions, increasing the overall complexity to up to ten constituents. The origin of each added species and reaction is detailed in  Sect. \ref{sect:simple-mol} and Sect. \ref{sect:COMs}. Our extended network was then used to model eleven complex organic molecules (COMs), i.e., molecules consisting of at least 6 atoms and containing carbon \citep{herbst2009}, as well as thirteen smaller species composed of two to five atoms. The choice of these selected modeled species was motivated by the observations carried out with the ACES survey, and as such we tried to address as many observable molecules as possible within its covered frequency range, which in total covers 85.96 to 101.44 GHz. The complete list of these molecules can be found in Tab. \ref{tab:molecules}. 

\subsection{Grid of Physical Parameters}\label{sect:methodology-grid}

We examined a diverse range of physical parameters in our study to match the possible environments in both the CMZ and external galaxies. We concentrated on the most fundamental physical parameters that can shape chemistry, such as density, temperature, CRIR, and UV irradiation. However, while CRIR and UV irradiation are varied as input parameters, \texttt{UCLCHEM} does not self-consistently model the associated heating and cooling, but instead adopts parameterized temperature profiles. 

For protostellar objects, we focused on modeling cores with a radius of 0.5~pc and with final temperatures ranging from 100~K to 500~K. In the case of C-shocks, we explored low to moderate shock velocities spanning from $10-40\,\mathrm{km\,s}^{-1}$ \citep{martin-pintado1997,zeng2018,zeng2020} in a medium with a strong magnetic field. The transverse magnetic field strength, $B_0$, is defined as $\dfrac{B_0}{\mu\mathrm{G}}=bm_0\sqrt{n_\mathrm{H}}$, where $bm_0$ is a magnetic parameter \citep{draine1983}. Following \citet{jimenez-serra2008}, we adopt a value of 1.4 for $bm_0$ to ensure that the resulting magnetic field is strong enough for the C-type shocks to emerge. We provide a detailed overview of the full parameter space in Tab. \ref{tab:UCLCHEM}. Below, we detail the setup of each modeled structure. 

\begin{table*}[t!]
\caption{Parameter space covered with \texttt{UCLCHEM} models.}
\label{tab:UCLCHEM}
\centering
\setlength{\extrarowheight}{2pt}
\begin{tabular}{c|l|l|l|l}
\hline
\hline
\addlinespace[0.04cm]  
&Parameter & Unit & Values & Notes \\
\addlinespace[0.04cm]  
\hline
\addlinespace[0.04cm]   
\multirow{7}{*}{\shortstack{Stage 1:\\Collapse}} &$n_\mathrm{H,\,final}$$^{a}$     & cm$^{-3}$  & $10^4, 10^5, 10^6, 10^7, 10^8$   & $n_\mathrm{H,\, init} = 10^2\,\mathrm{cm}^{-3}$  \\
&$T_\mathrm{init\,(gas,\,dust)}$  & K          & $15, 20, 25, 30, 35$                        & ...                              \\
&$\zeta / \zeta_0$                & $-$        & $10^1, 10^2, 10^3, 10^4$         & $\zeta_0 = 1.31\times10^{-17}\,\mathrm{s}^{-1}$ \\
&$G_0$                  & Habing     & $10^1, 10^2, 10^3, 10^4$   & ...                                              \\
&$A_\mathrm{V,\, start}$                   & mag        & 2                                & ...                                              \\
&$X(\mathrm{Si})$$^{b}$           & $-$        & $1.78\times10^{-8}, 1.78\times10^{{-6}\dag}$     & $^\dag X(\mathrm{Si})_\odot$ \citep{jenkins2009} \\
&$R_\mathrm{final}^c$                & pc        & 0.5                     & ... \\
\addlinespace[0.04cm]  
\hline
\addlinespace[0.04cm]  
\multirow{4}{*}{\shortstack{Stage 2:\\Protostellar\\Objects}} &$n_\mathrm{H}$$^{a}$              & cm$^{-3}$ & $10^6, 10^7, 10^8$ & ...               \\
&$G_0$                   & Habing    & $10^3, 10^4$             & ...               \\
&$T_\mathrm{final\,(gas,\,dust)}$  & K         & $100, 150, 200, 250, 300, 350, 400, 450, 500$               &  ...\\
&$M$            &   M$_\odot$        & 10, 25 & ...\\
\addlinespace[0.04cm]  
\hline
\addlinespace[0.04cm]  
\multirow{4}{*}{\shortstack{Stage 2:\\Shocks}} &$n_\mathrm{H,\,pre-shock}$$^{a}$ & cm$^{-3}$                  & $10^4, 10^5, 10^6$     & ...                                  \\
&$G_0$                  & Habing     & $10^1, 10^2, 10^3, 10^4$   & ...       \\
&$v_\mathrm{s}$                   & $\mathrm{km\,s}^{-1}$     & $10, 15, 20, 25, 30, 35, 40$                & ... \\
&$B_0^{d}$                            & $\mu\mathrm{G}$       & $140, 450, 1400$   & $B_0=bm_0\sqrt{n_\mathrm{H}}$                                   \\
\addlinespace[0.04cm]  
\hline
\end{tabular}
\tablefoot{$^{(a)}$Proton density, $n_\mathrm{H}$, is defined as $n(\mathrm{H}) + 2n(\mathrm{H}_2)$; $^{(b)}$The models of protostellar objects utilize collapse models with depleted Si abundance with respect to the Solar abundance \citep[e.g.,][]{savage1996}, while the shock models rely on those with the standard value of $1.78\times10^{-6}$ \citep{jenkins2009}; $^{(c)}$For consistency across models, we adopted a single radius. However, to evaluate the full impact of $R_\mathrm{final}$ on the chemistry of protostellar objects, we conducted an additional test with $R_\mathrm{final} = 0.05$~pc (see Sect. \ref{sect:radius}). The results of this test suggest that our protostellar object models are also applicable to smaller objects, with $R \gtrsim 0.05$~pc. $^{(d)}$Values of $B_0$ were not varied, as they depend on pre-shock densities. This work adopts $bm_0$ of 1.4 \citep{jimenez-serra2008}.} 
\end{table*}

\subsubsection{Stage 1 - collapsing cloud}

All models begin with a diffuse cloud with a number density of $10^2\,\mathrm{cm}^{-3}$. At the start, the abundance of all species, except for atomic elements, is set to zero, while elemental abundances follow their solar values \citep{asplund2009, jenkins2009}, with the exception of Si, which is varied by a factor of 100 to account for its observed depletion in star-forming regions \citep[e.g.][]{savage1996}. A cloud of radius 0.5 pc undergoes free-fall collapse \citep[as described in][]{rawlings1992}, reaching final densities in the range of $10^4-10^8\,\mathrm{cm}^{-3}$, depending on the adopted physical structure in subsequent modeling. The cloud initial visual extinction, $A_\mathrm{V,base}$, is set to 2 mag (to account for the fact that we start from a gas density of 10$^2$\,cm$^{-3}$), with the time dependent visual extinction rapidly exceeding 10 mag as the cloud collapses.  

To reflect the elevated gas and dust temperatures characteristic of the CMZ, the starting temperature is set higher than typical Galactic disk conditions, ranging from $15-35$~K \citep[e.g.,][]{krieger2017, tang2021}. For a further discussion of this choice, we refer the reader to Sect. \ref{sect:low-COMs}. Furthermore, we expose the cloud to varying interstellar UV field strengths, $G_0$, and CRIRs. For the CRIR, we adopt values spanning $1.31\times(10^{-16}-10^{-13})\,\mathrm{s}^{-1}$ \citep{goto2013, lepetit2016, oka2019}, expressed in the models as $\zeta / \zeta_0$, where $\zeta_0$ denotes the canonical Galactic value of $1.31\times10^{-17}\,\mathrm{s}^{-1}$ typically used in chemical modeling and found in dense clouds \citep{caselli1998}. The $G_0$ is varied between $10^1-10^4$ Habing\footnote{The FUV interstellar radiation field is expressed in the Habing unit, where $G_0=1.6\times10^{-3}\,\mathrm{erg\,s}^{-1}\,\mathrm{cm}^{-2}$ \citep{habing1968}} \citep{lis2001, goicoechea2004, clark2013}. The cloud evolves over $6.5\times10^6$\,yr, which gives the chemistry time to evolve further after the cloud reaches its final density.

\subsubsection{Stage 2 - protostellar objects and shocks}\label{sect:stage2}

Following the chemical composition of the collapsed cloud, we model two physical scenarios to recover the chemistry of the CMZ. The widespread presence of COMs observed in the CMZ \citep[e.g.,][]{requena-torres2006,requena-torres2008,jones2011,zeng2018,zeng2023} is likely associated with a presence of ubiquitous low-velocity shocks \citep[e.g.,][]{hasegawa1994,zeng2020,yang2025} at cloud scales and hot core-like chemistry \citep{requena-torres2006}. Therefore, considering the typical conditions in some of the most representative regions within the CMZ, we constructed a grid of protostellar objects and low-to-mid velocity shocks models. 

We select from the collapse stage the models with a silicon depletion factor of 100, and then apply a protostellar heating profile to it. Since this study aims to account for regions like the massive star-forming complex Sgr B2(N), where cores of varying masses have been observed \citep[e.g.,][]{sanchezmonge2017, bonfand2019}, we included core mass as a variable. We modeled two cases: a typical high-mass core of 10~$\mathrm{M}_\odot$ and a more massive core of 25~$\mathrm{M}_\odot$. The warm-up of these protostellar objects with densities ranging from $10^6-10^8\,\mathrm{cm}^{-3}$, follow the time and radially dependent temperature profiles given in \citet{viti2004} \citep[with modifications by][]{awad2010}, resembling the heat up of the dust and gas around a hot core. Heated up to temperatures in the range $100-500$~K, the modeled objects evolve for 1\,Myr. They are exposed to the same CRIRs as their natal clouds, but the UV radiation at this stage is set to vary between $10^3-10^4$ Habing to account for the increased UV radiation field coming from the central protostar.

Considering the strong magnetic fields in the CMZ \citep[e.g.,][]{chuss2003, pillai2015,butterfield2024,karoly2025}, we model C-type shocks propagating in a medium with $B_0$ varying between $140-1400\,\mu\mathrm{G}$ for pre-shock densities in the range of $10^4-10^6\,\mathrm{cm}^{-3}$, respectively. These densities are aligned with densities of the CMZ, which are often orders of magnitude higher than those in the Galactic disk \citep[e.g.,][]{guesten1983, mills2018}. Similarly, as for protostellar objects, the pre-shock gas in the shock models is subject to the same levels of CRIR, while the impinging UV field varies in the same range as in the collapsing cloud models. The shock models are constrained to follow-up on collapse models with no Si depletion to account for the fact that Si is released from dust grains into the gas phase by sputtering \citep{caselli1997,jimenez-serra2008}. C-shocks in this study propagate with low-to-mid velocities in the range of $10-40\,\mathrm{km\,s}^{-1}$ \citep[e.g.,][]{requena-torres2006,requena-torres2008,martin2008,zeng2020}. 

\subsubsection{Size of protostellar objects}\label{sect:radius}
In our models, we adopt a fixed radius of 0.5~pc for both the protostellar objects and the clouds in which the shocks propagate. While this may be somewhat larger than typical protostellar object sizes and appear more consistent with clump scales, our primary aim is to compare chemical abundances under different physical scenarios rather than to reproduce detailed spatial structures. Nevertheless, it is important to assess how the assumption of a fixed radius impacts the results, particularly for more compact protostellar objects with a radius of 0.05~pc. 

In our models, UV radiation plays only a minor role in shaping the chemical evolution. This is because, at 0.5 pc, in stage~1, when the cloud collapses as a function of time, the visual extinction quickly reaches the threshold where UV photons can no longer penetrate the cloud. At smaller scales, however, extinction may remain below this threshold for longer, especially at the highest considered $G_0$ values.  

Any potential differences would primarily arise from the choice of $G_0$ during the beginning of the collapse stage, which could have lasting effects on protostellar object models. However, when we compared models with $R_\mathrm{final}=0.5$~pc to those with one order of magnitude smaller radius, we did not find any significant changes in the chemical evolution of species, nor did it change the level of importance of $G_0$ in our models. This suggests that the influence of $G_0$ on the overall model results is likely secondary and that the outcomes for the considered parameter space are not strongly sensitive to variations in radius. 

This can be understood by considering the $A_\mathrm{V}$ values associated with the densities of the protostellar objects. Since higher density leads to greater extinction, we can focus on the lowest-density model ($n_\mathrm{H}=10^6$~cm$^{-3}$) as a conservative case. For this model, $A_\mathrm{V}$ reaches approximately 98~mag at $R_\mathrm{final} = 0.05$~pc, and  $\approx966$~mag at $R_\mathrm{final} = 0.5$~pc. Consequently, even for the highest assumed $G_0$, UV photons should be fully attenuated at $A_\mathrm{V}$ values significantly lower than those in our models and are quickly surpassed in all cases during the collapse. Therefore, assuming a constant $R_\mathrm{final}$ does not compromise the applicability of the results presented here.

\subsection{Molecular tracers}\label{sect:molecules}

Molecules offer a wealth of information about the history and properties of a given region. Some chemical species are good tracers of temperatures  e.g., H$_2$CO \citep[e.g.,][]{schoier2004}, CH$_3$CN \citep[e.g.,][]{giannetti2017}, CH$_3$CCH \citep[e.g.,][]{askne1984}, some are prime shock tracers, e.g., SiO \citep[e.g.,][]{martin-pintado1992,lu2021,walker2021,yang2025}, others trace UV irradiation, e.g., C$_2$H, CN \citep{sternberg1995,aalto2002}. Many of these species are now routinely observed in the Galactic Center. It is in this region where a true richness of COMs is being uncovered \citep{belloche2013, jimenez-serra2025}, including prebiotic molecules known to be essential building blocks of life \citep[e.g.,][]{jimenez-serra2020,jimenez-serra2022,rivilla2020,rivilla2021,rivilla2022,rivilla2023}.  

However, our understanding of complex chemistry under extreme conditions, like those observed in the Galactic Center, is still limited. This poses big challenges, as we now observe more of these complex molecules in external galaxies, e.g., in NGC 253 \citep[e.g.,][]{mangum2019,martin2021} or PKS 1830-211 \citep{muller2011}. With this work, we aim to provide a broad range of chemical templates that can be used to interpret chemical signatures of extreme environments similar to the CMZ. Our choice of chemical species is motivated by the requirements of the most recent large-scale mosaics of the region, specifically by the avenues opened with the ACES survey \citep{longmore2025}.

\subsubsection{Species composed of less than six atoms}\label{sect:simple-mol}

The first group of species we consider, hereafter referred to as ``simple molecules'', includes those composed of up to five atoms (see Tab. \ref{tab:molecules} for the full list). All selected species are part of the default \texttt{UCLCHEM} chemical network, except for the carbon-chain molecule thioxoethenylidene (C$_2$S, also known as CCS).  

Thioxoethenylidene (C$_2$S) was first detected in TMC-1 and Sgr B2 by \citet{saito1987} and has since been used as a tracer of dark molecular cloud cores \citep{suzuki1992}, making it a valuable probe of high-density environments and cloud evolutionary stages \citep[e.g.,][]{pineda2020}. To model its chemistry, we utilize gas-phase reactions from UMIST 2022 \citep{umist2022} and include C$_2$S on grains only through freeze-out. This choice is motivated by existing uncertainties in sulfur chemistry and the fact that a comprehensive investigation of carbon-chain sulfur-bearing species is beyond the scope of this work. A more detailed treatment of C$_2$S, however, would require an expanded sulfur reaction network to account for potential grain-surface formation pathways, as suggested by \citet{laas2019}.

\subsubsection{Complex organic molecules}\label{sect:COMs}

The complex organic chemistry considered in this work goes beyond what is typically required in studies using \texttt{UCLCHEM}, and as such six out of eleven species (see Tab. \ref{tab:molecules} for the full list) had to be added specifically for the purpose of this work. 

The five new COMs added in the network -- CH$_3$NCO, HCOOCH$_3$, C$_2$H$_5$CN, C$_2$H$_5$OH, and CH$_3$OCH$_3$ -- were incorporated based on the work of \citet{quenard2018}. The only fully novel species is methanethiol (methyl mercaptan, CH$_3$SH). It was first detected in the CMZ, specifically in the same cloud complex as C$_2$S, i.e., Sgr B2 \citep{linke1979} and has since been observed in warm, dense regions associated with both low- and high-mass star formation \citep[e.g.,][]{lee2017,vastel2018}.

For this study, we adopted the gas-phase chemistry of CH$_3$SH from the UMIST 2022 database. Additionally, we incorporated CH$_3$SH onto grain surfaces following \citet{perrero2022} and included several possible grain-surface reactions based on \citet{kerr2015}, \citet{gorai2017}, and \citet{lamberts2018}. To ensure completeness, we extended the gas-phase network with selected reactions from KIDA 2024 \citep{kida2024}, incorporating the necessary reactants.

\begin{figure}[t!]
     \centering
     \includegraphics[width=1\linewidth]{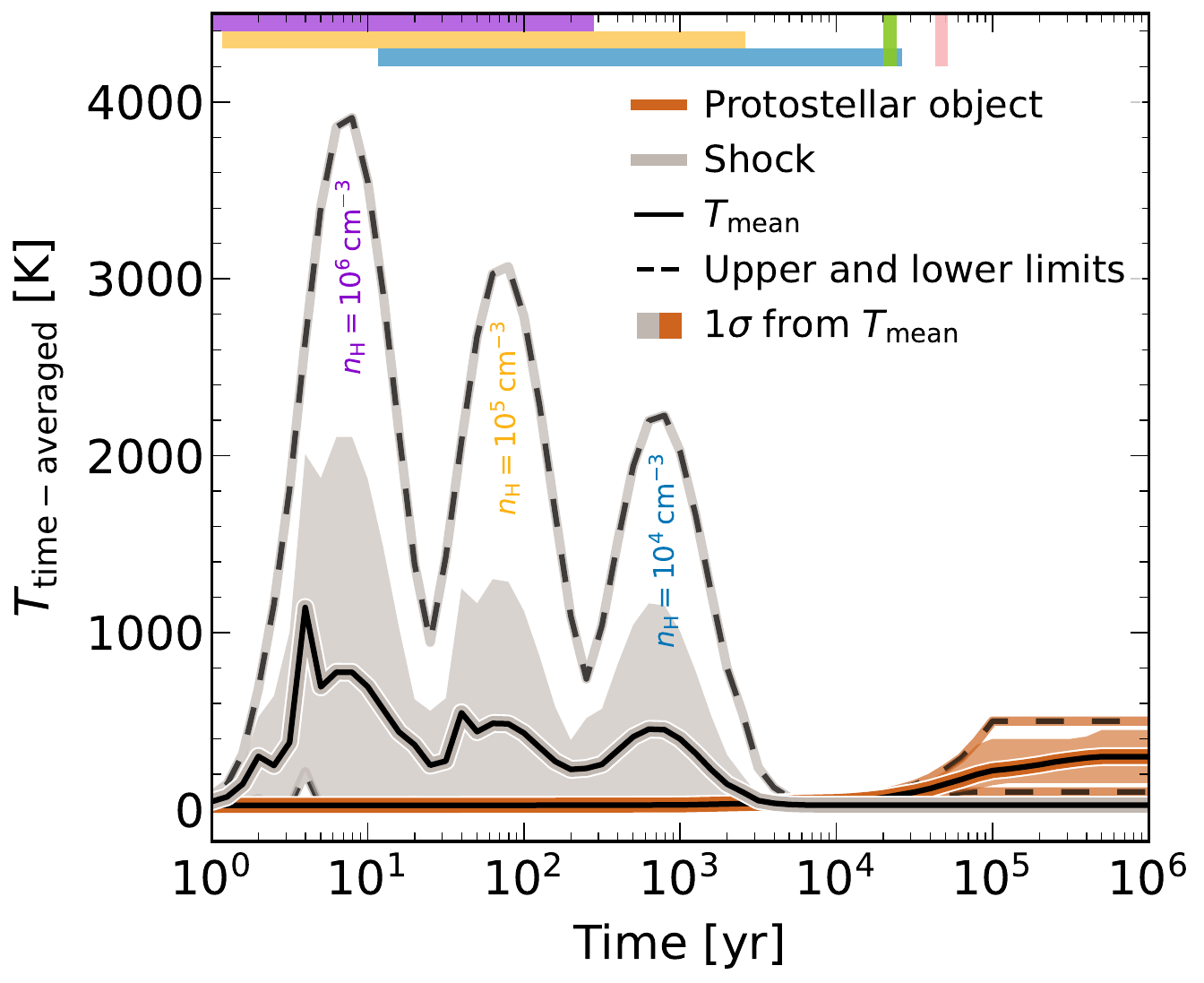}
     \caption{Stage 2 time-averaged temperature profiles for all models of shocks and protostellar objects defined in Tab. \ref{tab:UCLCHEM}. ``Time-averaged'' refers to averaging the temperature across all models of a given type at fixed logarithmic time steps. The solid black line represents the mean coupled gas and dust temperature at each logarithmic time step, while shaded regions indicate the 1$\sigma$ confidence interval. Dashed lines mark the upper and lower temperature limits. Since these profiles are averaged across all models, multiple shock events appear, though each individual model experiences only one. The timing of each shock event is dictated by the pre-shock medium density, with higher densities leading to shorter timescales, while the temperature is primarily determined by shock velocity. The densities associated with each shock event are annotated, with color-matched rectangular patches at the top of the figure indicating the full possible timing of those events. In contrast, protostellar heating operates on significantly longer timescales. Once the heating stage concludes, the temperature stabilizes at a plateau, set by the maximum protostellar object temperature. Since more massive cores warm up faster, influencing the onset of the temperature plateau, we mark with vertical lines at the top of the figure the earliest plateau onset times: pink for 10~M$_\odot$ objects and green for 25~M$_\odot$.}\label{fig:temp_evolution}
 \end{figure}
 
\section{Results}\label{sect:results}

\begin{figure*}[htbp]
     \centering
     \includegraphics[width=1\linewidth]{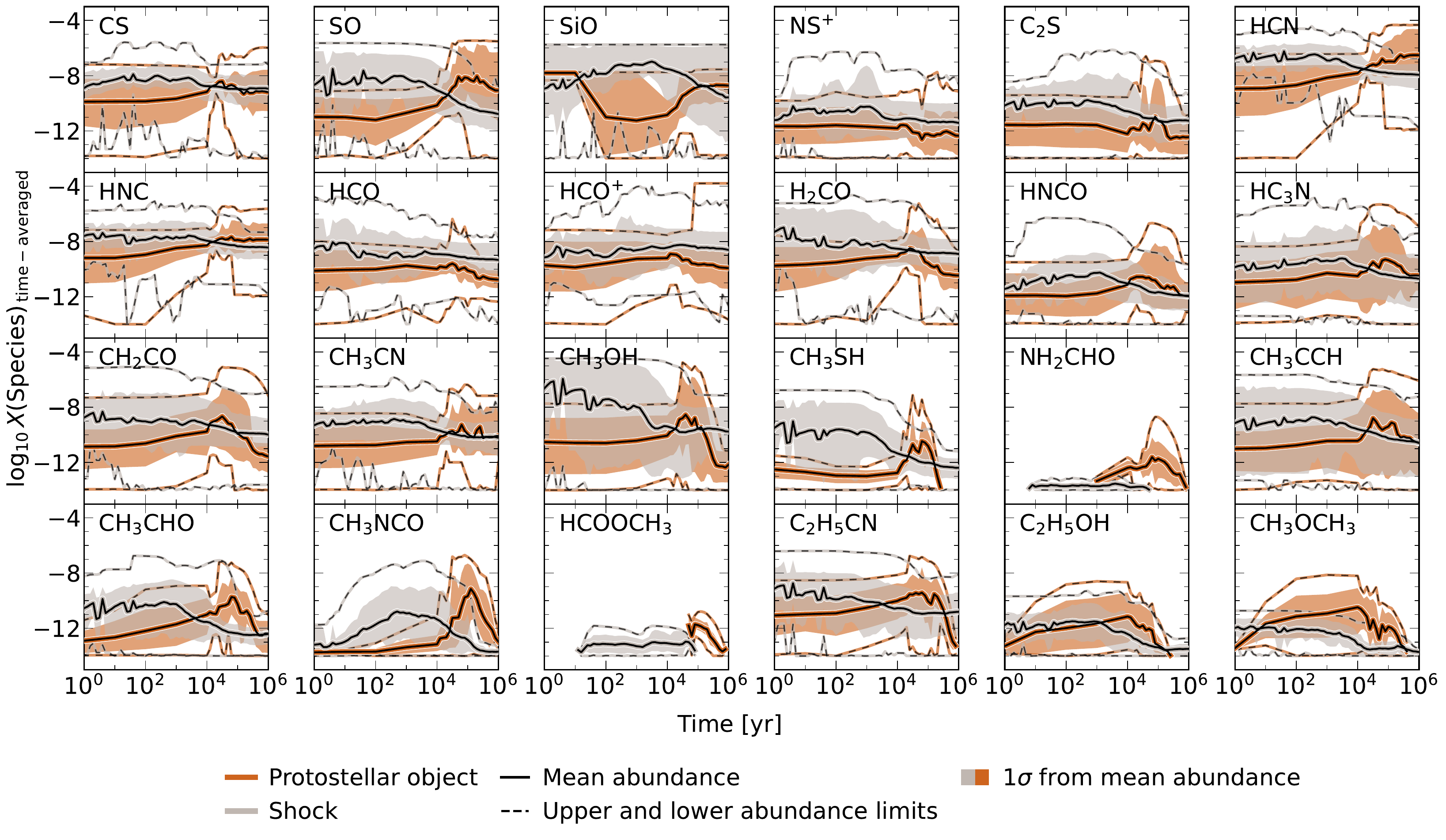}
     \caption{Time-averaged chemical evolution of the studied species across all second stage models, as defined in Tab. \ref{tab:UCLCHEM}. Line styles and shading follow those in Fig. \ref{fig:temp_evolution}. Only gas-phase abundances above the detectable threshold ($X > 10^{-14}$) are considered. Changes in abundance, including both enhancement and depletion, generally follow the timescales seen in the average temperature profiles (see Fig. \ref{fig:temp_evolution}). In models of protostellar objects, the species that most frequently dominates over the others is HCN, whereas in shocks, CH$_3$OH is the most dominant during the shock phase, while SiO takes over in the post-shock phase.}
     \phantomsection
     \label{fig:time_evolution}
 \end{figure*}

Each physical process considered in this study operates on distinct timescales (see Fig. \ref{fig:temp_evolution}). While shocks and protostellar heating represent separate mechanisms primarily responsible for the desorption of species from icy grain surfaces, there exists a potential overlap in their effects \citep[e.g.,][]{viti2001}. Shocks induce sputtering, whereby molecules are ejected from ice mantles through collisions with neutrals and ions. In contrast, protostellar heating facilitates molecular release predominantly via thermal desorption. However, when these processes occur on comparable timescales, or when the observed gas-phase abundances suggest contributions from both mechanisms -- a scenario not uncommon in the CMZ \citep[e.g.,][]{busch2024} -- it becomes essential to examine the temporal evolution of individual species and assess how their chemistry responds to varying physical conditions.

To provide a comprehensive overview of general trends and characterize the behavior of each modeled species, we calculated the time-averaged chemical evolution for each species. Time-averaging was introduced to enable a direct comparison of abundance trends at identical time steps for both shocks and protostellar objects, as discussed further in Sect. \ref{sect:chemical-evolution}. Next, we analyzed the distributions of abundances (relative to H) across different initial densities, temperatures, cosmic-ray ionization rates, and representative properties of each object (shock velocities and protostellar object temperatures). These distributions are examined in detail in Sect. \ref{sect:distributions}. For this study, we focus solely on the detectable abundance regime of gas-phase species, empirically defined as $>10^{-14}$, and perform all calculations only for values above this threshold.

Throughout this paper, we adopt the following naming conventions to describe the evolutionary stages in our models. The term ``warm-up'' refers to the stage during which the temperature of a modeled protostellar object is increasing but has not yet reached its final value, $T_\mathrm{final\,(gas,\,dust)}$. Once the temperature reaches $T_\mathrm{final\,(gas,\,dust)}$, the object is referred to as a ``(fully warmed-up) protostellar object''. In this stage, we follow the chemical evolution over the first $10^5$ years after reaching the final temperature, consistent with the typical lifetime of a hot core \citep{mckee2003}. The term ``shock'' refers to the stage in shock models when $T_\mathrm{(gas,\,dust)} > T_\mathrm{init\,(gas,\,dust)}$, thereby including the subsequent relaxation period during which the medium cools but has not yet returned to its initial temperature. The ``post-shock'' stage begins once the medium has reached a steady state, i.e., when both temperature and density have stabilized and $T_\mathrm{(gas,\,dust)} = T_\mathrm{init\,(gas,\,dust)}$. This stage is followed for up to $10^5$ years, in line with the timescale considered for protostellar objects for a more accurate comparison between the effects of protostellar heating and shocks. Lastly, when comparing the chemistry associated with shock models to that in protostellar object models, we refer to them as the ``shocked'' and ``non-shocked'' cases, respectively.

\subsection{Chemical evolution}\label{sect:chemical-evolution}

 \begin{figure*}[t!]
     \centering
     \includegraphics[width=1\linewidth]{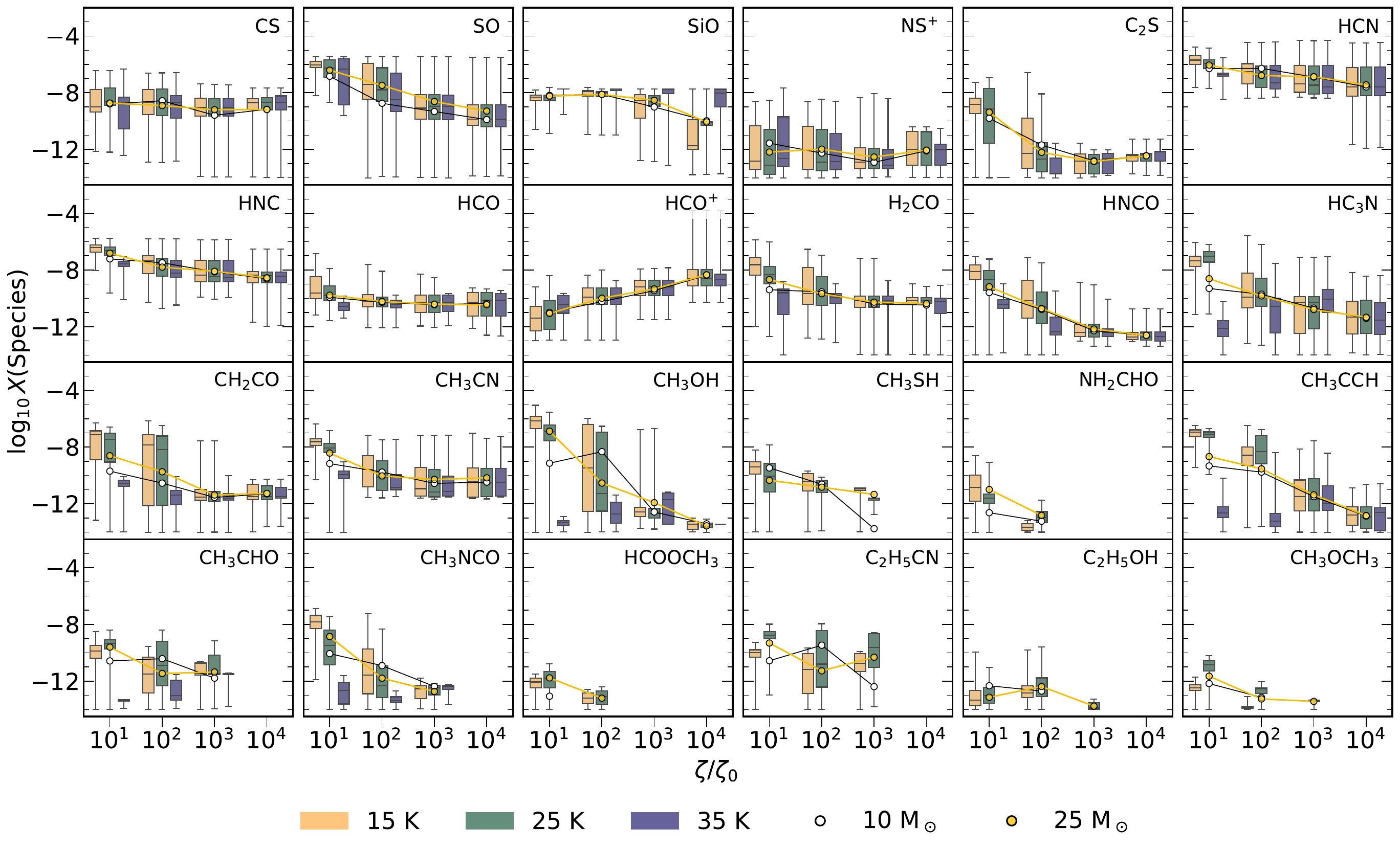}
     \caption{Box plots of the fractional abundances of each species in the protostellar objects (fully warmed-up phase), as a function of the cosmic-ray ionization rate (x-axis) and initial temperature of gas and dust (indicated by different colors). The interquartile ranges (IQRs) represent the middle 50\% of the abundance distribution, while the whiskers extend to the minimum and maximum values, covering the full range of the data. Additionally, the evolution of mean abundances for different object masses is overplotted, with circles denoting the mean values for each mass. If no distribution is observed at a given $\zeta/\zeta_0$, it suggests the species remained below the observational threshold throughout.}\label{fig:CRIR_evolution_PO}
 \end{figure*}

Tracking the chemical evolution of species over time reveals differences in chemical timescales and highlights key points of differentiation. To ensure consistency across models and avoid issues arising from non-uniform time-stepping in the \texttt{UCLCHEM} models, we generated a logarithmically spaced time sequence for both shocks and protostellar objects. At each generated time step, we computed the minimum and maximum fractional abundances of each species for each of the objects, along with the mean fractional abundance over time and the corresponding 1$\sigma$ confidence interval (defined as the 15.9 and 84.1 percentiles of the abundance distribution; see Fig. \ref{fig:time_evolution}) The evolution of surface and bulk species is presented in Sect. \ref{sect:surface-and-bulk}.

It is useful to begin by contrasting the key physical conditions in shocks and protostellar objects. Aside from the obvious density mismatch, shocks with velocities $>10\;\mathrm{km\,s}^{-1}$  exhibit a sharp temperature rise, with peak values exceeding 500~K. This surpasses the maximum temperatures found in the protostellar objects considered in this study and occurs over a very short timescale. As a result, all shocks examined here, except for the lowest velocity case at $10\;\mathrm{km\,s}^{-1}$, reach significantly higher temperatures than protostellar sources at some stage of their evolution. As the temperature rises, it eventually decreases after reaching a peak. This occurs in the relaxation zone of the shock, before the gas transitions into the post-shocked stage, where the temperature returns to its pre-shock value. In contrast, protostellar objects experience a gradual temperature increase followed by a plateau once the final value is reached. That said, there are intervals during the shock evolution where the temperature passes through values comparable to those in protostellar sources. However, they are relatively short-lived. 

Both the average temperature profiles (Fig. \ref{fig:temp_evolution}) and chemical evolution (Fig. \ref{fig:time_evolution}) indicate that shock signatures are primarily associated with timescales $\lesssim10^4$ years. Additionally, the denser the medium the shorter the shock timescale, as shown by the three peaks in the shock temperature in Fig. \ref{fig:temp_evolution}. The velocity of the shock determines the maximum temperature, with higher velocities producing greater temperature jumps. In our models, the highest temperature of 3941 K occurs in a 40~km$\,\mathrm{s}^{-1}$ shock propagating through a medium with a pre-shock density of $10^6\mathrm{cm}^{-3}$. 

In contrast to the rapid temperature changes in shocks, the warming of protostellar objects occurs over significantly longer timescales, typically exceeding $10^4$ years. The core mass plays a crucial role: a more massive 25 M$_\odot$ core warms up at least twice as fast as a lower-mass core at lower temperatures and up to five times faster in the highest temperature case. Thus, for the protostellar object defined in Tab. \ref{tab:UCLCHEM}, the timescales for reaching the final core temperature range from $4.7\times10^4$ to $4.9\times10^5$ years for a 10 M$_\odot$ core and from $2.2\times10^4$ to $9.9\times10^4$ years for a 25 M$_\odot$ core. 

Even though the timing of temperature conditions is crucial for understanding the differences between the two types of processes, shocks also induce sputtering. This process depends on the average energy transferred on grains via collisions with the shocked material. However, it occurs so rapidly that it is effectively instantaneous, quickly releasing all ices into the gas phase for the gas densities and shock velocities explored in our modeling (for details, see \citealt{jimenez-serra2008}).

\subsubsection{Evolution in protostellar objects}
For most species, the contribution from the protostellar object is easily distinguishable, as it aligns with the warm-up timescales and the final temperature plateau (see solid and dashed orange lines and orange-shaded areas in Fig. \ref{fig:time_evolution}). In most cases, this manifests as an increase in fractional abundance at timescales corresponding to the cores reaching their final temperatures.

However, for certain species, instead of a simple abundance jump, a significant narrowing of the abundance spread occurs, creating an almost tilted hourglass shape in the abundance evolution. The narrowest point corresponds to the phase when the cores have just reached their maximum temperature, followed by a widening of the abundance distribution. Species exhibiting this behavior include CS, HCN, HNC, H$_2$CO, and CH$_3$CN. In all cases except for H$_2$CO, the mean abundance does not decrease after the narrowing. Additionally, two species -- C$_2$H$_5$OH and CH$_3$OCH$_3$ -- show a steady increase in abundance over time, but instead of stabilizing, their abundances drop sharply after $10^4$ years. A similar rapid decline beyond $10^4-10^5$ years is observed for several other species, including C$_2$S, CH$_3$OH, CH$_3$SH, NH$_2$CHO, CH$_3$CHO, CH$_3$NCO, HCOOCH$_3$, and C$_2$H$_5$CN. Among these, the abundances of CH$_3$OH, CH$_3$SH, and CH$_3$NCO remain relatively stable before $10^4$ years, then rise sharply by several orders of magnitude, before declining again after $10^5$ years.

Across all the models of protostellar objects, the highest abundance was recorded for HCO$^+$ in a protostellar object with a density of $10^6$ cm$^{-3}$, a temperature of 450~K, and a mass of 10 M$_\odot$, where the natal cloud has a temperature of 35~K. This occurred in a setup with $\zeta$ of the order of $10^{-13}$ s$^{-1}$ and $G_0 = 10^3$ Habing during both the collapse and protostellar heating stages. Notably, the species that most consistently maintain the highest abundance in the models of protostellar objects, regardless of core mass, at the fully warmed-up protostellar object stage is HCN. This is followed by SO and HCO$^+$. The consistent high abundance for HCN and HCO$^+$ may indeed be what makes these two species so consistently abundant in energetic galaxies such as starbursts and AGN-dominated galaxies \citep[e.g.,][]{izumi2013,izumi2016,butterworth2022,nishimura2024}. 

However, we point out that the extremely high HCO$^+$ abundances reported in this work are also influenced by updates in UMIST 2022 \citep{umist2022} relative to UMIST 2012 \citep{umist2012}. When using the same modeling setup that produces the highest $X(\mathrm{HCO}^+)$ in this study, but with UMIST 2012, the resulting abundance is nearly three orders of magnitude lower. This may help to reconcile discrepancies between observations and models of HCO$^+$ \citep[e.g.,][]{viti2002,rawlings2004}, where sustaining high HCO$^+$ abundances required invoking extreme chemistry or periodic chemical cycles. Nevertheless, a detailed analysis of the chemistry of HCO$^+$ is beyond the scope of this paper and will be addressed in a forthcoming study.

\subsubsection{Evolution in shocks}

Just as in protostellar objects, we observe an increase in abundance occurring at a characteristic timescale, which in the case of shocks is $\lesssim10^4$ years (see solid and dashed gray lines and gray-shaded areas in Fig. \ref{fig:time_evolution}). The rise in temperature and density begins almost immediately, leading to abundance enhancements within the first years. However, in the post-shock stage, abundances generally decline steadily, as expected. Once the shock has passed, gas and dust cool back to their initial temperatures, while the density increases, promoting the freeze-out of species onto the icy mantles of dust grains.

Only two species -- formamide (NH$_2$CHO) and methyl formate (HCOOCH$_3$) -- exhibit relatively minor enhancements in shock models. This appears to be primarily related to their initial reservoirs in dust and the gas-phase conditions. The temperature rise in shocks is more abrupt and short-lived compared to protostellar heating. Combined with the near-instantaneous sputtering of icy mantles, this limits the formation of COMs on grain surfaces, which can only resume once freeze-out becomes efficient again. However, efficient freeze-out occurs only in the post-shock phase, where no significant desorption mechanisms exist. We discuss these low abundances in Sect. \ref{sect:low-COMs}, where we further explore the chemical mechanisms at play.

In shock models, the highest recorded abundance also belongs to HCO$^+$, occurring in a shock propagating at 15 km$\,\mathrm{s}^{-1}$ through a medium with an initial temperature of 20~K and density of $n_\mathrm{H}= 10^4$ cm$^{-3}$. At peak abundance, the gas and dust reached 298 K, with a density of $2.8 \times 10^4$ cm$^{-3}$. This setup featured a cosmic-ray ionization rate of $\zeta \sim 10^{-13}$ s$^{-1}$ and an interstellar radiation field of $G_0 = 10^1$ Habing, maintained throughout both the collapse and shock stages.

Again, as in the models of protostellar objects, the species exhibiting the highest peak abundance at any point in time is not necessarily the one that maintains the highest abundance consistently. In shocks, methanol (CH$_3$OH) dominates the abundance distribution, followed by HCN and SiO. In the post-shock phase, SiO takes over as the most abundant species, followed by HCN and HCO$^+$.

\begin{figure*}[t]
     \centering
     \includegraphics[width=1\linewidth]{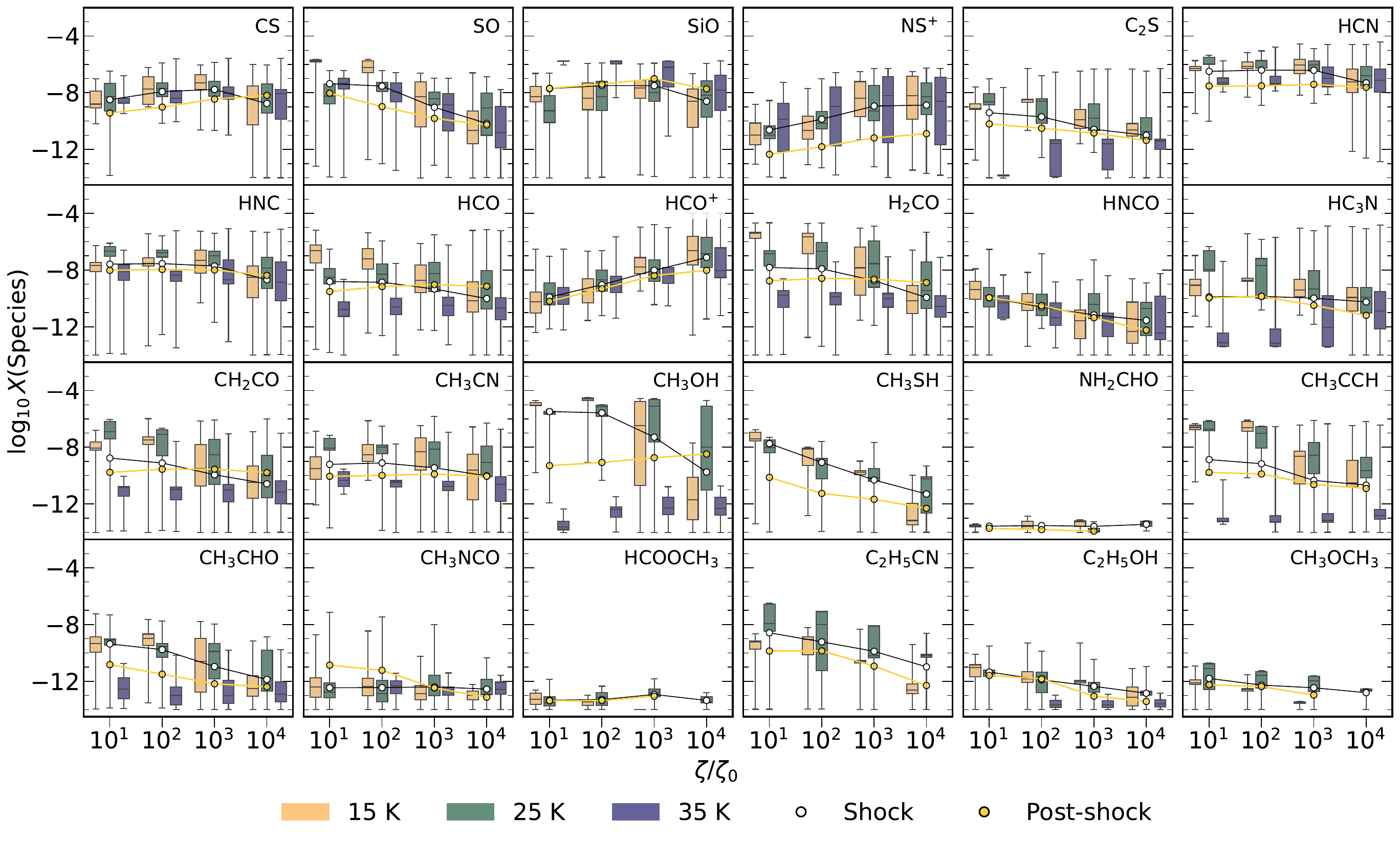}
     \caption{As in Fig. \ref{fig:CRIR_evolution_PO}, but for shock models. Here, the mean abundance evolution is calculated for different shock sub-stages, i.e., shock and post-shock.}\label{fig:CRIR_evolution_shocks}
 \end{figure*}

\subsection{Distribution of fractional abundances}\label{sect:distributions}

Another way to understand chemical trends and predict the behavior of each species is by closely examining the distribution of abundances as a function of varying physical parameters. We focused on the most influential parameters that drive significant changes in species behavior and, therefore, excluded the effects of $G_0$ from sub-distributions due to its negligible impact in stage 2 models with high extinction ($A_\mathrm{V}\gg10$). Given the multidimensional nature of the data, we structured the distributions into logically related subgroups. As always, our analysis is based on the observable part of the models.

First, we examined the distributions of abundances as a function of cosmic-ray ionization rates and initial temperatures of the medium (Sect. \ref{sect:crir_Tinit}). For models of protostellar objects, we considered only the fully warmed-up protostellar-object stage and analyzed variations across different masses (see Fig. \ref{fig:CRIR_evolution_PO}). In shock models, we instead examined variations based on substage, shock and post-shock, since they differ in timescales and physical conditions (see Fig. \ref{fig:CRIR_evolution_shocks}).

\begin{figure*}[t!]
     \centering
     \includegraphics[width=1\linewidth]{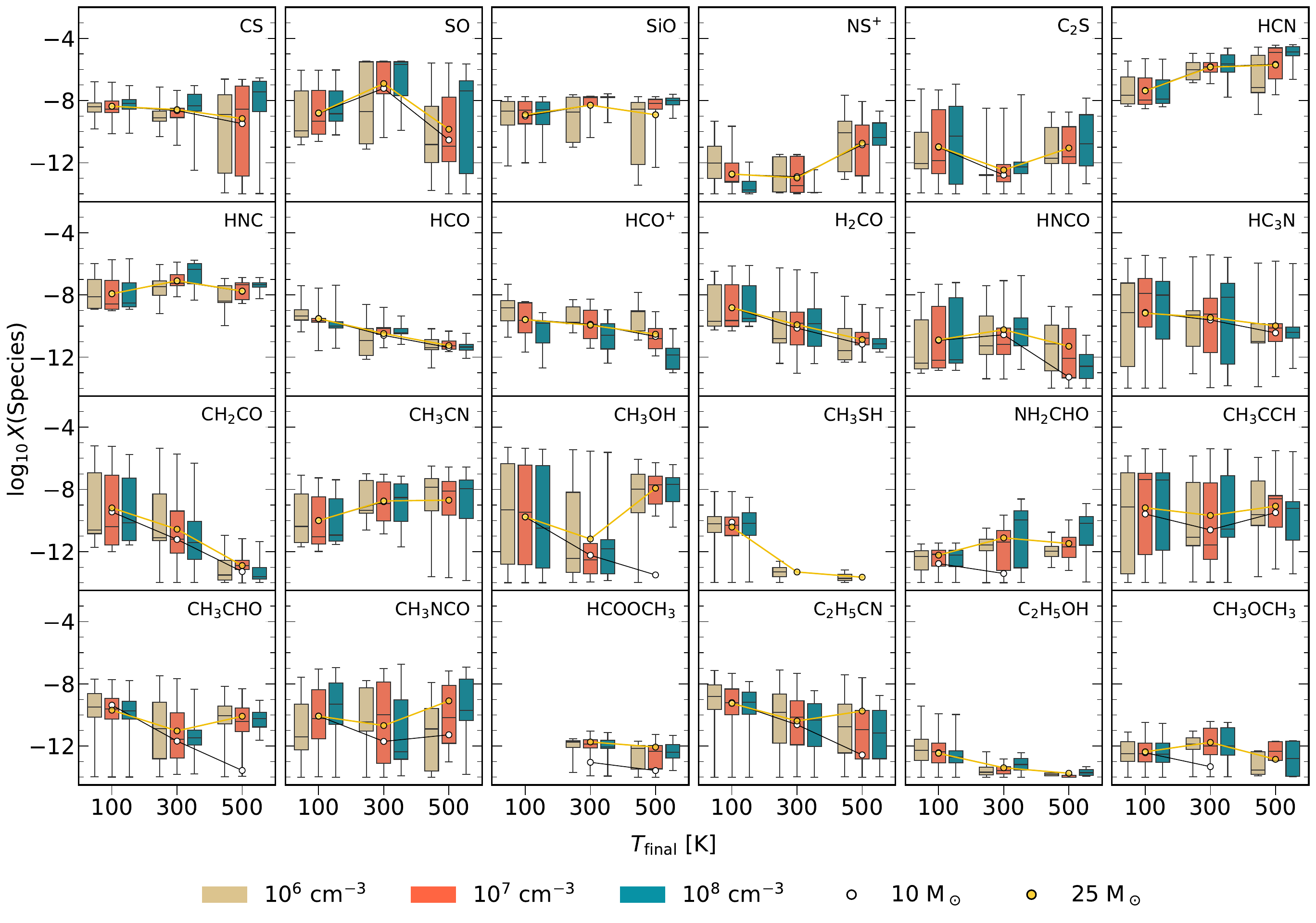}
     \caption{As in Fig. \ref{fig:CRIR_evolution_PO}, but with the mean abundance evolution as a function of final temperature and object density.}\label{fig:Tfinal_initialDens}
 \end{figure*}

Second, we analyzed the co-dependence of model-specific properties and medium densities. For protostellar objects, we focused on three final temperature groups: 100, 300, and 500~K (Sect. \ref{sect:temp_dens}, Fig. \ref{fig:VS_initialDens}). For shocks, we explored distributions at velocities of 10, 20, 30, and 40~km$\,\mathrm{s}^{-1}$ (Sect. \ref{sect:vel_dens}, Fig. \ref{fig:Tfinal_initialDens}). As in the previous distributions, we further examined trends within individual protostellar object masses and across shock and post-shock sub-stages. 

\subsubsection{Cosmic rays and initial temperature of the medium}\label{sect:crir_Tinit}

The intensity of the cosmic ray ionization flux can strongly impact the abundances of the modeled species, to the extent that some COMs in the models of protostellar objects cannot withstand certain ionization rates and drop below the detectable threshold (see Fig. \ref{fig:CRIR_evolution_PO}). In case of protostellar objects there are species which appear to favor only lower ionization levels, i.e., NH$_2$CHO and HCOOCH$_3$ are exclusively associated with regions where $\zeta<1.31\times10^{-14}\;\mathrm{s}^{-1}$. Other species can withstand higher levels of ionization, but CH$_3$SH, CH$_3$CHO, CH$_3$NCO, C$_2$H$_5$CN, C$_2$H$_5$OH, and CH$_3$OCH$_3$ will eventually fall below the detectability threshold in all protostellar object models once $\zeta$ reaches $1.31\times10^{-13}\;\mathrm{s}^{-1}$. 

In shock models, the behavior is more complex (see Fig. \ref{fig:CRIR_evolution_shocks}), with detectability strongly depending on both temperature and evolutionary stage. However, it is worth noting that there is no species which will be completely below the detectability threshold at any considered ionization rate. While all the simpler species will remain detectable no matter the temperature or the stage, the differences emerge among COMs.

Broadly, two main trends emerge when we consider the temperature effects. Firstly, the enhancement of CH$_3$SH and C$_2$H$_5$CN is most apparent at temperatures $\leq30$~K. Secondly, NH$_2$CHO, HCOOCH$_3$, and CH$_3$OCH$_3$ show a similar behavior, with their strongest enhancement occurring up to $\leq25$~K, with the exception of the $15$~K pre-shocked medium at the highest ionization level, at which point they become undetectable. For the remaining species, the temperature effects are closely linked to the evolutionary stage, i.e., whether the gas is in a shocked or post-shocked stage.

This dependence on evolutionary stage is well reflected in the detectability of certain species. A notable example is C$_2$H$_5$OH, which appears detectable across all stages; however, at the highest $\zeta$ it remains detectable in both shock and post-shock only if $T=20-30$~K. Moreover, for $\zeta=1.31\times(10^{-14}$--$10^{-15})\;\mathrm{s}^{-1}$, if the initial temperature is 35~K, significant enhancement occurs only during the shock phase.

Meanwhile, NH$_2$CHO, HCOOCH$_3$, and CH$_3$OCH$_3$ reach detectable levels only in the shock phase at the highest ionization level, provided that the temperature is $T=20-25$~K. Additionally, HCOOCH$_3$ and CH$_3$OCH$_3$ remain undetectable in the post-shock stage at $\zeta=1.31\times10^{-14}\;\mathrm{s}^{-1}$ and $T=15$~K.

More broadly, cosmic rays tend to have a detrimental effect on the chemistry, a trend that is particularly pronounced in the models of protostellar objects. HCO$^+$ is the only species whose abundance consistently increases with CRIR, both in shocks and in protostellar environments. NS$^+$, by contrast, shows a more environment-dependent behavior: its abundance is enhanced with increasing CRIR in shocks, but remains largely unaffected in protostellar objects.
 
\subsubsection{Temperature and density of protostellar objects}\label{sect:temp_dens}

  \begin{figure*}[t!]
     \centering
     \includegraphics[width=1\linewidth]{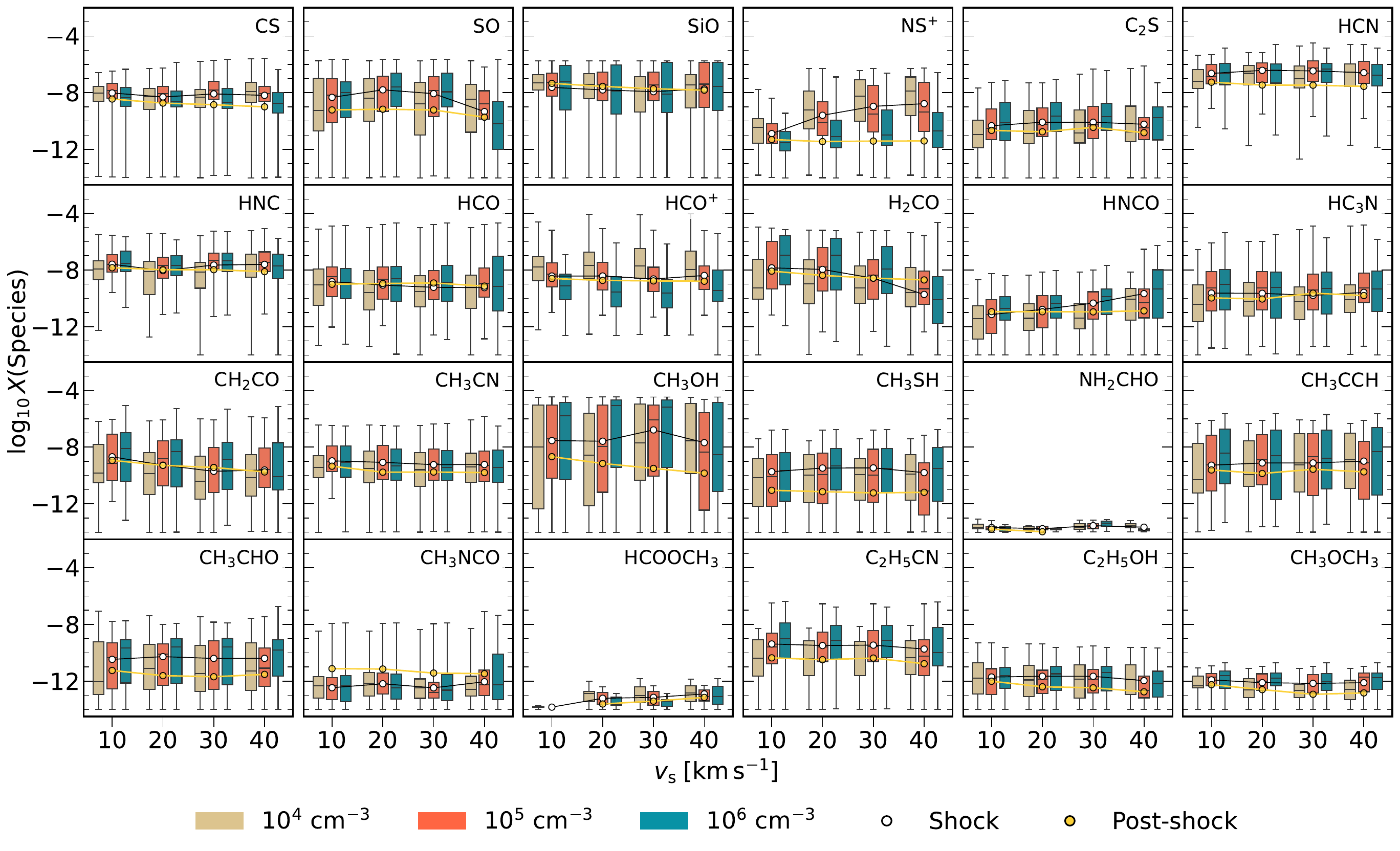}
     \caption{As in Fig. \ref{fig:CRIR_evolution_shocks}, but with the mean abundance evolution as a function of shock velocity and cloud density.}\label{fig:VS_initialDens}
 \end{figure*}

The key physical parameters in our protostellar object models are core mass, density, and temperature, with the latter playing the most prominent role. Understanding how these parameters influence molecular abundances under CMZ-like conditions is essential for interpreting the underlying chemistry in such environments. By examining abundance distributions across a range of temperatures and densities, and by analyzing how they vary with core mass, we aim to identify characteristic trends in the molecular behavior of protostellar objects.

The primary effect of core mass is the timescale of the warm-up phase, which progresses significantly faster in more massive cores. While lower masses generally result in lower abundances, the impact is relatively minimal for most modeled species. However, some exhibit strong deviations (see Fig. \ref{fig:Tfinal_initialDens}), particularly at temperatures above 300~K, where mean abundance trends deteriorate significantly. In the case of CH$_3$OH, differences can reach up to five orders of magnitude at the highest temperatures. Other species, including HNCO, CH$_3$CHO, CH$_3$NCO, and C$_2$H$_5$CN, also show notable abundance differences under these conditions. Species such as NH$_2$CHO and CH$_3$OCH$_3$, however, begin to decline significantly starting from the lowest temperature in the lower-mass core. Eventually, CH$_3$OCH$_3$ drops below detectability in models with $T_\mathrm{final} \geq 400$~K, whereas NH$_2$CHO becomes unobservable in all cases at 500~K.

Two other molecules strongly influenced by core mass are HCOOCH$_3$ and CH$_3$SH. For HCOOCH$_3$, the models predict consistently higher abundances across the observable range in the higher-mass core. Additionally, across densities, HCOOCH$_3$ tends to be enhanced only at temperatures of $T > 250$~K. In contrast, CH$_3$SH remains detectable at all temperatures in the higher-mass core, but only appears at 100~K in the lower-mass core. Furthermore, in high-density objects ($n_\mathrm{H} > 10^7\,\mathrm{cm}^{-3}$), the enhancement of CH$_3$SH appears linked to lower temperatures, at $T < 250$~K, indicating its sensitivity to both density and temperature.

Overall, species such as CS, HCO, HCO$^+$, H$_2$CO, HC$_3$N, CH$_2$CO, CH$_3$SH, and C$_2$H$_5$OH decrease in mean abundance with increasing final temperature across both core masses. Conversely, HCN and CH$_3$CN tend to increase in abundance as the final temperature rises. The remaining species follow non-monotonic trends or show behavior dependent on core mass, increasing in abundance up to a certain temperature before declining. As shown in Fig.~\ref{fig:Tfinal_initialDens}, this turnover typically occurs around 300~K.

\subsubsection{Shock velocity and density of the pre-shock gas}\label{sect:vel_dens}

As previously mentioned, shock velocity, along with the density of the ambient medium, determines the maximum temperature reached during a shock event. Consequently, increasing shock velocities influence the molecules not only through sputtering but also via temperature effects. The maximum temperatures reached in shocks with velocities of 10, 20, 30, and 40~km$\,\mathrm{s}^{-1}$ are 323~K, 916~K, 2137~K, and 3941~K, respectively. 

Interestingly, despite the substantial differences introduced by shock velocities, the overall trends (see Fig.~\ref{fig:VS_initialDens}) are more stable than those observed in models of protostellar objects. However, a clear divergence emerges at the density level. As expected, the two ions considered in this study (namely, HCO$^+$ and NS$^+$) follow a general trend of decreasing abundance with increasing density, whereas neutral species typically show the opposite behavior. A more detailed examination reveals a nuanced picture: species such as SO, C$_2$S, H$_2$CO, CH$_3$OH, CH$_3$SH, CH$_3$CHO, and C$_2$H$_5$CN predominantly get enhanced at higher densities, but this pattern breaks down for shocks with velocities exceeding 30~km\,s$^{-1}$. Other species, including CS, HCN, and CH$_3$CN, tend to have abundances that benefit most from intermediate densities across most velocity cases. Finally, a small subset of species, such as HNC, NH$_2$CHO, and CH$_3$NCO, do not fit neatly into either category, displaying more complex behavior.

Shock velocity typically introduces non-monotonic effects on abundance distributions for most of the species. However, some species exhibit relatively steady trends with changing shock velocity during the shock phase. Among those that show a general enhancement are NS$^+$, HNCO, and HCOOCH$_3$, while H$_2$CO and CH$_2$CO tend to decline. Moreover, at the post-shock stage, the mean abundances of most species show little evolution with increasing shock velocity, including those that were enhanced at higher velocities. There are exceptions: CH$_2$CO continues to follow its shock-phase trend, while CH$_3$OH and C$_2$H$_5$OH experience an almost steady decrease. Additionally, the post-shock stage, where the medium cools down and freeze-out becomes efficient, generally results in lower average abundances, except for CH$_3$NCO, which stands out as an exception. 

Similar to the case of protostellar objects, most species remain observable at some level across the shock velocity–density parameter space. However, NH$_2$CHO and HCOOCH$_3$ make exceptions. NH$_2$CHO is detectable across all densities and stages only for shock velocities below $20\,\mathrm{km\,s}^{-1}$. Above this threshold, its behavior becomes more complex: NH$_2$CHO is sufficiently enhanced primarily during the shock stage in most configurations, but at the highest density and velocities exceeding $35\,\mathrm{km\,s}^{-1}$, it becomes fully undetectable. Taken together, these physical conditions suggest that NH$_2$CHO may favor low-velocity shocks in lower-density environments.

For HCOOCH$_3$, the abundance appears to be primarily governed by the temperature attained in shocks, consistent with results from the protostellar object models. At $10\,\mathrm{km\,s}^{-1}$ and densities $\gtrsim 10^5\,\mathrm{cm}^{-3}$, it is completely undetectable. In the post-shock stage, HCOOCH$_3$ remains undetectable in the low-density case, as well as at mid-to-high densities when subjected to $15\,\mathrm{km\,s}^{-1}$ shocks.

\section{Discussion}
\label{sect:discussion}

Although this study primarily provides an overview of chemical templates, it remains essential to identify which species emerge as the most promising tracers of the modeled environments. To do this, we have selected molecules that are both abundant and exhibit significant variation across different gas types. We discuss our approach in detail and present the most viable tracers in Sect.~\ref{sect:tracers}. We also propose several molecular ratios in Sect. \ref{sect:ratios}. Then in Sect.~\ref{sect:observations}, we compare abundances of selected species with observational data to evaluate how well the models capture CMZ-like conditions. Finally, we address the relatively low abundances predicted for some COMs in shock models in Sect.~\ref{sect:low-COMs}, and explore scenarios that could lead to their enhancement.

\subsection{Most Viable Tracers}\label{sect:tracers}

Distinguishing the chemical signatures of non-shocked gas in protostellar objects from shock-influenced gas is essential for understanding the nature of the observed medium. To achieve this, we examined the maximum abundances of all species across different gas environments associated with protostellar objects and shocks to assess their detectability limits under varying ionization rates. We identified species that exhibit significant differences between these environments, defined as a difference of at least one order of magnitude between shocks and fully warmed-up protostellar objects. We also highlighted species capable of reaching high abundances in specific ionization regimes ($X(\mathrm{Species}) > 10^{-7}$). We discuss this further in Sect.~\ref{sect:CRIR-upper-limit}.

Following a similar approach to our cosmic-ray ionization rate analysis, we examined species abundances in shocked gas and fully warmed-up protostellar objects across temperatures of 100, 200, 300, 400, and 500~K, focusing on trends within the overlapping temperature range (Sect. \ref{sect:loc-temp}). We also analyzed highly abundant species at each pre-shock and protostellar object density, identifying those expected to be prominent under specific conditions (Sect. \ref{sect:density}).

\subsubsection{Tracers of cosmic-ray ionization rate}
\label{sect:CRIR-upper-limit}

Our analysis reveals that cosmic-ray ionization rate variations create distinct molecular abundance patterns that differ systematically between the studied environments. As shown in Sect.~\ref{sect:crir_Tinit}, most species exhibit decreasing abundances with increasing $\zeta$, but their responses vary significantly between the studied environments. We identify species that show particularly consistent ionization trends -- HCO$^+$, H$_2$CO, and CH$_3$SH -- making them promising tracers for constraining cosmic ray ionization rates. In the following analysis, we categorize molecular species based on their ionization sensitivity and detectability patterns across shock-driven environments and those associated with protostellar objects.

\begin{figure*}[!t]
    \centering
    \includegraphics[width=1\linewidth]{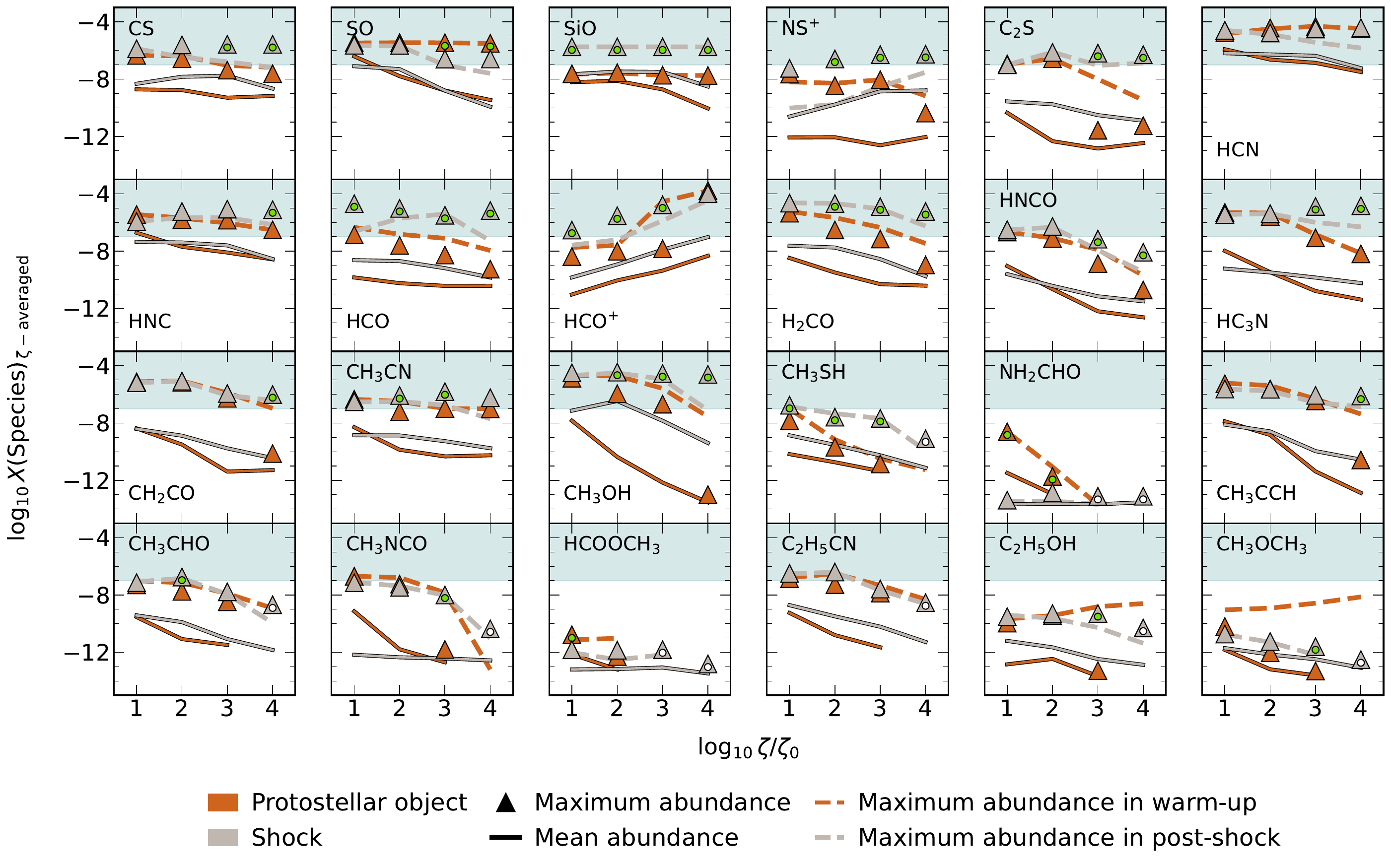}
    \caption{Maximum and mean abundances of species at different evolutionary stages across models with varying cosmic-ray ionization rates. Orange and gray triangles show maximum abundances in fully warmed-up protostellar objects and shocks, respectively, while lines of corresponding colors represent mean abundances. Dashed lines indicate maximum abundances during the warm-up stage of protostellar objects (orange) and the post-shock stage (gray). Green circles mark species where the maximum abundance differs by at least one order of magnitude between shock and fully warmed-up protostellar objects, indicating potential effectiveness as tracers of cosmic-ray ionization rate in the respective environments. White circles identify cases where one model type produces detectable abundances while the other falls below detection limits. Gray shading highlights the abundance region where species are considered highly abundant ($X$(Species) $> 10^{-7}$) under a given ionization rate. For detailed explanation of each evolutionary stage, we refer the reader to Sect. \ref{sect:results}.}
    \label{fig:abundance_vs_zeta}
\end{figure*}

In models of both protostellar objects and shocks, species that show at least an order of magnitude abundance difference between the two gas types at a given cosmic-ray ionization rate generally reach higher maximum abundances in shocked gas (Fig.~\ref{fig:abundance_vs_zeta}). This pattern is clearly seen for SiO, HCO, H$_2$CO, CH$_3$OH, and CH$_3$SH. We also find that COMs display a much broader spread of maximum abundances, while simpler species tend to concentrate around higher values (Fig.~\ref{fig:abundance_vs_zeta}). The more constrained abundance ranges of simpler species suggest they may serve as more reliable probes of the extreme physical conditions found in different Galactic Center environments.

The abundance patterns also vary between evolutionary stages within each environment type. Shocked gas typically exhibits similar or higher maximum abundances than post-shock gas, whereas the relationship between evolutionary stages in protostellar objects is less predictable. In protostellar models, the warm-up stage can exhibit higher maximum abundances than fully warmed-up objects, as clearly seen for CH$_3$OCH$_3$, which shows significantly higher abundances during warm-up at all ionization rates. This behavior likely reflects differences in chemical timescales and the degree of thermal processing between evolutionary stages.

Furthermore, all species considered in this study, except for HCN, exhibit a meaningful difference in their maximum abundance limit between the two environments for at least one ionization rate. The highest number of species showing significant differences occurs at $\zeta \gtrsim 10^{-14}\:\mathrm{s}^{-1}$. However, the difference is sometimes driven by one species falling below the detectability threshold rather than a direct abundance comparison, as can be seen for the majority of COMs (cases denoted with white dots in Fig. \ref{fig:abundance_vs_zeta}).

Only two species exhibit significant differences between shocks and protostellar objects in their maximum abundance limits across the whole ionization range, with higher values achievable in shocks: SiO and HCO. Of these two species, SiO shows no significant variation in maximum abundances across different ionization rates. Only at the highest $\zeta$ its mean abundance decrease by approximately one order of magnitude. This relative insensitivity to ionization rate variations explains why the SiO abundance remain relatively constant across Giant Molecular Clouds (GMCs) in the Galactic Center, independent of their location within the CMZ \citep[see][]{martin-pintado1997}.

On the other hand, HCO shows a more pronounced sensitivity to ionization levels. While it remains highly abundant in shocks across all $\zeta$ values, suggesting that shock-driven processes sustain its abundance even under strong ionization, its abundance in protostellar objects drops sharply with increasing CRIR. Specifically, the maximum abundance declines from $1.35\times10^{-7}$ at $\zeta=10^{-16}\:\mathrm{s}^{-1}$ to $5.46\times10^{-10}$ at $\zeta=10^{-13}\:\mathrm{s}^{-1}$, indicating a nearly three-order-of-magnitude drop and a strong dependence on ionization. This suggests that HCO may help constrain the CRIR in a given medium. However, its diagnostic value is complicated by the fact that HCO is also a well-known tracer of PDRs \citep[see][]{schenewerk1988, gerin2009}. Additionally, as shown by \citet{armijos-abendano2020}, both elevated CRIRs and shocks appear necessary to reproduce the observed HCO abundances in the Sgr B2 molecular cloud.

To systematically categorize these diverse behaviors, we identify four distinct groups based on their ionization trends and detectability limits across different environments. When species exhibit intermediate behavior and defy easy classification, we assign them to multiple groups. We summarize these classifications in Tab.~\ref{tab:species_classification}.

The first group, showing minimal dependence on ionization across all environments. This group consists of CS, SO, SiO, HCN, HNC, and CH$_3$CN. The stable behavior of these species makes them ideal as reference standards for normalizing molecular ratios, as their abundances remain largely independent of ionization conditions. 

The second group shows relatively systematic ionization trends across all environments, making them potentially robust tracers of cosmic-ray intensity, and includes HCO$^+$, H$_2$CO, and  and CH$_3$SH. We find that H$_2$CO and CH$_3$SH undergo systemic decrease of abundances as ionization increases in every environment, while HCO$^+$ exhibits a systematic increase. To further evaluate whether these species can serve as diagnostics of CRIRs, we calculated their relative mean abundance ratios, which we discuss in detail in Sect. \ref{sect:ratios}. 

The third group consists of HCO, HNCO, and CH$_3$CHO. While all of these species show consistent trends in protostellar objects-related gas, shocked and post-shocked maxima and means are nearly consistent, as they appear to undergo enhancements at some point once $\zeta>10^{-16}\:\mathrm{s}^{-1}$, which breaks that consistency. However, this group consists of species potentially prone to being tracers of ionization rate in protostellar objects.

The fourth group exhibits selective sensitivity to high cosmic-ray ionization rates in protostellar environments while showing consistent trends in shocked gas. Within this group, we distinguish between species based on their detectability behavior. Some species show significant abundance changes in protostellar environments at $\zeta \geq 10^{-14}\:\mathrm{s}^{-1}$ but remain detectable: NS$^+$, C$_2$S, HC$_3$N, CH$_2$CO, CH$_3$OH, and CH$_3$CCH. Among these, C$_2$S and HC$_3$N show abundance drops already at $\zeta = 10^{-14}\,\mathrm{s}^{-1}$. Other species in this group experience such severe abundance declines in high-ionization protostellar environments that they fall below detectability thresholds, while remaining detectable in shocked gas: CH$_3$SH, NH$_2$CHO, CH$_3$CHO, CH$_3$NCO, HCOOCH$_3$, C$_2$H$_5$CN, C$_2$H$_5$OH, and CH$_3$OCH$_3$. Among these, NH$_2$CHO and HCOOCH$_3$ become undetectable in protostellar objects once $\zeta \gtrsim 10^{-14}\,\mathrm{s}^{-1}$, while NH$_2$CHO, HCOOCH$_3$, and CH$_3$OCH$_3$ become undetectable in post-shocked gas at $\zeta = 10^{-13}\,\mathrm{s}^{-1}$. The contrasting behavior between protostellar and shocked environments makes these species valuable discriminators of heating mechanisms.

Several species exhibit nuanced behavior that places them at their group boundaries. CS remains virtually unchanged in shocks but shows moderate changes in protostellar objects, positioning it between groups one and two. SO displays stable maximum abundances but declining mean values, suggesting behavior intermediate between different classification criteria.

\begin{table*}[!t]
    \centering
    \caption{Classification of molecular species based on their response to cosmic-ray ionization rates, temperatures, and densities.}
    \begin{tabular}{p{6cm}p{5.5cm}p{5cm}}
    \hline
    \hline
    \textbf{Species group} & \textbf{Abundance Behavior} & \textbf{Diagnostic Use} \\
    \hline
    \addlinespace[0.2cm]
    CS, SO, SiO, HCN, HNC, CH$_3$CN & Minimal variations across $\zeta$ in all environments & Potentially can act as normalizers for abundance ratios \\
    \addlinespace[0.2cm]
    HCO$^+$, H$_2$CO, CH$_3$SH & Systematic changes as function of $\zeta$ across all environments; HCO$^+$ increases, others decrease with $\zeta$  & Best suited as $\zeta$ tracers \\
    \addlinespace[0.2cm]
    HCO, HNCO, CH$_3$CHO & Systematic, decreasing trends in protostellar objects as function of $\zeta$ &  Potentially useful as $\zeta$ tracers in protostellar environments\\
    \addlinespace[0.2cm]
    NS$^+$, C$_2$S, HC$_3$N, CH$_2$CO, CH$_3$OH, CH$_3$SH$^\star$, NH$_2$CHO$^\star$, CH$_3$CCH, CH$_3$CHO$^\star$, CH$_3$NCO$^\star$, HCOOCH$_3$$^\star$, C$_2$H$_5$CN$^\star$, C$_2$H$_5$OH$^\star$, CH$_3$OCH$_3$$^\star$ & Significant abundance changes in protostellar objects at $\zeta \gtrsim 10^{-14}$ s$^{-1}$; consistent behavior in shocks & Differentiate shocked vs. protostellar environments at high $\zeta$ \\
    \addlinespace[0.2cm]
    NS$^+$, CH$_2$CO, CH$_3$OH, CH$_3$SH, NH$_2$CHO HCOOCH$_3$ & $\sim300$~K is a turning point - more pronounced difference between protostellar objects and shocks or behavioral patterns shifts & Selective sensitivity to temperature \\
    \addlinespace[0.2cm]
    HCO, HCO$^+$, CH$_3$SH, CH$_3$NCO, HCOOCH$_3$ & Consistent trends across a broad range of temperatures discriminating the gas types & Differentiate shocked vs. protostellar environments based on temperature \\
    \addlinespace[0.2cm]
    NS$^+$, CH$_3$SH, CH$_3$CHO, CH$_3$NCO & Abundance always $<10^{-7}$: NS$^+$ in protostellar objects and post-shocked gas; CH$_3$SH, CH$_3$CHO in protostellar objects; CH$_3$NCO in shocks & Environmental abundance constraints \\  
    \hline
    \label{tab:species_classification}
    \end{tabular}
    \tablefoot{$^\star$Species, for which the significant abundance changes lead to falling below the detectability threshold ($X<10^{-14}$). We discuss this in more detail in Sect. \ref{sect:CRIR-upper-limit}.}
\end{table*}

\subsubsection{Temperature as a tracer distinguishing protostellar objects and shocks}
\label{sect:loc-temp}

\begin{figure*}[t!]
     \centering
     \includegraphics[width=1\linewidth]{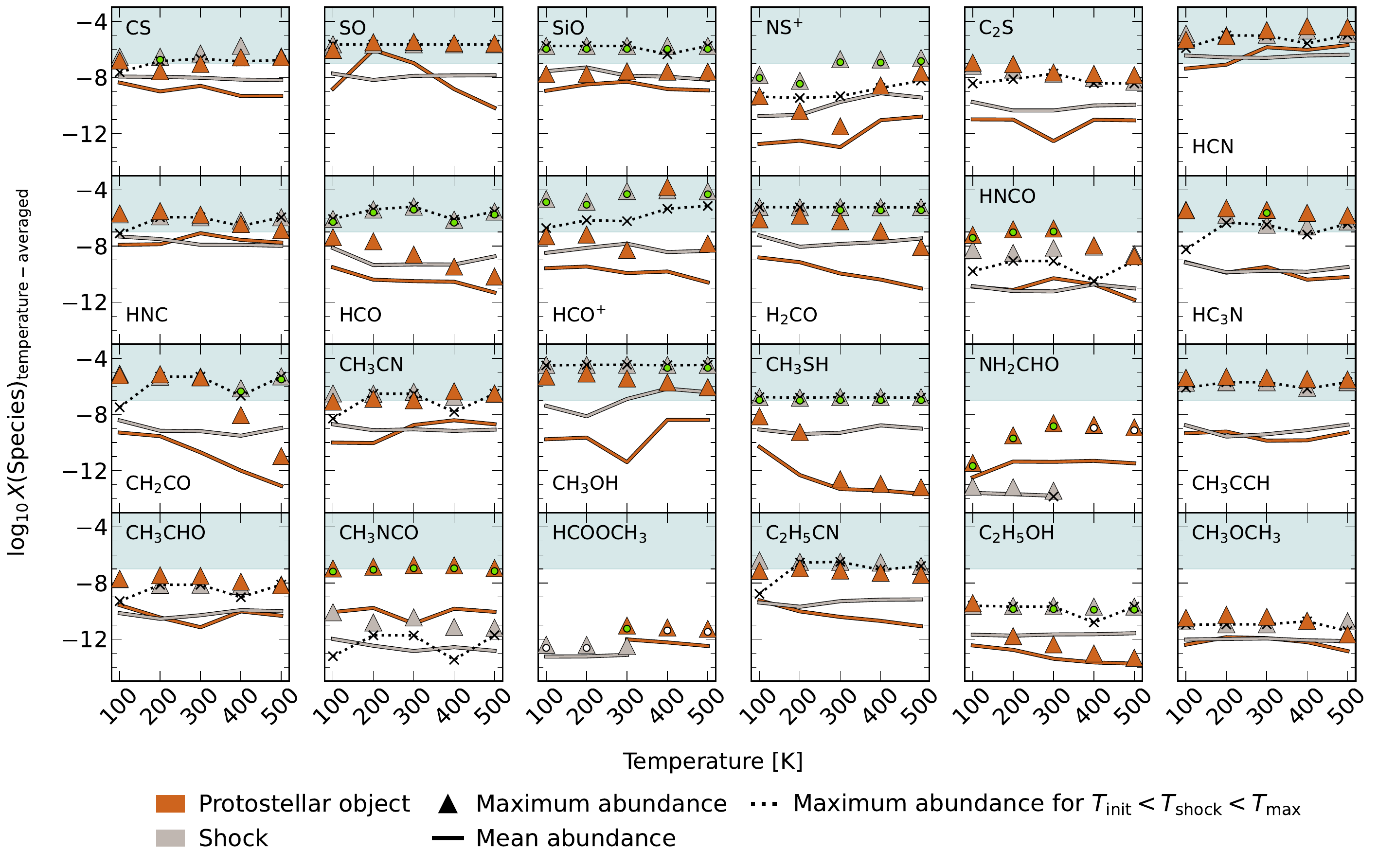}
     \caption{Maximum and mean abundances of species across different environment types as a function of temperature. Orange and gray triangles show maximum abundances in fully warmed-up protostellar objects and shocks, respectively, while solid lines of corresponding colors represent mean abundances. Dotted black lines indicate maximum abundances in the heating phase of shocks, before peak temperature is reached. Circles indicate species showing significant abundance differences between environments at a given temperature, following the same criteria as Fig.~\ref{fig:abundance_vs_zeta}. For shocks, temperatures are binned with $\pm0.5$~K tolerance around target values.}\label{fig:loc_temp}
 \end{figure*}

Temperature plays a crucial role in shaping molecular abundances, but to reliably distinguish between shock-driven processes and protostellar heating, comparisons must be made within a common temperature range. We focus on 100--500~K, where conditions in shocked gas and protostellar environments overlap, ensuring that observed abundance differences arise from underlying physical processes, such as dynamical timescales or desorption mechanisms. Our analysis reveals that maximum abundances in shocked gas remain relatively constant with temperature, consistent with the transient nature of shock heating that provides insufficient time for significant chemical evolution.

Because the temperature evolution differs between these environments, careful sampling within this range is necessary to enable meaningful comparisons. In fully warmed-up protostellar objects, temperatures remain constant within our target range. In contrast, shocks experience rapid and complex temperature changes that require binning. To achieve this, we selected model points within $\pm0.5$~K of each target temperature. The $\pm0.5$~K tolerance was chosen by examining the number of model points per bin to ensure adequate statistical sampling. We also tested wider tolerances and found no significant impact on the results, indicating that our conclusions are stable against reasonable variations in bin width. This fixed tolerance balances sampling accuracy with statistical robustness, avoids the uneven bin widths and potential biases introduced by percentage-based criteria, and ensures a consistent basis for comparing molecular abundances across both environments.

Figure~\ref{fig:loc_temp} shows that several species -- including NH$_2$CHO, CH$_3$CHO, HCOOCH$_3$, C$_2$H$_5$OH, and CH$_3$OCH$_3$ -- do not reach the high-abundance regime in either gas type. Among these, NH$_2$CHO is detectable in shocked gas only below 300~K. In contrast, HCOOCH$_3$ is detectable in shocked gas up to 300~K but becomes detectable in protostellar objects only above this temperature.

We also identify species for which distinguishing between shocked and protostellar gas based on abundance behavior is challenging. This group includes HCN, HNC, CH$_3$CN, CH$_3$CCH, CH$_3$CHO, and CH$_3$OCH$_3$, which show similar trends in both upper limits and mean abundances across the temperature range. Although HCN and HNC exhibit some variations, these are insufficient to draw firm conclusions.

Similarly, CS and HC$_3$N display significant differences in abundance in a single temperature bin but otherwise behave similarly across environments. SO, C$_2$S, and C$_2$H$_5$CN show comparable trends in maximum abundance but differ in mean abundances, suggesting modest environmental effects. Lastly, CH$_3$OH exhibits significantly higher mean abundances in shocked gas at 400 and 500~K; however, the differences in maximum abundance remain relatively modest compared to species like CH$_2$CO, which exhibit clearer environmental contrasts across this temperature range.

Some of these trends are consistent with findings from Sect. \ref{sect:CRIR-upper-limit}, where we showed that shocks tend to sustain higher abundances of certain species under broader conditions (e.g., NS$^+$, HCO, HCO$^+$, H$_2$CO, CH$_2$CO, CH$_3$OH, CH$_3$SH, and C$_2$H$_5$OH). For several molecules, 300~K appears to mark a turning point where abundance differences become more pronounced or behavioral patterns shift. This is true for NH$_2$CHO and HCOOCH$_3$, as well as for NS$^+$, CH$_2$CO, CH$_3$OH, and CH$_3$SH. In the case of NS$^+$, the maximum abundance difference at 300~K reaches nearly five orders of magnitude; CH$_3$OH shows a similarly large contrast in mean abundance. Except for NS$^+$ and the specific detectability patterns of NH$_2$CHO and HCOOCH$_3$, most other species begin to diverge meaningfully between shocked and protostellar gas above 300~K, offering a temperature regime where the heating mechanism becomes distinguishable.

Additionally, as discussed in Sect. \ref{sect:CRIR-upper-limit}, certain abundance effects become prominent only under specific conditions. At 400~K, for example, HCO$^+$ can reach significantly higher abundances in non-shocked gas than typically expected, but this strongly depends on ionization. Therefore, we emphasize that abundance trends should not be considered in isolation, as multiple interdependent factors shape the observed behavior.

Overall, we find that HCO, HCO$^+$, CH$_3$SH, CH$_3$NCO, and HCOOCH$_3$ are best-suited for discriminating between environments based on their consistent trends across the whole considered temperature range. We exclude NH$_2$CHO because its discrepancy between shocks and protostellar objects is primarily driven by very low abundances in shocks, a behavior we discuss further in Sect. \ref{sect:low-COMs}. We also summarize this in Tab. \ref{tab:species_classification}.

While our conclusions are based on the overall behavior of shocks, examining the individual shock phases, heating and cooling, reveals additional nuances worth considering. Shocks comprise two main stages: a heating phase, where the temperature rises, and a cooling phase, where it gradually declines. The cooling phase generally reflects the trends seen for shocks as a whole, with one notable exception: at 200~K, CH$_3$CHO shows a significantly higher maximum abundance in protostellar objects than in shocked gas.

In contrast, the heating phase can diverge substantially from the overall shock behavior. As shown by the dotted lines in Fig.\ref{fig:loc_temp}, the upper abundance limits in this phase are often significantly lower. This effect is particularly evident for species like NS$^+$ and CH$_3$NCO. Furthermore, if only the heating phase were considered, HCOOCH$_3$ would not be detected at all, and NH$_2$CHO would be detectable only at 300~K.

\subsubsection{Density of pre-shock medium and protostellar objects}
\label{sect:density}

A direct comparison between shocked and protostellar regions at matched densities is only feasible in one case ($n_\mathrm{H} = 10^6$~cm$^{-3}$), where we find no significant differences for most species. Despite this, it remains informative to examine how species behave across the full range of densities considered in each environment. Specifically, we focus on which species become highly abundant and whether notable differences in their detectability limits emerge. Since both environments were modeled at three densities (see Tab.~\ref{tab:UCLCHEM}), we refer to them throughout this section as low, mid, and high density cases.

At the overlapping density ($n_\mathrm{H} = 10^6$~cm$^{-3}$) only NS$^+$, HCO, HCO$^+$, HNCO, and CH$_3$SH show meaningful variations. NS$^+$ becomes highly abundant ($X > 10^{-7}$) only in shocked gas, while reaching significantly lower maxima in post-shock gas ($\sim10^{-10}$) and protostellar objects ($\sim10^{-8}$). HCO$^+$ shows the largest discrepancy, though this is partly influenced by ionization effects: it reaches abundances of $\sim10^{-4}$ in protostellar objects, $\sim3 \times 10^{-6}$ in shocks, and $\sim6 \times 10^{-8}$ in post-shock gas. HCO, on the other hand, is abundant across all environments, but peaks in shocked gas ($\sim2 \times 10^{-6}$) and is lowest in protostellar objects ($\sim10^{-7}$). Finally, HNCO and CH$_3$SH become highly abundant ($\sim10^{-7}$) in shocked and post-shock gas, but are an order of magnitude less abundant in protostellar environments.

Among the three gas types -- shocked, post-shocked, and non-shocked gas associated with protostellar objects -- we identify four species that never reach high abundances: NH$_2$CHO, HCOOCH$_3$, C$_2$H$_5$OH, and CH$_3$OCH$_3$. In contrast, NS$^+$ becomes highly abundant only in shocks, while SiO never achieves high abundance in protostellar objects. However, the behavior of SiO reflects our modeling assumptions, i.e., we varied the initial Si abundance to accurately represent sputtering effects in shocks (see Tab.~\ref{tab:UCLCHEM}). Beyond these consistent patterns, we find that a subset of species exhibit abundance shifts between abundance regimes depending on density within specific gas types, making them potential density tracers as described below.

Several species exhibit environment-specific abundance patterns that make them potential diagnostic tracers (see also Tab. \ref{tab:species_classification}). CH$_3$SH and CH$_3$CHO never reach high abundances in protostellar environments under any density condition, yet both become highly abundant in shocked and post-shocked gas at high densities (CH$_3$SH also exceeds the threshold in shocks at mid density). This behavior makes them potential tracers of shocks propagating through dense gas.

Conversely, CH$_3$NCO shows the opposite pattern, never becoming highly abundant in shocked or post-shocked gas while exceeding the threshold in protostellar environments at mid and high densities. Other species display more complex environmental dependencies: CS and C$_2$S remain consistently highly abundant across all densities in both shocks and protostellar objects but drop below the high-abundance threshold in post-shocked gas at low density. HCO$^+$ maintains high abundances in shocks at each density but fails to reach this threshold in post-shocked gas at high density and in protostellar objects at both mid and high densities.

There is also a group of species showing stronger density sensitivity. C$_2$H$_5$CN reaches high abundance in protostellar objects only at low density, while requiring at least mid density in shocked and post-shocked environments, suggesting optimal enhancement conditions at densities of $\sim10^5-10^6$~cm$^{-3}$. HNCO becomes highly abundant in protostellar objects only at the highest density, while remaining abundant in shocked and post-shocked gas except at the lowest density.

For species that can reach high abundance at each density in gas associated with protostellar objects, we find little variation in their detectability limits. Once a species becomes highly abundant in this environment, its maximum abundance remains relatively stable across different densities, with no significant shifts that would alter its detectability. The only modest changes, if any, occur in shocked and post-shocked gas. This suggests that in protostellar environments, the conditions leading to high abundances for most studied species depend less on density than they do in shock-dominated regions.

\subsection{Important abundance ratios}\label{sect:ratios}

Molecular abundance ratios provide a practical and insightful bridge between chemical models and observations. While absolute abundances can reveal broad chemical trends, ratios help mitigate uncertainties, such as assumptions about column density, by emphasizing relative chemical behavior. This allows for more robust comparisons across environments and between models and observations. In this section, we identify a set of molecular ratios from our models that are particularly sensitive to cosmic-ray ionization rates, temperature, and underlying energetic processes. These ratios highlight contrasts that are both chemically meaningful and potentially observable, making them promising diagnostics for distinguishing between physical regimes in protostellar and shocked gas.

We begin by examining species that could potentially carry important diagnostic information about cosmic-ray ionization rates. We chose to consider CH$_3$SH/HCO$^+$ and H$_2$CO/HCO$^+$ because both CH$_3$SH and H$_2$CO display inverse abundance trends relative to HCO$^+$ with increasing ionization (for more see Sect. \ref{sect:CRIR-upper-limit}). 

We also considered the ion ratio NS$^+$/HCO$^+$ to explore its potential for complementary diagnostics. While HCO$^+$ consistently increases with CRIR, the behavior of NS$^+$ is more environment-dependent: it increases with $\zeta$ in shocked and post-shock gas but shows a decreasing trend in warm-up and protostellar models. Additionally, we tested NS$^+$/H$_2$CO and NS$^+$/CS ratios. The NS$^+$/H$_2$CO ratio could be relevant in shocked gas, where both species exhibit inverse abundance trends across the $\zeta$ range in which they diverge significantly from protostellar conditions. Similarly, NS$^+$/CS ratio may serve as a robust tracer of ionization-driven chemistry in shocks, since NS$^+$ increases with CRIR while CS remains relatively stable.

We find that CH$_3$SH/HCO$^+$ and H$_2$CO/HCO$^+$ decline by approximately 3–4 orders of magnitude as $\log_{10}(\zeta/\zeta_0)$ increases from 1 to 4 across all gas types. There is a consistent offsets between key environment pairs: protostellar objects vs. warm-up, and shocks vs. post-shock. Moreover, the H$_2$CO/HCO$^+$ ratio is greater than one at $\log_{10}(\zeta/\zeta_0) \leq 2$, and drops below unity for higher ionization rates. The NS$^+$/H$_2$CO ratio increases with $\log_{10}(\zeta/\zeta_0)$ in both environments. However, the increase is more pronounced in shocked gas once ionization increases beyond $\log_{10}(\zeta/\zeta_0) \sim 2$, whereas in post-shock gas, it aligns more closely with the behavior seen in protostellar-object-related conditions. This divergence suggests that once the NS$^+$/H$_2$CO ratio approaches or exceeds unity at elevated CRIRs it could be a signature of a recently shocked medium. Nonetheless, these three ratios appear to be promising at tracing CRIRs and may additionally help distinguish between dominant energetics.

\begin{figure}
    \centering
    \includegraphics[width=1\linewidth]{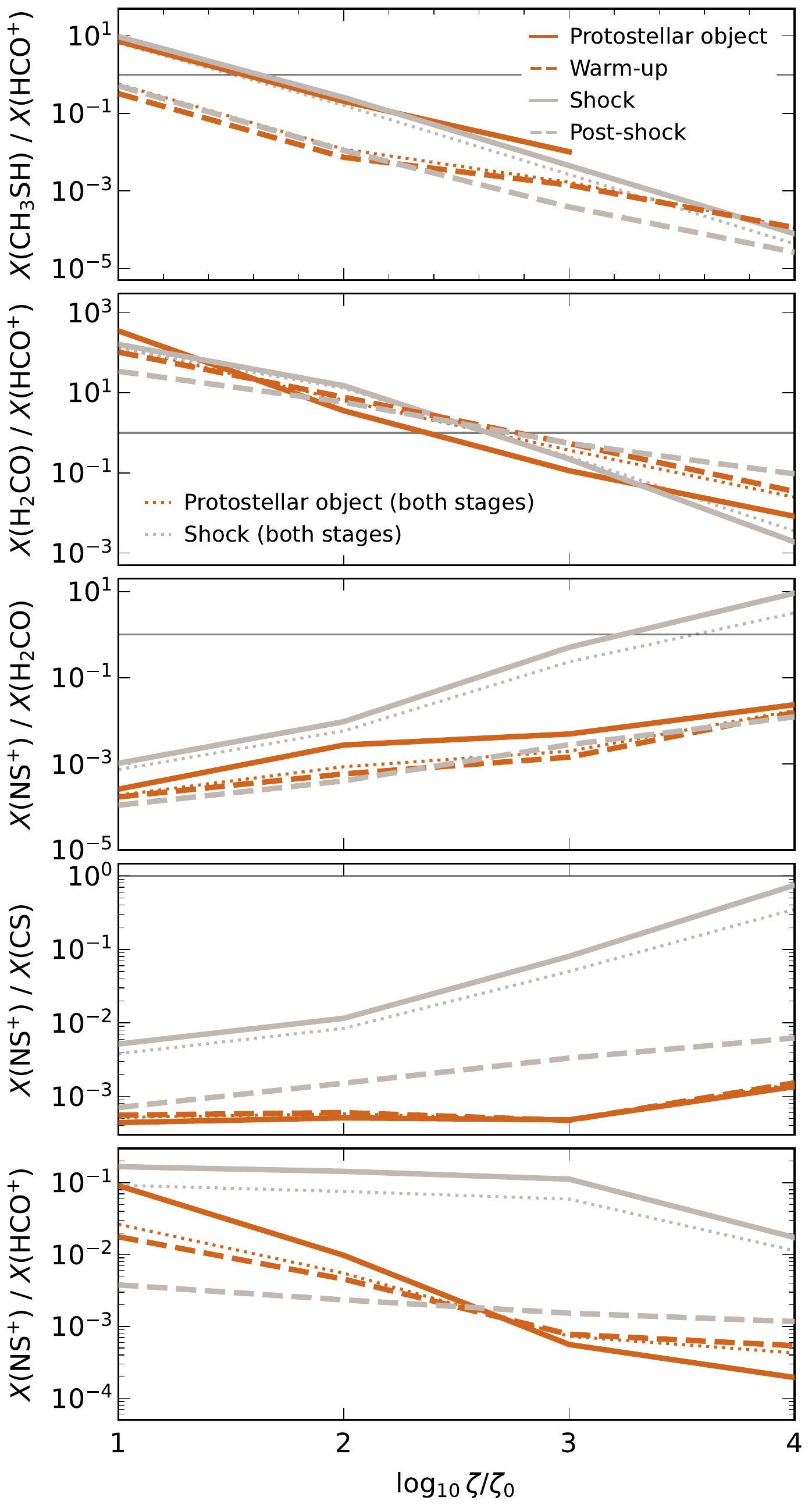}
    \caption{Mean abundance ratios as a function of cosmic-ray ionization rate for five diagnostic species: CS, CH$_3$SH, H$_2$CO, HCO$^+$, and NS$^+$. Each panel shows ratios for four physical environments: protostellar objects (solid orange lines), warm-up (dashed orange lines), shocks (solid gray lines), and post-shock (dashed gray lines). Neutral-to-ion ratios (top two panels) generally decline by 3–4 orders of magnitude as $\log_{10}(\zeta/\zeta_0)$ increases from 1 to 4, but show distinct slopes between environments. The ion-to-neutral ratios (third and fourth panel) exhibits a clear trend in shock-related gas, increasing with increasing ionization rate. However, NS$^+$/H$_2$CO also exhibits consistent increasing trend in protostellar objects. The ion-to-ion ratio NS$^+$/HCO$^+$ (bottom panel) exhibits a shallower response, remaining relatively flat in shock-related gas while declining more clearly in thermally processed gas.}
    \label{fig:zeta-ratios}
\end{figure}

The NS$^+$/CS ratio reflects the distinct ionization sensitivity of its constituent species, showing a clear increasing trend with ionization, but only in shocked gas. There is also a significant distinction between shocked and post-shocked gas: in the latter, the ratio increases monotonically but modestly, whereas in shocked gas it spans over two orders of magnitude. In contrast, the ion-to-ion ratio NS$^+$/HCO$^+$ responds more similarly to neutral-to-ion ratios, as it decreases with ionization. However, it has a shallower slopes in shock-related gas, where it remains relatively flat for $\log_{10}(\zeta/\zeta_0) \leq3$, while in protostellar environments it declines significantly with increasing ionization, decreasing from $\sim10^{-1}$ to $\sim10^{-4}$. Overall, the two ratios seem to be sensitive to ionization in different environments, with NS$^+$/CS potentially probing cosmic-ray ionization in shocks, and NS$^+$/HCO$^+$ serving as a diagnostic in protostellar objects.

\begin{figure*}[!t]
\centering
\includegraphics[width=0.9\linewidth]{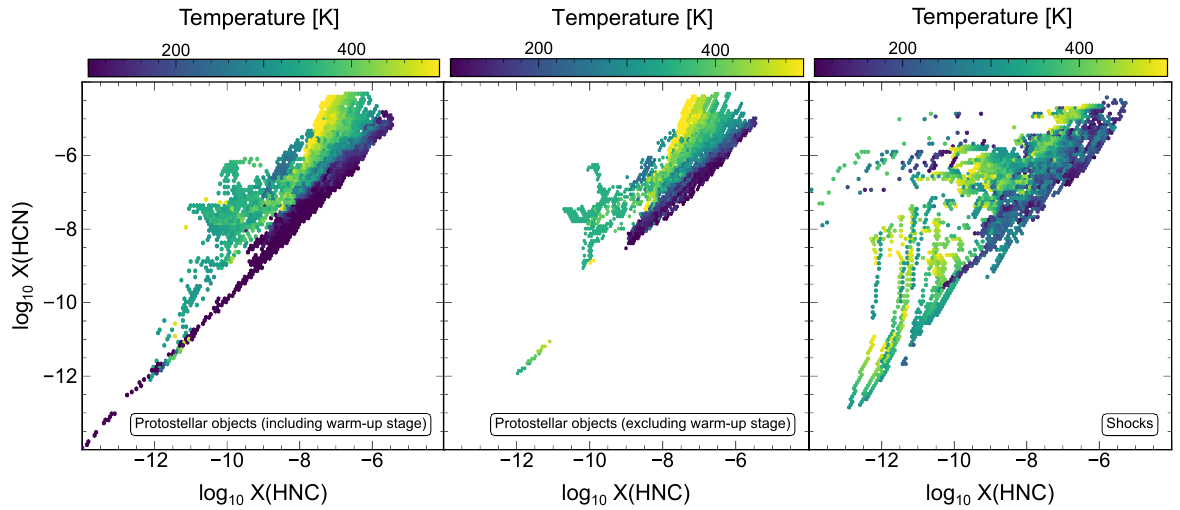}
\caption{Hexagonal plots comparing the relationship between HNC and HCN abundances, with the mean temperature per hexagonal bin ($T$) as the color scale, for the models of (left and middle panels) protostellar objects and (right panel) shocks. The temperature range is consistently set to 100--500 K in both panels to enable a direct comparison. The color mapping represents the mean temperature within each hexagonal bin, computed using 100 bins for both axes. These figures illustrate how distinct chemical and physical processes shape the HCN-to-HNC ratio, with the ratio serving as a useful temperature tracer in protostellar objects, whereas in shocks, it appears highly variable and less predictive of temperature. The lower-abundance tail in protostellar objects appears to be primarily shaped by the warm-up phase, as illustrated by the contrast between the leftmost and center panels.}
\label{fig:HNC-HCN}
\end{figure*}

Exploring the interdependence of species abundances and their relation to physical parameters revealed an interesting trend for HCN and HNC, which is worth mentioning given that these species alone were not strongly indicative of the underlying processes. Observationally, the ratio of these two molecules is frequently used as chemical thermometers for molecular gas  at low temperature \citep[e.g.][]{graninger2014,colzi2018,hacar2020} but its use at higher temperature is debated \citep[e.g.,][]{behrens2022}. Comparing their behavior over the 100--500~K range in both shocked and non-shocked gas in protostellar objects, we find that the HCN-to-HNC abundance ratio does exhibit a 
temperature dependence, albeit with a large spread of abundances at higher temperatures, in non-shocked gas, as shown in Fig. \ref{fig:HNC-HCN}. In shocked gas, however, this dependence is no longer apparent. Additionally, in protostellar objects, we find a strong linear correlation between the two species, with a Pearson correlation coefficient of 0.87, further supporting the idea that this ratio is primarily influenced by protostellar heating rather than shock-related processes. 

Although the HCN-to-HNC ratio appears to reflect properties characteristic of protostellar environments, caution is warranted, as our relation is derived across a broad model grid spanning a wide parameter space. This ratio is also dependent on the ionization fraction of the gas and hence on the collisional excitation of HCN with electrons, since HCN probes lower density gas than HNC \citep[see][]{goicoechea2022}. Additionally, \citet{behrens2022, behrens2024} suggest that cosmic rays may influence this ratio, while \citet{harada2024} argue it can act as a FUV thermometer in the GMCs of M83.

\subsection{Observational comparison}\label{sect:observations}

To evaluate the relevance of our chemical models and assess how well they capture the complexity of CMZ-like environments, we place them in an observational context. By comparing selected modeled abundances with those derived from observations, we aim to gauge the robustness of current chemical networks and physical assumptions, and assess the diagnostic value of the templates presented in this paper. We compare our modeled abundances with observations towards molecular clouds in the Galactic Center from \citet{requena-torres2006} and \citet{rodriguez-almeida2021}, and compare our protostellar object models with observations and models reported by \citet{belloche2025}. Throughout these comparisons, we refer to shock, post-shock, and protostellar object phases as defined in Sect. \ref{sect:results}.

\begin{figure}[htbp]
    \centering
    \includegraphics[width=0.99\linewidth]{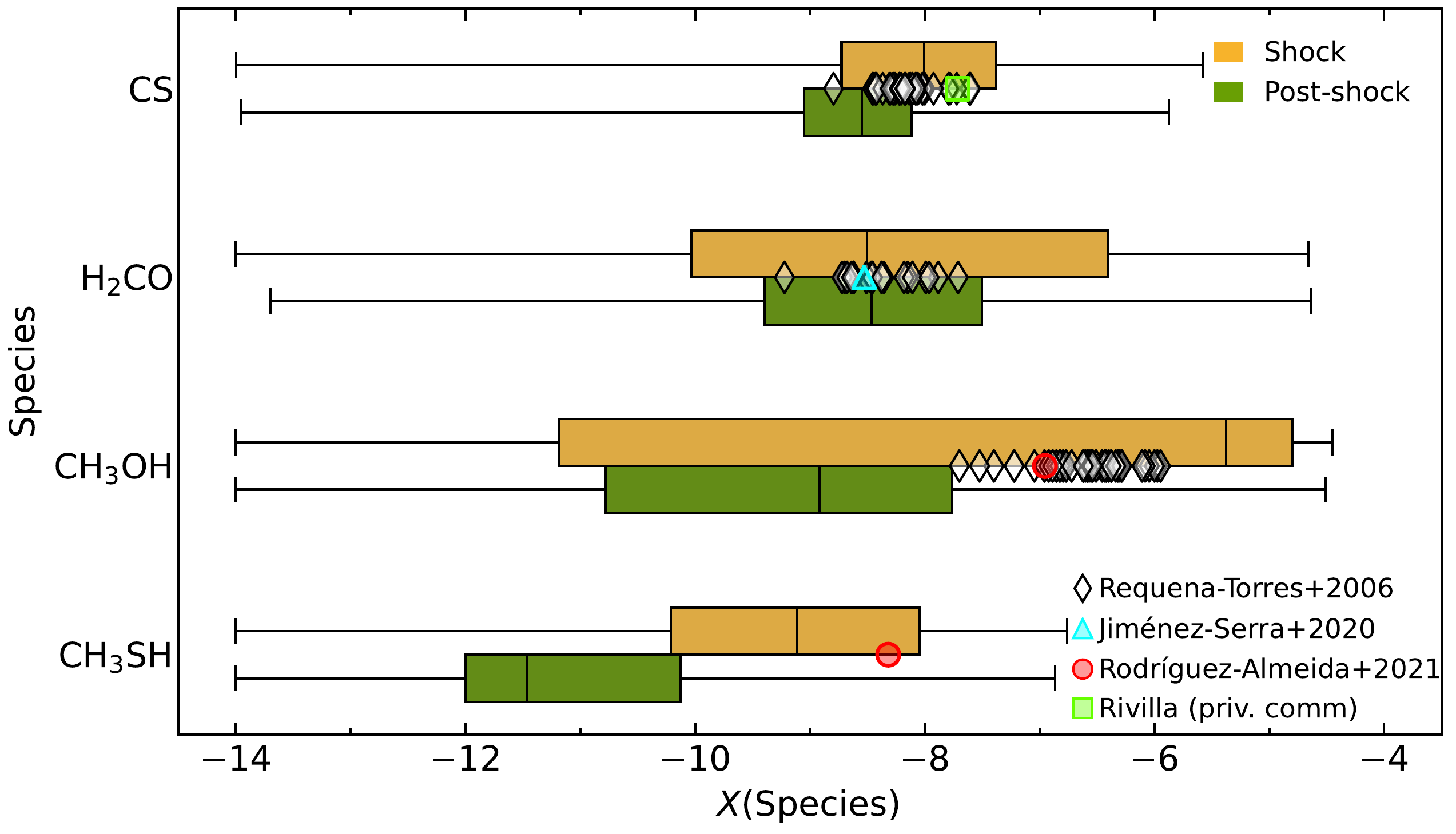}
    \vspace{1em}

    \includegraphics[width=0.99\linewidth]{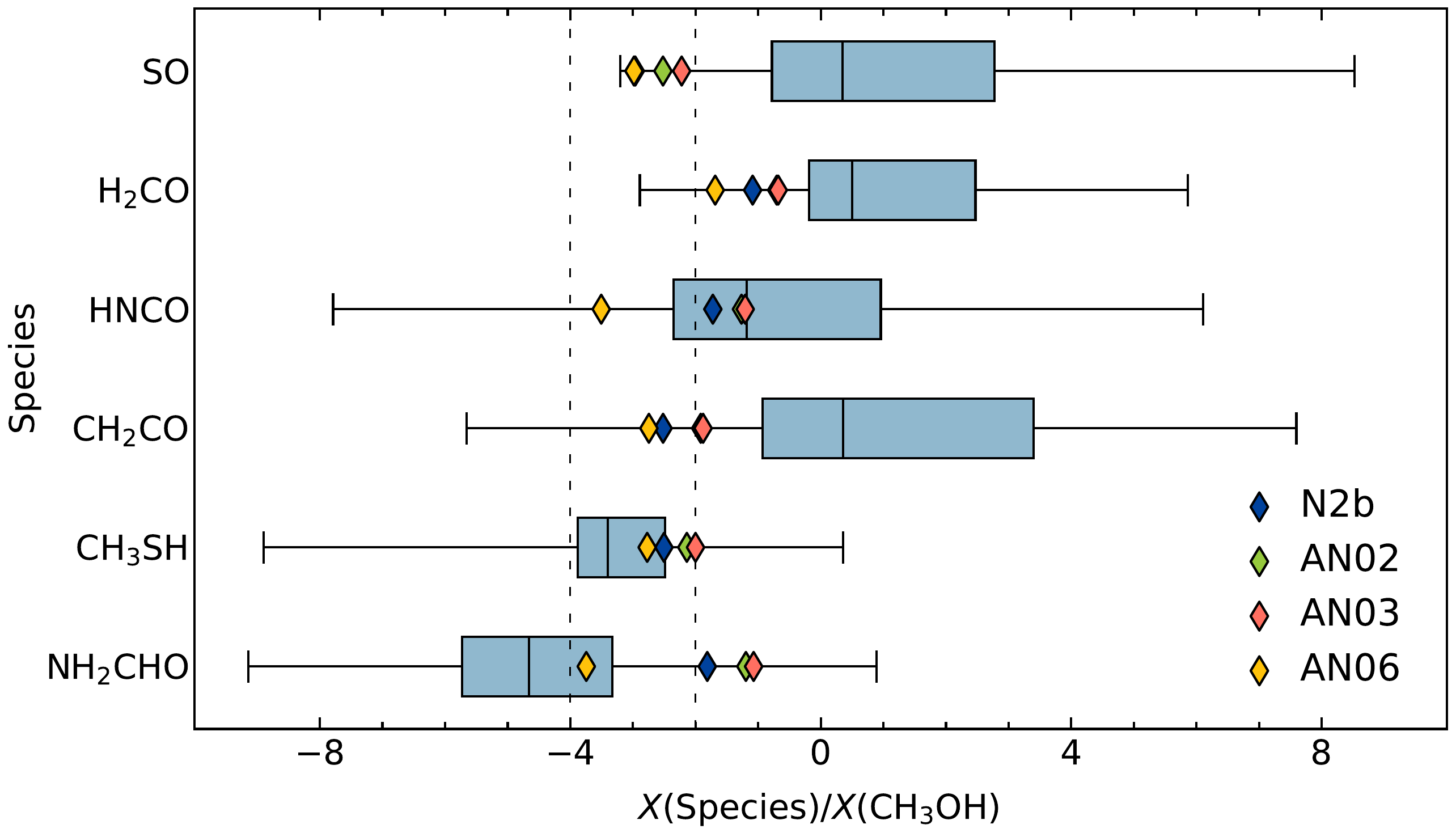}

    \caption{Comparison of modeled abundances with observations toward Galactic Center molecular clouds and the high-mass star-forming region Sgr B2(N2). The top panel shows abundance distributions for shocked (yellow) and post-shocked (green) gas, while the bottom panel presents those for protostellar objects. Shock models are compared with observations of 40 molecular clouds from \citet{requena-torres2006}, including data toward data toward G+0.693–0.027 for H$_2$CO from \citet{jimenez-serra2020}, CH$_3$SH and CH$_3$OH from \citet{rodriguez-almeida2021}, and CS from Rivilla (priv. comm). Models are also compared with recent data toward Sgr B2(N2) from \citet{belloche2025}. Following Figure 7 in \citet{belloche2025}, modeled protostellar abundances are shown relative to CH$_3$OH and with dashed lines we indicate the 1\% and 0.01\% levels with respect to methanol.}
    \label{fig:observ-compar-both}
\end{figure}

In the top panel of Fig. \ref{fig:observ-compar-both}, we present the comparison between our model predictions with the molecular abundances of CS, H$_2$CO, CH$_3$OH, and CH$_3$SH measured toward a sample of 40 molecular clouds observed by \citet{requena-torres2006} across the whole CMZ. From Fig. \ref{fig:observ-compar-both}, we find that the shock abundances match better the observed abundances of these species. This implies that short-lived ($<10^4$ yr) shocks must be continuously occurring in the Galactic Center, replenishing the material back into the gas phase since the post-shock abundances dramatically drop due to the efficient freeze-out onto the surface of dust grains. This is an effect that was already noted by \citet{requena-torres2006} but with a much basic desorption/freeze-out model. Here, we confirm this hypothesis and therefore, the high abundances seen for simple species and COMs in molecular clouds in the Galactic Center are due to the presence of recurring shocks that prevent gas from reaching the post-shock stage.

We note that \citet{belloche2025} recently presented observations of up to 58 species in hot cores associated with Sgr~B2(N2) and compared these with chemical models based on work of \citet{shope2024}, which employs the gas-grain chemical kinetics code MAGICKAL \citep{garrod2013,garrod2022}. Their study explores a narrower range of H$_2$ gas densities, temperatures, and cosmic-ray ionization rates, representing a subset of the broader parameter space considered in our model grid. Nevertheless, it is worth comparing our predictions with the values reported in \citet{belloche2025}.

In the bottom panel of Fig. \ref{fig:observ-compar-both} we show abundances of SO, H$_2$CO, HNCO, CH$_2$CO, CH$_3$SH, and NH$_2$CHO relative to CH$_3$OH based on our models and overplotted values extracted from \citet{belloche2025}. Here, we consider the data for protostellar objects as a whole, including the warm-up stage. We chose mostly simple species, as their chemistry is generally better understood than that of COMs. However, we included CH$_3$SH to test whether it can also match observations in protostellar environments, and NH$_2$CHO, which is significantly underpredicted in our shock models (discussed further in Sect. \ref{sect:low-COMs}).

We find that all observations fall within our predicted range, including NH$_2$CHO, with HNCO and CH$_3$SH showing the best agreement in at least two out of four sources. However, the question remains whether this agreement is consistent with expectations from the \citet{shope2024} models used by \citet{belloche2025}. Before addressing this, we must first consider several important modeling differences between our approach and that of \citet{shope2024}.

First, during the collapse phase, their gas temperature is fixed at 10~K, while dust temperature varies with visual extinction, ranging from $\sim$8 to 16~K. By contrast, we adopt elevated gas and dust temperatures from the outset, which -- as discussed in Sect.~\ref{sect:low-COMs} -- has a fundamental impact on COM chemistry. Second, their models begin at a hydrogen nucleus density of $n_\mathrm{H}=3\times10^{3}\,\mathrm{cm}^{-3}$, more than an order of magnitude higher than our starting density of $10^2\,\mathrm{cm}^{-3}$, resulting in potential significant differences in chemical initial conditions. Third, the cosmic-ray ionization rates they explore range from $0.1$ to $100\,\zeta_0$, whereas our models span $10-10000\,\zeta_0$, reflecting environments with more extreme ionization levels. Finally, their models incorporate non-diffusive surface chemistry \citep{garrod2022}, which is not included in \texttt{UCLCHEM} but is expected to be especially important at low dust temperatures \citep[$T_\mathrm{dust} \sim 10$~K; ][]{jimenez-serra2025b}, such as those used in their models.

While \citet{belloche2025} employs a detailed modeling and fitting routine to compare observations with MAGICKAL models, our goal here is to perform a more direct, quantitative check of whether our models reproduce the observed abundance ratios. We focus only on a subset of the simple species -- SO, H$_2$CO, HNCO, and CH$_2$CO -- and evaluate whether the models can simultaneously match all observed abundance ratios within one order of magnitude. 

For source N2b, we find that protostellar-object models with $\zeta = 10^{-16}\,\mathrm{s}^{-1}$ and ages $\lesssim 6\times10^4$ years provide a good match across all selected species. In the case of AN02 and AN03, the best agreement is obtained for higher ionization rates of $\zeta = 10^{-14}\,\mathrm{s}^{-1}$, though models with $\zeta$ between $10^{-16}$ and $10^{-14}\,\mathrm{s}^{-1}$ also reproduce the observed ratios. Additionally, here a fair number of models with shorter warm-up durations of $\sim2\times10^4$ years, corresponding to more massive protostellar objects, also reproduce the observed ratios (see Sect.~\ref{sect:chemical-evolution} for details on the timescales). Finally, for AN06, the data are best matched by models involving higher-mass protostellar objects and shorter evolutionary timescales, again showing a preference for $\zeta = 10^{-16}\,\mathrm{s}^{-1}$. In all cases, the preferred gas density is $n_\mathrm{H} = 10^8\,\mathrm{cm}^{-3}$.

These predicted values are in reasonable agreement with the conclusions of \citet{belloche2025}. In their models, which explore a range of warm-up timescales, those with the shortest duration ($\sim2\times10^4$ years) appear to fit the data better. They also find that models with $\zeta \approx 10^{-15}\,\mathrm{s}^{-1}$ are better at matching the observed abundances. While our best-fitting ionization rate varies by source, it consistently falls within the $\zeta = 10^{-16}-10^{-14}\,\mathrm{s}^{-1}$ range, supporting the idea of enhanced ionization in SgrB2(N2). Finally, \citet{belloche2025} identify $n_\mathrm{H} = 2\times10^{8}\,\mathrm{cm}^{-3}$ as the best-matching density, which is consistent with our findings. Overall, our grid aligns reasonably well with results reported for SgrB2(N2), highlighting the potential utility of the model templates developed in this work.
 
\subsection{Low abundances of certain COMs in shocks}\label{sect:low-COMs}

Our shock models predict relatively low abundances of formamide (NH$_2$CHO), methyl formate (HCOOCH$_3$), ethanol (C$_2$H$_5$OH), and dimethyl ether (CH$_3$OCH$_3$) compared to both observational data and predictions by other modeling works \citep[e.g.,][]{requena-torres2006, requena-torres2008, quenard2018, lopez-sepulcre2019, lopez-sepulcre2024}. In this section, we briefly address possible reasons for this discrepancy, while a more detailed investigation will be the subject of a forthcoming study.

Firstly, we assume gas and dust temperatures of at least 15~K in the collapse stage. This choice is physically justified by the conditions prevalent in the CMZ. Unlike the Galactic disk, where dense, shielded regions can reach temperatures as low as 10~K, the CMZ environment is shaped by elevated background radiation fields and dynamic gas flows that prevent such low temperatures, even in quiescent regions. Assuming colder dust would be inconsistent with these conditions and could misrepresent the initial chemical state of the gas and dust before shocks propagate through it. Furthermore, three-dimensional astrochemical simulations and synthetic observations of turbulent molecular clouds by \citet{bisbas2021}, which incorporate enhanced $G_0$ and $\zeta$, demonstrate that low-temperature conditions are unlikely to exist in the CMZ. Additionally, observations confirm that gas temperatures in the CMZ may be even higher and appear to have a ``pedestal'' near 60~K \citep{ginsburg2016}, which further strengthens the need for higher temperatures as initial conditions for the models.

Secondly, we choose a threshold of 30 K for the reactions on the dust grains, above this temperature there is no freeze-out and ice chemistry ceases to evolve on the grain, allowing for the ice species to be desorbed into the gas phase. This means that for the 30 and 35 K stage 1 models, the evolution of the chemistry occurs primarily in the gas phase. This is reflected in the fact that species with efficient hydrogenation pathways, such as methanol and acetaldehyde are less abundant with the higher initial stage 1 temperatures for both protostellar and shock modeling.

While the assumed temperature conditions are consistent with expectations for the CMZ and yield reasonable abundances for most species, achieving observable levels of gas-phase COMs from a single shock event may require initial dust temperatures below 15 K. One of the key roles of shocks is to induce sputtering, suggesting that temperature constraints could be linked to the need for substantial reservoirs of these molecules on grain surfaces. We have assessed the temperature dependence for species with lower-than-expected abundances and found that there exists an upper initial temperature limit of $\sim$13 K, above which we cannot recover the observed levels of these COMs (i.e., $X$(Species) $\approx 10^{-10}-10^{-8}$). More details will follow in a forthcoming study. Additionally, \cite{tram2025} found that modeling C$_2$H$_5$OH in protostellar objects may pose additional challenges related to proximity to the radiation source and effects of assumed grain sizes.

We have also gauged the effects of cyclic shocks. Given the widespread nature of large-scale shocks in the CMZ, this scenario is likely to occur. We modeled three consecutive shock events in a medium with an initial density of $n_\mathrm{H} = 10^5$~cm$^{-3}$, a starting temperature of $T_\mathrm{init} = 15$~K, and shock velocities of $v_\mathrm{S} = 10\,\mathrm{km\,s}^{-1}$. We found that different species respond in distinct ways. NH$_2$CHO shows only a modest increase, with a maximum enhancement below one order of magnitude. In contrast, HCOOCH$_3$, C$_2$H$_5$OH, and CH$_3$OCH$_3$ respond much more strongly: C$_2$H$_5$OH and CH$_3$OCH$_3$ can be enhanced by about an order of magnitude, and HCOOCH$_3$ by nearly three. Although this does not fully explain the discrepancy between our models and observations in this setup, it highlights the potential importance of recurring shocks. Future work should explore whether multiple shock cycles under fine-tuned physical conditions can build up enough COMs to be efficiently frozen back onto grains.

\section{Conclusions}\label{sect:conclusions}

In this study, we set out to explore the chemical fingerprints of 24 species in protostellar- and shock-related environments in the Central Molecular Zone, one of the most physically and chemically extreme regions of the Milky Way. By modeling a wide range of conditions representative of the CMZ-like environments, we identified key molecular tracers and diagnostic abundance patterns that offer insight into the underlying energetics and evolutionary stages of dense gas in this environment. Our goal was to establish a set of versatile chemical templates that can guide the interpretation of current and future spectral line surveys. Below, we summarize the main takeaways:

\begin{enumerate}
    \item Average shock signatures are predominantly associated with short timescales ($\lesssim10^4$ years), while those linked to protostellar objects emerge at later stages ($\gtrsim10^4$ years). 
    \item Species most frequently achieving maximum abundances across our model grid are: CH$_3$OH, HCN, SiO, and HCO$^+$ in shocked environments and HCN, SO, and HCO$^+$ in protostellar objects.
    \item Cosmic-ray ionization rates and temperatures have the biggest influence on the predicted abundances. We identified HCO$^+$, H$_2$CO, and CH$_3$SH as promising tracers for constraining cosmic-ray ionization rates, as they show particularly consistent ionization trends in each considered gas type.
    \item We propose five ratios that could provide diagnostics of cosmic-ray ionization rates: 
    \begin{itemize}
        \item CH$_3$SH/HCO$^+$, H$_2$CO/HCO$^+$, and NS$^+$/H$_2$CO in both environments at different evolutionary stages,
        \item NS$^+$/CS for shock-driven regions, and
        \item NS$^+$/HCO$^+$ in protostellar object environments. 
    \end{itemize}
    \item The maximum attainable abundances in shocked gas remain relatively stable with temperature, suggesting that temperature-driven chemical changes in shocks have insufficient time to lead to significant chemical evolution.
    \item 300~K marks a noticeable shift either in abundance behavior or level of differentiation between protostellar and shocked environments for NS$^+$, CH$_2$CO, CH$_3$OH, CH$_3$SH, NH$_2$CHO, and HCOOCH$_3$.
    \item The HCN-to-HNC abundance ratio exhibits temperature dependence in protostellar objects, but not in shocks. 
    \item Several species never reach high abundances of $\geq10^{-7}$ in certain environments:
    \begin{itemize}
        \item NS$^+$ in protostellar objects and post-shocked gas,
        \item CH$_3$SH and CH$_3$CHO in protostellar objects, and
        \item CH$_3$NCO in shocks.
    \end{itemize}
\end{enumerate}

Our analysis also reveals significant shifts in predicted HCO$^+$ abundances when using updated chemical networks, with differences reaching nearly three orders of magnitude at cosmic-ray ionization rates around $\sim10^{-13}\:\mathrm{s}^{-1}$. These shifts result in HCO$^+$ becoming the most abundant species across the whole grid.

We find that our models match the observational data for the majority of species, supporting scenarios involving the presence of recent and/or cyclic shocks in the molecular clouds of the Galactic Center and enhanced ionization in Sgr B2(N2). However, our shock models fail to predict the observed high abundances of formamide, methyl formate, ethanol, and dimethyl ether. This discrepancy appears to be related to initial gas and dust conditions and will be addressed in a forthcoming study.

Future studies should also assess the importance of recurring shock events, as tests indicate they can have a significant impact on some COMs. Moreover, the effects of increased metallicity should be evaluated, as it could potentially affect CMZ chemistry where metallicity can be largely super-solar \citep[e.g.,][and references therein]{giveon2002,garcia2021}. However, the work presented here provides a basis for future, more in-depth and finely tuned modeling efforts.

\begin{acknowledgements}
The research of KMD and SV is funded by the European Research Council (ERC) Advanced Grant MOPPEX 833460.vii.

I.J-S., L.C., and V.M.R. acknowledge support from the grant PID2022-136814NB-I00 by the Spanish Ministry of Science, Innovation and Universities/State Agency of Research MICIU/AEI/10.13039/501100011033 and by ERDF, UE. 

I.J-.S also acknowledges funding from the ERC Consolidator Grant OPENS (project number 101125858) funded by the European Union.

V.M.R. also acknowledges support from the grant RYC2020-029387-I funded by MICIU/AEI/10.13039/501100011033 and by "ESF, Investing in your future", and from the Consejo Superior de Investigaciones Cient{\'i}ficas (CSIC) and the Centro de Astrobiolog{\'i}a (CAB) through the project 20225AT015 (Proyectos intramurales especiales del CSIC); and from the grant CNS2023-144464 funded by MICIU/AEI/10.13039/501100011033 and by “European Union NextGenerationEU/PRTR”. 

X.L.\ acknowledges support from the National Key R\&D Program of China (No.\ 2022YFA1603101), the Strategic Priority Research Program of the Chinese Academy of Sciences (CAS) Grant No.\ XDB0800300, State Key Laboratory of Radio Astronomy and Technology, the National Natural Science Foundation of China (NSFC) through grant Nos.\ 12273090 and 12322305, the Natural Science Foundation of Shanghai (No.\ 23ZR1482100), and the CAS ``Light of West China'' Program No.\ xbzg-zdsys-202212.

A.S.-M.\ acknowledges support from the RyC2021-032892-I grant funded by MCIN/AEI/10.13039/501100011033 and by the European Union `Next GenerationEU’/PRTR, as well as the program Unidad de Excelencia María de Maeztu CEX2020-001058-M, and support from the PID2023-146675NB-I00 (MCI-AEI-FEDER, UE).

M.G.S.-M.\ thank the Spanish MICINN for funding support under grant PID2023-146667NB-I00 and acknowledges support from the NSF under grant CAREER 2142300.

\end{acknowledgements}
\bibliographystyle{aa}
\bibliography{references}

\begin{appendix}
\section{Time-averaged evolution of surface and bulk} \label{sect:surface-and-bulk}

\begin{figure*}[htbp]
    \centering
    \includegraphics[width=0.9\linewidth]{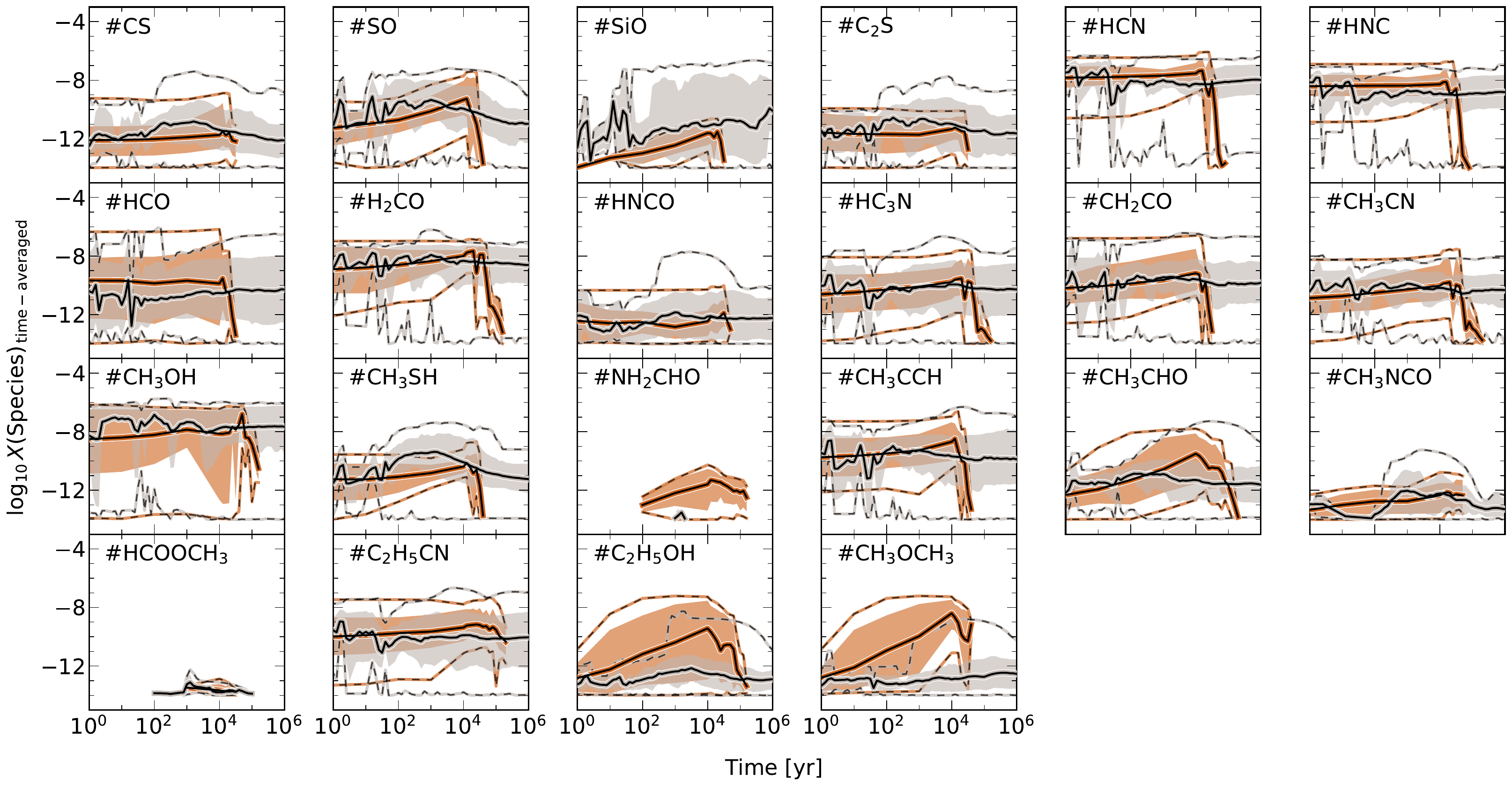}

    \vspace{1em} 

    \includegraphics[width=0.9\linewidth]{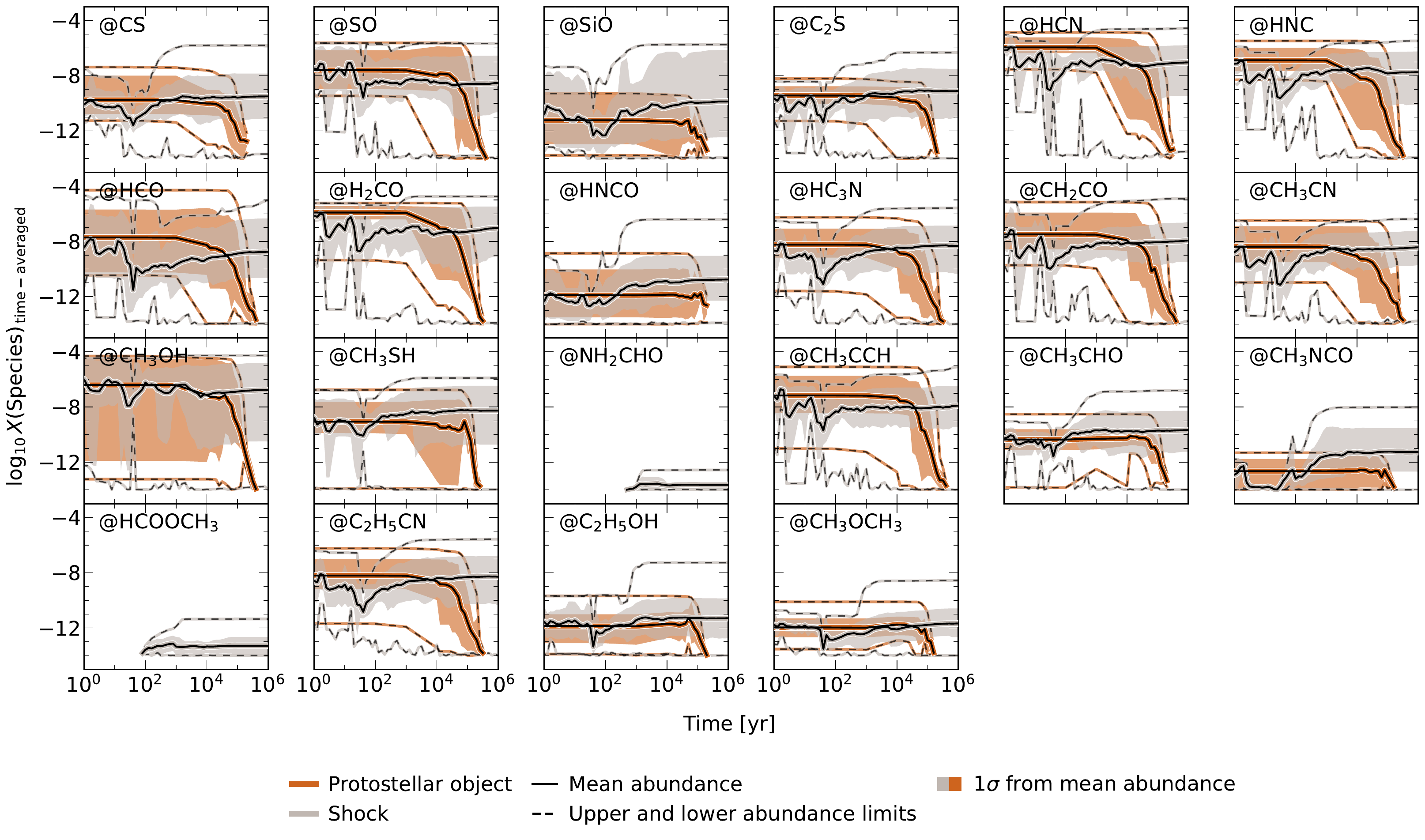}

    \caption{Time-averaged chemical evolution of selected species on the surface (top) and in the bulk (bottom). ``Time-averaging'' refers to averaging abundances across all models of a given type at fixed logarithmic time intervals. The solid black line shows the mean abundances at each time step, with shaded areas representing the 1$\sigma$ confidence interval. Dashed lines indicate the upper and lower temperature bounds. }
    \label{fig:time_evolution_surface_bulk}
\end{figure*}

Chemical modeling with \texttt{UCLCHEM} includes three phases: gas, surface, and bulk. Both the surface and bulk phases provide valuable insight into the chemistry occurring on dust grains. The surface refers to the outermost monolayer of ice, where species can adsorb from or desorb into the gas phase. The bulk lies beneath the surface and represents the deeper ice mantle. Species in the bulk can diffuse to the surface and may either desorb into the gas phase or be destroyed in fast shocks. Therefore, the total ice composition is the combined abundance of species in both the surface and bulk. In this section, we present the time-averaged chemical evolution of species on the surface and in the bulk (Fig. \ref{fig:time_evolution_surface_bulk}).

We observe that, in protostellar objects, virtually all surface species are eventually desorbed during the warm-up phase, followed by gradual depletion of the bulk reservoir as well. In contrast, shocked environments display more variability in surface and bulk composition. This variability is primarily due to the differing timing and frequency of shock events, which tends to dilute the averaged chemical signatures compared to the more systematic evolution seen in protostellar objects.

\end{appendix}

\end{document}

%% file: author.tex
\author{
         Katarzyna~M.~Dutkowska \inst{\ref{STRW}} \fnmsep\thanks{E-mail: \url{dutkowska@strw.leidenuniv.nl}}
    \and Gijs Vermari\"en \inst{\ref{STRW}}
    \and Serena Viti\inst{\ref{STRW},\ref{TRA},\ref{UCL}}
    \and Izaskun Jim\'enez-Serra\inst{\ref{CAB}}
    \and Laura Colzi\inst{\ref{CAB}}
    \and Laura~A.~Busch\inst{\ref{MPE}}
    \and Víctor~M.~Rivilla\inst{\ref{CAB}}
    \and Elisabeth~A.C.~Mills\inst{\ref{KU}}
    \and Sergio Mart\'{i}n \inst{\ref{ESO.Chile},\ref{JAO}}
    \and Christian Henkel \inst{\ref{MPIfR}}
    \and Pablo Garc\'ia\inst{\ref{UCN},\ref{NAOC}}
    \and Xing Lu\inst{\ref{SHAO},\ref{KLRAT}}
    \and Miriam~G.~Santa-Maria\inst{\ref{UF}}
    \and Jairo Armijos-Abenda\~no\inst{\ref{OAQ}}
    \and Yue Hu\inst{\ref{IAS}}
    \and J\"urgen Ott\inst{\ref{NRAO-SOC}}
    \and Kai Smith\inst{\ref{KU}}
    \and Fengwei Xu \inst{\ref{KIAA}}
    \and Shaoshan Zeng \inst{\ref{RIKEN}}
    \and \'Alvaro S\'anchez-Monge\inst{\ref{ICE},\ref{IEEC}}
    \and Anika Schmiedeke\inst{\ref{GBO}}
    \and Jaime~E.~Pineda\inst{\ref{MPE}}
    \and Steven~N.~Longmore\inst{\ref{ljmu}}
    \and Thanja Lamberts\inst{\ref{LIC},\ref{STRW}}
    }
\institute{
             Leiden Observatory, Leiden University, P.O. Box 9513, 2300 RA Leiden, The Netherlands\label{STRW}
        \and Transdisciplinary Research Area (TRA) ‘Matter’/Argelander-Institut für Astronomie, University of Bonn, Bonn, Germany\label{TRA}
        \and Department of Physics and Astronomy, University College London, Gower Street, London, UK\label{UCL}
        \and Centro de Astrobiología (CAB), CSIC-INTA, Ctra. de Ajalvir Km. 4, 28850, Torrejón de Ardoz, Madrid, Spain\label{CAB}
        \and Max-Planck-Institut f\"ur extraterrestrische Physik, Gie\ss enbachstra\ss e 1, 85748 Garching bei M\"unchen, Germany\label{MPE}
        \and Department of Physics and Astronomy, University of Kansas, 1251 Wescoe Hall Drive, Lawrence, KS 66045, USA\label{KU}
        \and European Southern Observatory, Alonso de Córdova, 3107, Vitacura, Santiago 763-0355, Chile \label{ESO.Chile}
        \and Joint ALMA Observatory, Alonso de Córdova, 3107, Vitacura, Santiago 763-0355, Chile \label{JAO}
        \and  Max-Planck-Institut f\"ur Radioastronomie, Auf dem H{\"u}gel 69, 53121 Bonn, Germany \label{MPIfR}
        \and Chinese Academy of Sciences South America Center for Astronomy, National Astronomical Observatories, CAS, Beijing 100101, China\label{NAOC}
        \and Instituto de Astronomía, Universidad Católica del Norte, Av.\ Angamos 0610, Antofagasta, Chile\label{UCN}
        \and Shanghai Astronomical Observatory, Chinese Academy of Sciences, 80 Nandan Road, Shanghai 200030, P.\ R.\ China\label{SHAO}
        \and State Key Laboratory of Radio Astronomy and Technology, A20 Datun Road, Chaoyang District, Beijing, 100101, P.\ R.\ China\label{KLRAT}
        \and Department of Astronomy, University of Florida, P.O. Box 112055, Gainesville, FL 32611, USA \label{UF}
        \and Observatorio Astronomico Nacional (OAN), Alfonso XII, 3, 28014, Madrid, Spain  \label{OAN}
        \and Observatorio Astron\'omico de Quito, Observatorio Astron\'omico Nacional, Escuela Polit\'ecnica Nacional, 170403, Quito, Ecuador \label{OAQ}
        \and Institute for Advanced Study, 1 Einstein Drive, Princeton, NJ 08540, USA\label{IAS} 
        \and National Radio Astronomy Observatory, 1011 Lopezville Road, Socorro, NM 87801, USA\label{NRAO-SOC}
        \and Kavli Institute for Astronomy and Astrophysics, Peking University, Beijing 100871, People's Republic of China\label{KIAA}
        \and Star and Planet Formation Laboratory, Cluster for Pioneering Research, RIKEN, 2-1 Hirosawa, Wako, Saitama, 351-0198, Japan\label{RIKEN}
        \and Institute of Space Sciences (ICE, CSIC), Campus UAB, Carrer de Can Magrans s/n, 08193, Bellaterra (Barcelona), Spain\label{ICE}
        \and Institute of Space Studies of Catalonia (IEEC), 08860, Barcelona, Spain\label{IEEC}
        \and Green Bank Observatory, P.O. Box 2, Green Bank, WV 24944, USA\label{GBO}
        \and  Astrophysics Research Institute, Liverpool John Moores University, IC2, Liverpool Science Park, 146 Brownlow Hill, Liverpool L3 5RF, UK\label{ljmu}         
        \and Leiden Institute of Chemistry, Leiden University, P.O. box 9502, 2300 RA Leiden, The Netherlands\label{LIC}
}

%% file: aanda.bbl
\begin{thebibliography}{144}
\expandafter\ifx\csname natexlab\endcsname\relax\def\natexlab#1{#1}\fi

\bibitem[{{Aalto} {et~al.}(2002){Aalto}, {Polatidis}, {H{\"u}ttemeister}, \&
  {Curran}}]{aalto2002}
{Aalto}, S., {Polatidis}, A.~G., {H{\"u}ttemeister}, S., \& {Curran}, S.~J.
  2002, \aap, 381, 783

\bibitem[{{Armijos-Abenda{\~n}o} {et~al.}(2020){Armijos-Abenda{\~n}o},
  {Mart{\'\i}n-Pintado}, {L{\'o}pez}, {Llerena}, {Harada}, {Requena-Torres},
  {Mart{\'\i}n}, {Rivilla}, {Riquelme}, \& {Aldas}}]{armijos-abendano2020}
{Armijos-Abenda{\~n}o}, J., {Mart{\'\i}n-Pintado}, J., {L{\'o}pez}, E.,
  {et~al.} 2020, \apj, 895, 57

\bibitem[{{Askne} {et~al.}(1984){Askne}, {Hoglund}, {Hjalmarson}, \&
  {Irvine}}]{askne1984}
{Askne}, J., {Hoglund}, B., {Hjalmarson}, A., \& {Irvine}, W.~M. 1984, \aap,
  130, 311

\bibitem[{{Asplund} {et~al.}(2009){Asplund}, {Grevesse}, {Sauval}, \&
  {Scott}}]{asplund2009}
{Asplund}, M., {Grevesse}, N., {Sauval}, A.~J., \& {Scott}, P. 2009, \araa, 47,
  481

\bibitem[{{Awad} {et~al.}(2010){Awad}, {Viti}, {Collings}, \&
  {Williams}}]{awad2010}
{Awad}, Z., {Viti}, S., {Collings}, M.~P., \& {Williams}, D.~A. 2010, \mnras,
  407, 2511

\bibitem[{{Barnes} {et~al.}(2019){Barnes}, {Longmore}, {Avison}, {Contreras},
  {Ginsburg}, {Henshaw}, {Rathborne}, {Walker}, {Alves}, {Bally}, {Battersby},
  {Beltr{\'a}n}, {Beuther}, {Garay}, {Gomez}, {Jackson}, {Kainulainen},
  {Kruijssen}, {Lu}, {Mills}, {Ott}, \& {Peters}}]{barnes2019}
{Barnes}, A.~T., {Longmore}, S.~N., {Avison}, A., {et~al.} 2019, \mnras, 486,
  283

\bibitem[{{Battersby} {et~al.}(2025){Battersby}, {Walker}, {Barnes},
  {Ginsburg}, {Lipman}, {Alboslani}, {Hatchfield}, {Bally}, {Glover},
  {Henshaw}, {Immer}, {Klessen}, {Longmore}, {Mills}, {Molinari}, {Smith},
  {Sormani}, {Tress}, \& {Zhang}}]{battersby2025}
{Battersby}, C., {Walker}, D.~L., {Barnes}, A., {et~al.} 2025, \apj, 984, 156

\bibitem[{{Bayet} {et~al.}(2011){Bayet}, {Williams}, {Hartquist}, \&
  {Viti}}]{bayet2011}
{Bayet}, E., {Williams}, D.~A., {Hartquist}, T.~W., \& {Viti}, S. 2011, \mnras,
  414, 1583

\bibitem[{{Behrens} {et~al.}(2022){Behrens}, {Mangum}, {Holdship}, {Viti},
  {Harada}, {Mart{\'\i}n}, {Sakamoto}, {Muller}, {Tanaka}, {Nakanishi},
  {Herrero-Illana}, {Yoshimura}, {Aladro}, {Colzi}, {Emig}, {Henkel}, {Huang},
  {Humire}, {Meier}, {Rivilla}, {van der Werf}, \& {Alma Comprehensive
  High-Resolution Extragalactic Molecular Inventory (Alchemi)
  Collaboration}}]{behrens2022}
{Behrens}, E., {Mangum}, J.~G., {Holdship}, J., {et~al.} 2022, \apj, 939, 119

\bibitem[{{Behrens} {et~al.}(2024){Behrens}, {Mangum}, {Viti}, {Holdship},
  {Huang}, {Bouvier}, {Butterworth}, {Eibensteiner}, {Harada}, {Mart{\'\i}n},
  {Sakamoto}, {Muller}, {Tanaka}, {Colzi}, {Henkel}, {Meier}, {Rivilla}, \&
  {van der Werf}}]{behrens2024}
{Behrens}, E., {Mangum}, J.~G., {Viti}, S., {et~al.} 2024, \apj, 977, 38

\bibitem[{{Belloche} {et~al.}(2019){Belloche}, {Garrod}, {M{\"u}ller},
  {Menten}, {Medvedev}, {Thomas}, \& {Kisiel}}]{belloche2019}
{Belloche}, A., {Garrod}, R.~T., {M{\"u}ller}, H.~S.~P., {et~al.} 2019, \aap,
  628, A10

\bibitem[{{Belloche} {et~al.}(2025){Belloche}, {Garrod}, {M{\"u}ller}, {Morin},
  {Willis}, \& {Menten}}]{belloche2025}
{Belloche}, A., {Garrod}, R.~T., {M{\"u}ller}, H.~S.~P., {et~al.} 2025, \aap,
  698, A143

\bibitem[{{Belloche} {et~al.}(2013){Belloche}, {M{\"u}ller}, {Menten},
  {Schilke}, \& {Comito}}]{belloche2013}
{Belloche}, A., {M{\"u}ller}, H.~S.~P., {Menten}, K.~M., {Schilke}, P., \&
  {Comito}, C. 2013, \aap, 559, A47

\bibitem[{{Bisbas} {et~al.}(2021){Bisbas}, {Tan}, \& {Tanaka}}]{bisbas2021}
{Bisbas}, T.~G., {Tan}, J.~C., \& {Tanaka}, K. E.~I. 2021, \mnras, 502, 2701

\bibitem[{{Bonfand} {et~al.}(2019){Bonfand}, {Belloche}, {Garrod}, {Menten},
  {Willis}, {St{\'e}phan}, \& {M{\"u}ller}}]{bonfand2019}
{Bonfand}, M., {Belloche}, A., {Garrod}, R.~T., {et~al.} 2019, \aap, 628, A27

\bibitem[{{Busch} {et~al.}(2024){Busch}, {Belloche}, {Garrod}, {M{\"u}ller}, \&
  {Menten}}]{busch2024}
{Busch}, L.~A., {Belloche}, A., {Garrod}, R.~T., {M{\"u}ller}, H. S.~P., \&
  {Menten}, K.~M. 2024, \aap, 681, A104

\bibitem[{{Butterfield} {et~al.}(2024){Butterfield}, {Chuss}, {Guerra},
  {Morris}, {Par{\'e}}, {Wollack}, {Dowell}, {Hankins}, {Karpovich}, {Siah},
  {Staguhn}, \& {Zweibel}}]{butterfield2024}
{Butterfield}, N.~O., {Chuss}, D.~T., {Guerra}, J.~A., {et~al.} 2024, \apj,
  963, 130

\bibitem[{{Butterworth} {et~al.}(2022){Butterworth}, {Holdship}, {Viti}, \&
  {Garc{\'\i}a-Burillo}}]{butterworth2022}
{Butterworth}, J., {Holdship}, J., {Viti}, S., \& {Garc{\'\i}a-Burillo}, S.
  2022, \aap, 667, A131

\bibitem[{{Caselli} {et~al.}(1997){Caselli}, {Hartquist}, \&
  {Havnes}}]{caselli1997}
{Caselli}, P., {Hartquist}, T.~W., \& {Havnes}, O. 1997, \aap, 322, 296

\bibitem[{{Caselli} {et~al.}(1998){Caselli}, {Walmsley}, {Terzieva}, \&
  {Herbst}}]{caselli1998}
{Caselli}, P., {Walmsley}, C.~M., {Terzieva}, R., \& {Herbst}, E. 1998, \apj,
  499, 234

\bibitem[{{Chen} {et~al.}(2016){Chen}, {Di Francesco}, {Johnstone}, {Sadavoy},
  {Hatchell}, {Mottram}, {Kirk}, {Buckle}, {Berry}, {Broekhoven-Fiene},
  {Currie}, {Fich}, {Jenness}, {Nutter}, {Pattle}, {Pineda}, {Quinn}, {Salji},
  {Tisi}, {Hogerheijde}, {Ward-Thompson}, {Bastien}, {Bresnahan}, {Butner},
  {Chrysostomou}, {Coude}, {Davis}, {Drabek-Maunder}, {Duarte-Cabral}, {Fiege},
  {Friberg}, {Friesen}, {Fuller}, {Graves}, {Greaves}, {Gregson}, {Holland},
  {Joncas}, {Kirk}, {Knee}, {Mairs}, {Marsh}, {Matthews}, {Moriarty-Schieven},
  {Mowat}, {Pezzuto}, {Rawlings}, {Richer}, {Robertson}, {Rosolowsky},
  {Rumble}, {Schneider-Bontemps}, {Thomas}, {Tothill}, {Viti}, {White},
  {Wouterloot}, {Yates}, \& {Zhu}}]{chen2016}
{Chen}, M. C.-Y., {Di Francesco}, J., {Johnstone}, D., {et~al.} 2016, \apj,
  826, 95

\bibitem[{{Chuss} {et~al.}(2003){Chuss}, {Davidson}, {Dotson}, {Dowell},
  {Hildebrand}, {Novak}, \& {Vaillancourt}}]{chuss2003}
{Chuss}, D.~T., {Davidson}, J.~A., {Dotson}, J.~L., {et~al.} 2003, \apj, 599,
  1116

\bibitem[{{Clark} {et~al.}(2013){Clark}, {Glover}, {Ragan}, {Shetty}, \&
  {Klessen}}]{clark2013}
{Clark}, P.~C., {Glover}, S. C.~O., {Ragan}, S.~E., {Shetty}, R., \& {Klessen},
  R.~S. 2013, \apjl, 768, L34

\bibitem[{{Colzi} {et~al.}(2018){Colzi}, {Fontani}, {Caselli}, {Ceccarelli},
  {Hily-Blant}, \& {Bizzocchi}}]{colzi2018}
{Colzi}, L., {Fontani}, F., {Caselli}, P., {et~al.} 2018, \aap, 609, A129

\bibitem[{{Colzi} {et~al.}(2024){Colzi}, {Mart{\'\i}n-Pintado}, {Zeng},
  {Jim{\'e}nez-Serra}, {Rivilla}, {Sanz-Novo}, {Mart{\'\i}n}, {Zhang}, \&
  {Lu}}]{colzi2024}
{Colzi}, L., {Mart{\'\i}n-Pintado}, J., {Zeng}, S., {et~al.} 2024, \aap, 690,
  A121

\bibitem[{{Cuppen} {et~al.}(2017){Cuppen}, {Walsh}, {Lamberts}, {Semenov},
  {Garrod}, {Penteado}, \& {Ioppolo}}]{cuppen2017}
{Cuppen}, H.~M., {Walsh}, C., {Lamberts}, T., {et~al.} 2017, \ssr, 212, 1

\bibitem[{{Dahmen} {et~al.}(1998){Dahmen}, {Huttemeister}, {Wilson}, \&
  {Mauersberger}}]{dahmen1998}
{Dahmen}, G., {Huttemeister}, S., {Wilson}, T.~L., \& {Mauersberger}, R. 1998,
  \aap, 331, 959

\bibitem[{{Draine} {et~al.}(1983){Draine}, {Roberge}, \&
  {Dalgarno}}]{draine1983}
{Draine}, B.~T., {Roberge}, W.~G., \& {Dalgarno}, A. 1983, \apj, 264, 485

\bibitem[{{Ferri{\`e}re} {et~al.}(2007){Ferri{\`e}re}, {Gillard}, \&
  {Jean}}]{ferriere2007}
{Ferri{\`e}re}, K., {Gillard}, W., \& {Jean}, P. 2007, \aap, 467, 611

\bibitem[{{Friesen} {et~al.}(2017){Friesen}, {Pineda}, {co-PIs}, {Rosolowsky},
  {Alves}, {Chac{\'o}n-Tanarro}, {How-Huan Chen}, {Chun-Yuan Chen}, {Di
  Francesco}, {Keown}, {Kirk}, {Punanova}, {Seo}, {Shirley}, {Ginsburg},
  {Hall}, {Offner}, {Singh}, {Arce}, {Caselli}, {Goodman}, {Martin}, {Matzner},
  {Myers}, {Redaelli}, \& {GAS Collaboration}}]{friesen2017}
{Friesen}, R.~K., {Pineda}, J.~E., {co-PIs}, {et~al.} 2017, \apj, 843, 63

\bibitem[{{Garc{\'\i}a} {et~al.}(2021){Garc{\'\i}a}, {Abel}, {R{\"o}llig},
  {Simon}, \& {Stutzki}}]{garcia2021}
{Garc{\'\i}a}, P., {Abel}, N., {R{\"o}llig}, M., {Simon}, R., \& {Stutzki}, J.
  2021, \aap, 650, A86

\bibitem[{{Garrod}(2013)}]{garrod2013}
{Garrod}, R.~T. 2013, \apj, 765, 60

\bibitem[{{Garrod} {et~al.}(2017){Garrod}, {Belloche}, {M{\"u}ller}, \&
  {Menten}}]{garrod2017}
{Garrod}, R.~T., {Belloche}, A., {M{\"u}ller}, H.~S.~P., \& {Menten}, K.~M.
  2017, \aap, 601, A48

\bibitem[{{Garrod} {et~al.}(2022){Garrod}, {Jin}, {Matis}, {Jones}, {Willis},
  \& {Herbst}}]{garrod2022}
{Garrod}, R.~T., {Jin}, M., {Matis}, K.~A., {et~al.} 2022, \apjs, 259, 1

\bibitem[{{Gerin} {et~al.}(2009){Gerin}, {Goicoechea}, {Pety}, \&
  {Hily-Blant}}]{gerin2009}
{Gerin}, M., {Goicoechea}, J.~R., {Pety}, J., \& {Hily-Blant}, P. 2009, \aap,
  494, 977

\bibitem[{{Giannetti} {et~al.}(2017){Giannetti}, {Leurini}, {Wyrowski},
  {Urquhart}, {Csengeri}, {Menten}, {K{\"o}nig}, \&
  {G{\"u}sten}}]{giannetti2017}
{Giannetti}, A., {Leurini}, S., {Wyrowski}, F., {et~al.} 2017, \aap, 603, A33

\bibitem[{{Ginsburg} {et~al.}(2016){Ginsburg}, {Henkel}, {Ao}, {Riquelme},
  {Kauffmann}, {Pillai}, {Mills}, {Requena-Torres}, {Immer}, {Testi}, {Ott},
  {Bally}, {Battersby}, {Darling}, {Aalto}, {Stanke}, {Kendrew}, {Kruijssen},
  {Longmore}, {Dale}, {Guesten}, \& {Menten}}]{ginsburg2016}
{Ginsburg}, A., {Henkel}, C., {Ao}, Y., {et~al.} 2016, \aap, 586, A50

\bibitem[{{Giveon} {et~al.}(2002){Giveon}, {Sternberg}, {Lutz}, {Feuchtgruber},
  \& {Pauldrach}}]{giveon2002}
{Giveon}, U., {Sternberg}, A., {Lutz}, D., {Feuchtgruber}, H., \& {Pauldrach},
  A.~W.~A. 2002, \apj, 566, 880

\bibitem[{{Goicoechea} {et~al.}(2013){Goicoechea}, {Etxaluze}, {Cernicharo},
  {Gerin}, {Neufeld}, {Contursi}, {Bell}, {De Luca}, {Encrenaz}, {Indriolo},
  {Lis}, {Polehampton}, \& {Sonnentrucker}}]{goicoechea2013}
{Goicoechea}, J.~R., {Etxaluze}, M., {Cernicharo}, J., {et~al.} 2013, \apjl,
  769, L13

\bibitem[{{Goicoechea} {et~al.}(2022){Goicoechea}, {Lique}, \&
  {Santa-Maria}}]{goicoechea2022}
{Goicoechea}, J.~R., {Lique}, F., \& {Santa-Maria}, M.~G. 2022, \aap, 658, A28

\bibitem[{{Goicoechea} {et~al.}(2004){Goicoechea},
  {Rodr{\'\i}guez-Fern{\'a}ndez}, \& {Cernicharo}}]{goicoechea2004}
{Goicoechea}, J.~R., {Rodr{\'\i}guez-Fern{\'a}ndez}, N.~J., \& {Cernicharo}, J.
  2004, \apj, 600, 214

\bibitem[{{Goicoechea} {et~al.}(2018){Goicoechea}, {Santa-Maria}, {Teyssier},
  {Cernicharo}, {Gerin}, \& {Pety}}]{goicoechea2018}
{Goicoechea}, J.~R., {Santa-Maria}, M.~G., {Teyssier}, D., {et~al.} 2018, \aap,
  616, L1

\bibitem[{{Gorai} {et~al.}(2017){Gorai}, {Das}, {Das}, {Sivaraman}, {Etim}, \&
  {Chakrabarti}}]{gorai2017}
{Gorai}, P., {Das}, A., {Das}, A., {et~al.} 2017, \apj, 836, 70

\bibitem[{{Goto} {et~al.}(2013){Goto}, {Indriolo}, {Geballe}, \&
  {Usuda}}]{goto2013}
{Goto}, M., {Indriolo}, N., {Geballe}, T.~R., \& {Usuda}, T. 2013, Journal of
  Physical Chemistry A, 117, 9919

\bibitem[{{Graninger} {et~al.}(2014){Graninger}, {Herbst}, {{\"O}berg}, \&
  {Vasyunin}}]{graninger2014}
{Graninger}, D.~M., {Herbst}, E., {{\"O}berg}, K.~I., \& {Vasyunin}, A.~I.
  2014, \apj, 787, 74

\bibitem[{{GRAVITY Collaboration} {et~al.}(2019){GRAVITY Collaboration},
  {Abuter}, {Amorim}, {Baub{\"o}ck}, {Berger}, {Bonnet}, {Brandner},
  {Cl{\'e}net}, {Coud{\'e} Du Foresto}, {de Zeeuw}, {Dexter}, {Duvert},
  {Eckart}, {Eisenhauer}, {F{\"o}rster Schreiber}, {Garcia}, {Gao}, {Gendron},
  {Genzel}, {Gerhard}, {Gillessen}, {Habibi}, {Haubois}, {Henning}, {Hippler},
  {Horrobin}, {Jim{\'e}nez-Rosales}, {Jocou}, {Kervella}, {Lacour},
  {Lapeyr{\`e}re}, {Le Bouquin}, {L{\'e}na}, {Ott}, {Paumard}, {Perraut},
  {Perrin}, {Pfuhl}, {Rabien}, {Rodriguez Coira}, {Rousset}, {Scheithauer},
  {Sternberg}, {Straub}, {Straubmeier}, {Sturm}, {Tacconi}, {Vincent}, {von
  Fellenberg}, {Waisberg}, {Widmann}, {Wieprecht}, {Wiezorrek}, {Woillez}, \&
  {Yazici}}]{GRAVITY2019}
{GRAVITY Collaboration}, {Abuter}, R., {Amorim}, A., {et~al.} 2019, \aap, 625,
  L10

\bibitem[{{Guesten} \& {Henkel}(1983)}]{guesten1983}
{Guesten}, R. \& {Henkel}, C. 1983, \aap, 125, 136

\bibitem[{{Guesten} {et~al.}(1985){Guesten}, {Walmsley}, {Ungerechts}, \&
  {Churchwell}}]{guesten1985}
{Guesten}, R., {Walmsley}, C.~M., {Ungerechts}, H., \& {Churchwell}, E. 1985,
  \aap, 142, 381

\bibitem[{{Habing}(1968)}]{habing1968}
{Habing}, H.~J. 1968, \bain, 19, 421

\bibitem[{{Hacar} {et~al.}(2020){Hacar}, {Bosman}, \& {van
  Dishoeck}}]{hacar2020}
{Hacar}, A., {Bosman}, A.~D., \& {van Dishoeck}, E.~F. 2020, \aap, 635, A4

\bibitem[{{Harada} {et~al.}(2015){Harada}, {Riquelme}, {Viti},
  {Jim{\'e}nez-Serra}, {Requena-Torres}, {Menten}, {Mart{\'\i}n}, {Aladro},
  {Martin-Pintado}, \& {Hochg{\"u}rtel}}]{harada2015}
{Harada}, N., {Riquelme}, D., {Viti}, S., {et~al.} 2015, \aap, 584, A102

\bibitem[{{Harada} {et~al.}(2024){Harada}, {Saito}, {Nishimura}, {Watanabe}, \&
  {Sakamoto}}]{harada2024}
{Harada}, N., {Saito}, T., {Nishimura}, Y., {Watanabe}, Y., \& {Sakamoto}, K.
  2024, \apj, 969, 82

\bibitem[{{Hasegawa} {et~al.}(1994){Hasegawa}, {Sato}, {Whiteoak}, \&
  {Miyawaki}}]{hasegawa1994}
{Hasegawa}, T., {Sato}, F., {Whiteoak}, J.~B., \& {Miyawaki}, R. 1994, \apjl,
  429, L77

\bibitem[{{Henshaw} {et~al.}(2023){Henshaw}, {Barnes}, {Battersby}, {Ginsburg},
  {Sormani}, \& {Walker}}]{henshaw2023}
{Henshaw}, J.~D., {Barnes}, A.~T., {Battersby}, C., {et~al.} 2023, in
  Astronomical Society of the Pacific Conference Series, Vol. 534, Protostars
  and Planets VII, ed. S.~{Inutsuka}, Y.~{Aikawa}, T.~{Muto}, K.~{Tomida}, \&
  M.~{Tamura}, 83

\bibitem[{{Herbst} \& {van Dishoeck}(2009)}]{herbst2009}
{Herbst}, E. \& {van Dishoeck}, E.~F. 2009, \araa, 47, 427

\bibitem[{{Holdship} {et~al.}(2017){Holdship}, {Viti}, {Jim{\'e}nez-Serra},
  {Makrymallis}, \& {Priestley}}]{uclchem3.0}
{Holdship}, J., {Viti}, S., {Jim{\'e}nez-Serra}, I., {Makrymallis}, A., \&
  {Priestley}, F. 2017, \aj, 154, 38

\bibitem[{{Indriolo} {et~al.}(2015){Indriolo}, {Neufeld}, {Gerin}, {Schilke},
  {Benz}, {Winkel}, {Menten}, {Chambers}, {Black}, {Bruderer}, {Falgarone},
  {Godard}, {Goicoechea}, {Gupta}, {Lis}, {Ossenkopf}, {Persson},
  {Sonnentrucker}, {van der Tak}, {van Dishoeck}, {Wolfire}, \&
  {Wyrowski}}]{indriolo2015}
{Indriolo}, N., {Neufeld}, D.~A., {Gerin}, M., {et~al.} 2015, \apj, 800, 40

\bibitem[{{Izumi} {et~al.}(2016){Izumi}, {Kohno}, {Aalto}, {Espada}, {Fathi},
  {Harada}, {Hatsukade}, {Hsieh}, {Imanishi}, {Krips}, {Mart{\'\i}n},
  {Matsushita}, {Meier}, {Nakai}, {Nakanishi}, {Schinnerer}, {Sheth},
  {Terashima}, \& {Turner}}]{izumi2016}
{Izumi}, T., {Kohno}, K., {Aalto}, S., {et~al.} 2016, \apj, 818, 42

\bibitem[{{Izumi} {et~al.}(2013){Izumi}, {Kohno}, {Mart{\'\i}n}, {Espada},
  {Harada}, {Matsushita}, {Hsieh}, {Turner}, {Meier}, {Schinnerer}, {Imanishi},
  {Tamura}, {Curran}, {Doi}, {Fathi}, {Krips}, {Lundgren}, {Nakai}, {Nakajima},
  {Regan}, {Sheth}, {Takano}, {Taniguchi}, {Terashima}, {Tosaki}, \&
  {Wiklind}}]{izumi2013}
{Izumi}, T., {Kohno}, K., {Mart{\'\i}n}, S., {et~al.} 2013, \pasj, 65, 100

\bibitem[{{James} {et~al.}(2020){James}, {Viti}, {Holdship}, \&
  {Jim{\'e}nez-Serra}}]{james2020}
{James}, T.~A., {Viti}, S., {Holdship}, J., \& {Jim{\'e}nez-Serra}, I. 2020,
  \aap, 634, A17

\bibitem[{{Jenkins}(2009)}]{jenkins2009}
{Jenkins}, E.~B. 2009, \apj, 700, 1299

\bibitem[{{Jimenez-Serra}(2025)}]{jimenez-serra2025}
{Jimenez-Serra}, I. 2025, arXiv e-prints, arXiv:2501.01782

\bibitem[{{Jim{\'e}nez-Serra} {et~al.}(2008){Jim{\'e}nez-Serra}, {Caselli},
  {Mart{\'\i}n-Pintado}, \& {Hartquist}}]{jimenez-serra2008}
{Jim{\'e}nez-Serra}, I., {Caselli}, P., {Mart{\'\i}n-Pintado}, J., \&
  {Hartquist}, T.~W. 2008, \aap, 482, 549

\bibitem[{{Jim{\'e}nez-Serra} {et~al.}(2020){Jim{\'e}nez-Serra},
  {Mart{\'\i}n-Pintado}, {Rivilla}, {Rodr{\'\i}guez-Almeida}, {Alonso Alonso},
  {Zeng}, {Cocinero}, {Mart{\'\i}n}, {Requena-Torres}, {Mart{\'\i}n-Domenech},
  \& {Testi}}]{jimenez-serra2020}
{Jim{\'e}nez-Serra}, I., {Mart{\'\i}n-Pintado}, J., {Rivilla}, V.~M., {et~al.}
  2020, Astrobiology, 20, 1048

\bibitem[{{Jim{\'e}nez-Serra} {et~al.}(2025){Jim{\'e}nez-Serra}, {Meg{\'\i}as},
  {Salaris}, {Cuppen}, {Taillard}, {Jin}, {Wakelam}, {Vasyunin}, {Caselli},
  {Pendleton}, {Dartois}, {Noble}, {Viti}, {Borshcheva}, {Garrod}, {Lamberts},
  {Fraser}, {Melnick}, {McClure}, {Rocha}, {Drozdovskaya}, \&
  {Lis}}]{jimenez-serra2025b}
{Jim{\'e}nez-Serra}, I., {Meg{\'\i}as}, A., {Salaris}, J., {et~al.} 2025, \aap,
  695, A247

\bibitem[{{Jim{\'e}nez-Serra} {et~al.}(2022){Jim{\'e}nez-Serra},
  {Rodr{\'\i}guez-Almeida}, {Mart{\'\i}n-Pintado}, {Rivilla}, {Melosso},
  {Zeng}, {Colzi}, {Kawashima}, {Hirota}, {Puzzarini}, {Tercero}, {de Vicente},
  {Rico-Villas}, {Requena-Torres}, \& {Mart{\'\i}n}}]{jimenez-serra2022}
{Jim{\'e}nez-Serra}, I., {Rodr{\'\i}guez-Almeida}, L.~F.,
  {Mart{\'\i}n-Pintado}, J., {et~al.} 2022, \aap, 663, A181

\bibitem[{{Jones} {et~al.}(2011){Jones}, {Burton}, {Tothill}, \&
  {Cunningham}}]{jones2011}
{Jones}, P.~A., {Burton}, M.~G., {Tothill}, N.~F.~H., \& {Cunningham}, M.~R.
  2011, \mnras, 411, 2293

\bibitem[{{J{\o}rgensen} {et~al.}(2020){J{\o}rgensen}, {Belloche}, \&
  {Garrod}}]{jorgensen2020}
{J{\o}rgensen}, J.~K., {Belloche}, A., \& {Garrod}, R.~T. 2020, \araa, 58, 727

\bibitem[{{Karoly} {et~al.}(2025){Karoly}, {Ward-Thompson}, {Pattle},
  {Longmore}, {Di Francesco}, {Whitworth}, {Johnstone}, {Sadavoy}, {Koch},
  {Yang}, {Furuya}, {Lu}, {Tamura}, {Debattista}, {Eden}, {Hwang}, {Poidevin},
  {Bijas}, {Chen}, {Chung}, {Coud{\'e}}, {Lin}, {Doi}, {Onaka}, {Fanciullo},
  {Liu}, {Li}, {Bastien}, {Hasegawa}, {Kwon}, {Lai}, \& {Qiu}}]{karoly2025}
{Karoly}, J., {Ward-Thompson}, D., {Pattle}, K., {et~al.} 2025, \apjl, 982, L22

\bibitem[{{Kerr} {et~al.}(2015){Kerr}, {Alecu}, {Thompson}, {Gao}, \&
  {Marshall}}]{kerr2015}
{Kerr}, K.~E., {Alecu}, I.~M., {Thompson}, K.~M., {Gao}, Y., \& {Marshall}, P.
  2015, Journal of Physical Chemistry A, 119, 7352

\bibitem[{{Krieger} {et~al.}(2017){Krieger}, {Ott}, {Beuther}, {Walter},
  {Kruijssen}, {Meier}, {Mills}, {Contreras}, {Edwards}, {Ginsburg}, {Henkel},
  {Henshaw}, {Jackson}, {Kauffmann}, {Longmore}, {Mart{\'\i}n}, {Morris},
  {Pillai}, {Rickert}, {Rosolowsky}, {Shinnaga}, {Walsh}, {Yusef-Zadeh}, \&
  {Zhang}}]{krieger2017}
{Krieger}, N., {Ott}, J., {Beuther}, H., {et~al.} 2017, \apj, 850, 77

\bibitem[{{Kruijssen} \& {Longmore}(2013)}]{kruijssen2013}
{Kruijssen}, J.~M.~D. \& {Longmore}, S.~N. 2013, \mnras, 435, 2598

\bibitem[{{Laas} \& {Caselli}(2019)}]{laas2019}
{Laas}, J.~C. \& {Caselli}, P. 2019, \aap, 624, A108

\bibitem[{{Lamberts}(2018)}]{lamberts2018}
{Lamberts}, T. 2018, \aap, 615, L2

\bibitem[{{Le Petit} {et~al.}(2016){Le Petit}, {Ruaud}, {Bron}, {Godard},
  {Roueff}, {Languignon}, \& {Le Bourlot}}]{lepetit2016}
{Le Petit}, F., {Ruaud}, M., {Bron}, E., {et~al.} 2016, \aap, 585, A105

\bibitem[{{Lee} {et~al.}(2017){Lee}, {Li}, {Ho}, {Hirano}, {Zhang}, \&
  {Shang}}]{lee2017}
{Lee}, C.-F., {Li}, Z.-Y., {Ho}, P. T.~P., {et~al.} 2017, \apj, 843, 27

\bibitem[{{Linke} {et~al.}(1979){Linke}, {Frerking}, \& {Thaddeus}}]{linke1979}
{Linke}, R.~A., {Frerking}, M.~A., \& {Thaddeus}, P. 1979, \apjl, 234, L139

\bibitem[{{Lis} {et~al.}(2001){Lis}, {Serabyn}, {Zylka}, \& {Li}}]{lis2001}
{Lis}, D.~C., {Serabyn}, E., {Zylka}, R., \& {Li}, Y. 2001, \apj, 550, 761

\bibitem[{{Longmore} {et~al.}(2025){Longmore}, {Bally}, {Barnes}, {Battersby},
  {Colzi}, {Ginsburg}, {Henshaw}, {Ho}, \& {Jimenez-Serra}}]{longmore2025}
{Longmore}, S.~N., {Bally}, J., {Barnes}, A.~T., {et~al.} 2025, \mnras,
  submitted

\bibitem[{{Longmore} {et~al.}(2013){Longmore}, {Bally}, {Testi}, {Purcell},
  {Walsh}, {Bressert}, {Pestalozzi}, {Molinari}, {Ott}, {Cortese}, {Battersby},
  {Murray}, {Lee}, {Kruijssen}, {Schisano}, \& {Elia}}]{longmore2013}
{Longmore}, S.~N., {Bally}, J., {Testi}, L., {et~al.} 2013, \mnras, 429, 987

\bibitem[{{L{\'o}pez-Sepulcre} {et~al.}(2019){L{\'o}pez-Sepulcre}, {Balucani},
  {Ceccarelli}, {Codella}, {Dulieu}, \& {Theul{\'e}}}]{lopez-sepulcre2019}
{L{\'o}pez-Sepulcre}, A., {Balucani}, N., {Ceccarelli}, C., {et~al.} 2019, ACS
  Earth and Space Chemistry, 3, 2122

\bibitem[{{L{\'o}pez-Sepulcre} {et~al.}(2024){L{\'o}pez-Sepulcre}, {Codella},
  {Ceccarelli}, {Podio}, \& {Robuschi}}]{lopez-sepulcre2024}
{L{\'o}pez-Sepulcre}, A., {Codella}, C., {Ceccarelli}, C., {Podio}, L., \&
  {Robuschi}, J. 2024, \aap, 692, A120

\bibitem[{{Lu} {et~al.}(2021){Lu}, {Li}, {Ginsburg}, {Longmore}, {Kruijssen},
  {Walker}, {Feng}, {Zhang}, {Battersby}, {Pillai}, {Mills}, {Kauffmann},
  {Cheng}, \& {Inutsuka}}]{lu2021}
{Lu}, X., {Li}, S., {Ginsburg}, A., {et~al.} 2021, \apj, 909, 177

\bibitem[{{Mangum} {et~al.}(2019){Mangum}, {Ginsburg}, {Henkel}, {Menten},
  {Aalto}, \& {van der Werf}}]{mangum2019}
{Mangum}, J.~G., {Ginsburg}, A.~G., {Henkel}, C., {et~al.} 2019, \apj, 871, 170

\bibitem[{{Marsh} {et~al.}(2016){Marsh}, {Ragan}, {Whitworth}, \&
  {Clark}}]{marsh2016}
{Marsh}, K.~A., {Ragan}, S.~E., {Whitworth}, A.~P., \& {Clark}, P.~C. 2016,
  \mnras, 461, L16

\bibitem[{{Mart{\'\i}n} {et~al.}(2021){Mart{\'\i}n}, {Mangum}, {Harada},
  {Costagliola}, {Sakamoto}, {Muller}, {Aladro}, {Tanaka}, {Yoshimura},
  {Nakanishi}, {Herrero-Illana}, {M{\"u}hle}, {Aalto}, {Behrens}, {Colzi},
  {Emig}, {Fuller}, {Garc{\'\i}a-Burillo}, {Greve}, {Henkel}, {Holdship},
  {Humire}, {Hunt}, {Izumi}, {Kohno}, {K{\"o}nig}, {Meier}, {Nakajima},
  {Nishimura}, {Padovani}, {Rivilla}, {Takano}, {van der Werf}, {Viti}, \&
  {Yan}}]{martin2021}
{Mart{\'\i}n}, S., {Mangum}, J.~G., {Harada}, N., {et~al.} 2021, \aap, 656, A46

\bibitem[{{Mart{\'\i}n} {et~al.}(2008){Mart{\'\i}n}, {Requena-Torres},
  {Mart{\'\i}n-Pintado}, \& {Mauersberger}}]{martin2008}
{Mart{\'\i}n}, S., {Requena-Torres}, M.~A., {Mart{\'\i}n-Pintado}, J., \&
  {Mauersberger}, R. 2008, \apj, 678, 245

\bibitem[{{Martin-Pintado} {et~al.}(1992){Martin-Pintado}, {Bachiller}, \&
  {Fuente}}]{martin-pintado1992}
{Martin-Pintado}, J., {Bachiller}, R., \& {Fuente}, A. 1992, \aap, 254, 315

\bibitem[{{Mart{\'\i}n-Pintado} {et~al.}(1997){Mart{\'\i}n-Pintado}, {de
  Vicente}, {Fuente}, \& {Planesas}}]{martin-pintado1997}
{Mart{\'\i}n-Pintado}, J., {de Vicente}, P., {Fuente}, A., \& {Planesas}, P.
  1997, \apjl, 482, L45

\bibitem[{{McElroy} {et~al.}(2013){McElroy}, {Walsh}, {Markwick}, {Cordiner},
  {Smith}, \& {Millar}}]{umist2012}
{McElroy}, D., {Walsh}, C., {Markwick}, A.~J., {et~al.} 2013, \aap, 550, A36

\bibitem[{{McKee} \& {Tan}(2003)}]{mckee2003}
{McKee}, C.~F. \& {Tan}, J.~C. 2003, \apj, 585, 850

\bibitem[{{Millar} {et~al.}(2024){Millar}, {Walsh}, {Van de Sande}, \&
  {Markwick}}]{umist2022}
{Millar}, T.~J., {Walsh}, C., {Van de Sande}, M., \& {Markwick}, A.~J. 2024,
  \aap, 682, A109

\bibitem[{{Mills} {et~al.}(2018){Mills}, {Ginsburg}, {Immer}, {Barnes},
  {Wiesenfeld}, {Faure}, {Morris}, \& {Requena-Torres}}]{mills2018}
{Mills}, E.~A.~C., {Ginsburg}, A., {Immer}, K., {et~al.} 2018, \apj, 868, 7

\bibitem[{{M{\"o}ller} {et~al.}(2025){M{\"o}ller}, {Schilke},
  {S{\'a}nchez-Monge}, \& {Schmiedeke}}]{moeller2025}
{M{\"o}ller}, T., {Schilke}, P., {S{\'a}nchez-Monge}, {\'A}., \& {Schmiedeke},
  A. 2025, \aap, 693, A160

\bibitem[{{Morris} {et~al.}(1983){Morris}, {Polish}, {Zuckerman}, \&
  {Kaifu}}]{morris1983}
{Morris}, M., {Polish}, N., {Zuckerman}, B., \& {Kaifu}, N. 1983, \aj, 88, 1228

\bibitem[{{Morris} \& {Serabyn}(1996)}]{morris1996}
{Morris}, M. \& {Serabyn}, E. 1996, \araa, 34, 645

\bibitem[{{Muller} {et~al.}(2011){Muller}, {Beelen}, {Gu{\'e}lin}, {Aalto},
  {Black}, {Combes}, {Curran}, {Theule}, \& {Longmore}}]{muller2011}
{Muller}, S., {Beelen}, A., {Gu{\'e}lin}, M., {et~al.} 2011, \aap, 535, A103

\bibitem[{{Nishimura} {et~al.}(2024){Nishimura}, {Aalto}, {Gorski},
  {K{\"o}nig}, {Onishi}, {Wethers}, {Yang}, {Barcos-Mu{\~n}oz}, {Combes},
  {D{\'\i}az-Santos}, {Gallagher}, {Garc{\'\i}a-Burillo},
  {Gonz{\'a}lez-Alfonso}, {Greve}, {Harada}, {Henkel}, {Imanishi}, {Kohno},
  {Linden}, {Mangum}, {Mart{\'\i}n}, {Muller}, {Privon}, {Ricci}, {Stanley},
  {van der Werf}, \& {Viti}}]{nishimura2024}
{Nishimura}, Y., {Aalto}, S., {Gorski}, M.~D., {et~al.} 2024, \aap, 686, A48

\bibitem[{{Oka} {et~al.}(2019){Oka}, {Geballe}, {Goto}, {Usuda}, {Benjamin},
  {McCall}, \& {Indriolo}}]{oka2019}
{Oka}, T., {Geballe}, T.~R., {Goto}, M., {et~al.} 2019, \apj, 883, 54

\bibitem[{{Padovani} {et~al.}(2022){Padovani}, {Bialy}, {Galli}, {Ivlev},
  {Grassi}, {Scarlett}, {Rehill}, {Zammit}, {Fursa}, \& {Bray}}]{padovani2022}
{Padovani}, M., {Bialy}, S., {Galli}, D., {et~al.} 2022, \aap, 658, A189

\bibitem[{{Padovani} {et~al.}(2020){Padovani}, {Ivlev}, {Galli}, {Offner},
  {Indriolo}, {Rodgers-Lee}, {Marcowith}, {Girichidis}, {Bykov}, \&
  {Kruijssen}}]{padovani2020}
{Padovani}, M., {Ivlev}, A.~V., {Galli}, D., {et~al.} 2020, \ssr, 216, 29

\bibitem[{{Perrero} {et~al.}(2022){Perrero}, {Enrique-Romero}, {Ferrero},
  {Ceccarelli}, {Podio}, {Codella}, {Rimola}, \& {Ugliengo}}]{perrero2022}
{Perrero}, J., {Enrique-Romero}, J., {Ferrero}, S., {et~al.} 2022, \apj, 938,
  158

\bibitem[{{Pillai} {et~al.}(2015){Pillai}, {Kauffmann}, {Tan}, {Goldsmith},
  {Carey}, \& {Menten}}]{pillai2015}
{Pillai}, T., {Kauffmann}, J., {Tan}, J.~C., {et~al.} 2015, \apj, 799, 74

\bibitem[{{Pineda} {et~al.}(2020){Pineda}, {Segura-Cox}, {Caselli},
  {Cunningham}, {Zhao}, {Schmiedeke}, {Maureira}, \& {Neri}}]{pineda2020}
{Pineda}, J.~E., {Segura-Cox}, D., {Caselli}, P., {et~al.} 2020, Nature
  Astronomy, 4, 1158

\bibitem[{{Qu{\'e}nard} {et~al.}(2018){Qu{\'e}nard}, {Jim{\'e}nez-Serra},
  {Viti}, {Holdship}, \& {Coutens}}]{quenard2018}
{Qu{\'e}nard}, D., {Jim{\'e}nez-Serra}, I., {Viti}, S., {Holdship}, J., \&
  {Coutens}, A. 2018, \mnras, 474, 2796

\bibitem[{{Rathborne} {et~al.}(2015){Rathborne}, {Longmore}, {Jackson},
  {Alves}, {Bally}, {Bastian}, {Contreras}, {Foster}, {Garay}, {Kruijssen},
  {Testi}, \& {Walsh}}]{rathborne2015}
{Rathborne}, J.~M., {Longmore}, S.~N., {Jackson}, J.~M., {et~al.} 2015, \apj,
  802, 125

\bibitem[{{Rawlings} {et~al.}(1992){Rawlings}, {Hartquist}, {Menten}, \&
  {Williams}}]{rawlings1992}
{Rawlings}, J.~M.~C., {Hartquist}, T.~W., {Menten}, K.~M., \& {Williams}, D.~A.
  1992, \mnras, 255, 471

\bibitem[{{Rawlings} {et~al.}(2004){Rawlings}, {Redman}, {Keto}, \&
  {Williams}}]{rawlings2004}
{Rawlings}, J.~M.~C., {Redman}, M.~P., {Keto}, E., \& {Williams}, D.~A. 2004,
  \mnras, 351, 1054

\bibitem[{{Requena-Torres} {et~al.}(2008){Requena-Torres},
  {Mart{\'\i}n-Pintado}, {Mart{\'\i}n}, \& {Morris}}]{requena-torres2008}
{Requena-Torres}, M.~A., {Mart{\'\i}n-Pintado}, J., {Mart{\'\i}n}, S., \&
  {Morris}, M.~R. 2008, \apj, 672, 352

\bibitem[{{Requena-Torres} {et~al.}(2006){Requena-Torres},
  {Mart{\'\i}n-Pintado}, {Rodr{\'\i}guez-Franco}, {Mart{\'\i}n},
  {Rodr{\'\i}guez-Fern{\'a}ndez}, \& {de Vicente}}]{requena-torres2006}
{Requena-Torres}, M.~A., {Mart{\'\i}n-Pintado}, J., {Rodr{\'\i}guez-Franco},
  A., {et~al.} 2006, \aap, 455, 971

\bibitem[{{Rivilla} {et~al.}(2022){Rivilla}, {Colzi}, {Jim{\'e}nez-Serra},
  {Mart{\'\i}n-Pintado}, {Meg{\'\i}as}, {Melosso}, {Bizzocchi},
  {L{\'o}pez-Gallifa}, {Mart{\'\i}nez-Henares}, {Massalkhi}, {Tercero}, {de
  Vicente}, {Guillemin}, {Garc{\'\i}a de la Concepci{\'o}n}, {Rico-Villas},
  {Zeng}, {Mart{\'\i}n}, {Requena-Torres}, {Tonolo}, {Alessandrini}, {Dore},
  {Barone}, \& {Puzzarini}}]{rivilla2022}
{Rivilla}, V.~M., {Colzi}, L., {Jim{\'e}nez-Serra}, I., {et~al.} 2022, \apjl,
  929, L11

\bibitem[{{Rivilla} {et~al.}(2021){Rivilla}, {Jim{\'e}nez-Serra},
  {Mart{\'\i}n-Pintado}, {Briones}, {Rodr{\'\i}guez-Almeida}, {Rico-Villas},
  {Tercero}, {Zeng}, {Colzi}, {de Vicente}, {Mart{\'\i}n}, \&
  {Requena-Torres}}]{rivilla2021}
{Rivilla}, V.~M., {Jim{\'e}nez-Serra}, I., {Mart{\'\i}n-Pintado}, J., {et~al.}
  2021, Proceedings of the National Academy of Science, 118, e2101314118

\bibitem[{{Rivilla} {et~al.}(2020){Rivilla}, {Mart{\'\i}n-Pintado},
  {Jim{\'e}nez-Serra}, {Mart{\'\i}n}, {Rodr{\'\i}guez-Almeida},
  {Requena-Torres}, {Rico-Villas}, {Zeng}, \& {Briones}}]{rivilla2020}
{Rivilla}, V.~M., {Mart{\'\i}n-Pintado}, J., {Jim{\'e}nez-Serra}, I., {et~al.}
  2020, \apjl, 899, L28

\bibitem[{{Rivilla} {et~al.}(2023){Rivilla}, {Sanz-Novo}, {Jim{\'e}nez-Serra},
  {Mart{\'\i}n-Pintado}, {Colzi}, {Zeng}, {Meg{\'\i}as}, {L{\'o}pez-Gallifa},
  {Mart{\'\i}nez-Henares}, {Massalkhi}, {Tercero}, {de Vicente}, {Mart{\'\i}n},
  {San Andr{\'e}s}, {Requena-Torres}, \& {Alonso}}]{rivilla2023}
{Rivilla}, V.~M., {Sanz-Novo}, M., {Jim{\'e}nez-Serra}, I., {et~al.} 2023,
  \apjl, 953, L20

\bibitem[{{Rodr{\'\i}guez-Almeida} {et~al.}(2021){Rodr{\'\i}guez-Almeida},
  {Jim{\'e}nez-Serra}, {Rivilla}, {Mart{\'\i}n-Pintado}, {Zeng}, {Tercero}, {de
  Vicente}, {Colzi}, {Rico-Villas}, {Mart{\'\i}n}, \&
  {Requena-Torres}}]{rodriguez-almeida2021}
{Rodr{\'\i}guez-Almeida}, L.~F., {Jim{\'e}nez-Serra}, I., {Rivilla}, V.~M.,
  {et~al.} 2021, \apjl, 912, L11

\bibitem[{{Roman-Duval} {et~al.}(2016){Roman-Duval}, {Heyer}, {Brunt}, {Clark},
  {Klessen}, \& {Shetty}}]{roman-duval2016}
{Roman-Duval}, J., {Heyer}, M., {Brunt}, C.~M., {et~al.} 2016, \apj, 818, 144

\bibitem[{{Saito} {et~al.}(1987){Saito}, {Kawaguchi}, {Yamamoto}, {Ohishi},
  {Suzuki}, \& {Kaifu}}]{saito1987}
{Saito}, S., {Kawaguchi}, K., {Yamamoto}, S., {et~al.} 1987, \apjl, 317, L115

\bibitem[{{S{\'a}nchez-Monge} {et~al.}(2017){S{\'a}nchez-Monge}, {Schilke},
  {Schmiedeke}, {Ginsburg}, {Cesaroni}, {Lis}, {Qin}, {M{\"u}ller}, {Bergin},
  {Comito}, \& {M{\"o}ller}}]{sanchezmonge2017}
{S{\'a}nchez-Monge}, {\'A}., {Schilke}, P., {Schmiedeke}, A., {et~al.} 2017,
  \aap, 604, A6

\bibitem[{{Santa-Maria} {et~al.}(2021){Santa-Maria}, {Goicoechea}, {Etxaluze},
  {Cernicharo}, \& {Cuadrado}}]{santa-maria2021}
{Santa-Maria}, M.~G., {Goicoechea}, J.~R., {Etxaluze}, M., {Cernicharo}, J., \&
  {Cuadrado}, S. 2021, \aap, 649, A32

\bibitem[{{Savage} \& {Sembach}(1996)}]{savage1996}
{Savage}, B.~D. \& {Sembach}, K.~R. 1996, \araa, 34, 279

\bibitem[{{Schenewerk} {et~al.}(1988){Schenewerk}, {Snyder}, {Hollis},
  {Jewell}, \& {Ziurys}}]{schenewerk1988}
{Schenewerk}, M.~S., {Snyder}, L.~E., {Hollis}, J.~M., {Jewell}, P.~R., \&
  {Ziurys}, L.~M. 1988, \apj, 328, 785

\bibitem[{{Schmiedeke} {et~al.}(2016){Schmiedeke}, {Schilke}, {M{\"o}ller},
  {S{\'a}nchez-Monge}, {Bergin}, {Comito}, {Csengeri}, {Lis}, {Molinari},
  {Qin}, \& {Rolffs}}]{schmiedeke2016}
{Schmiedeke}, A., {Schilke}, P., {M{\"o}ller}, T., {et~al.} 2016, \aap, 588,
  A143

\bibitem[{{Sch{\"o}ier} {et~al.}(2004){Sch{\"o}ier}, {J{\o}rgensen}, {van
  Dishoeck}, \& {Blake}}]{schoier2004}
{Sch{\"o}ier}, F.~L., {J{\o}rgensen}, J.~K., {van Dishoeck}, E.~F., \& {Blake},
  G.~A. 2004, \aap, 418, 185

\bibitem[{{Schw{\"o}rer} {et~al.}(2019){Schw{\"o}rer}, {S{\'a}nchez-Monge},
  {Schilke}, {M{\"o}ller}, {Ginsburg}, {Meng}, {Schmiedeke}, {M{\"u}ller},
  {Lis}, \& {Qin}}]{schworer2019}
{Schw{\"o}rer}, A., {S{\'a}nchez-Monge}, {\'A}., {Schilke}, P., {et~al.} 2019,
  \aap, 628, A6

\bibitem[{{Shope} {et~al.}(2024){Shope}, {El-Abd}, {Brogan}, {Hunter},
  {Willis}, {McGuire}, \& {Garrod}}]{shope2024}
{Shope}, B.~M., {El-Abd}, S.~J., {Brogan}, C.~L., {et~al.} 2024, \apj, 972, 146

\bibitem[{{Spilker} {et~al.}(2021){Spilker}, {Kainulainen}, \&
  {Orkisz}}]{spilker2021}
{Spilker}, A., {Kainulainen}, J., \& {Orkisz}, J. 2021, \aap, 653, A63

\bibitem[{{Sternberg} \& {Dalgarno}(1995)}]{sternberg1995}
{Sternberg}, A. \& {Dalgarno}, A. 1995, \apjs, 99, 565

\bibitem[{{Suzuki} {et~al.}(1992){Suzuki}, {Yamamoto}, {Ohishi}, {Kaifu},
  {Ishikawa}, {Hirahara}, \& {Takano}}]{suzuki1992}
{Suzuki}, H., {Yamamoto}, S., {Ohishi}, M., {et~al.} 1992, \apj, 392, 551

\bibitem[{{Tang} {et~al.}(2021){Tang}, {Wang}, {Wilson}, {Heyer}, {Gutermuth},
  {Schloerb}, {Yun}, {Bally}, {Loinard}, {Silich}, {Ch{\'a}vez}, {Haggard},
  {Monta{\~n}a}, {S{\'a}nchez-Arg{\"u}elles}, {Zeballos}, {Zavala}, \&
  {Le{\'o}n-Tavares}}]{tang2021}
{Tang}, Y., {Wang}, Q.~D., {Wilson}, G.~W., {et~al.} 2021, \mnras, 505, 2392

\bibitem[{{Tram} {et~al.}(2025){Tram}, {Viti}, {Dutkowska}, {Vermari{\"e}n},
  {Dijkhuis}, {Coutens}, {Csengeri}, \& {Hoang}}]{tram2025}
{Tram}, L.~N., {Viti}, S., {Dutkowska}, K.~M., {et~al.} 2025, \aap, submitted

\bibitem[{{Tsuboi} {et~al.}(2015){Tsuboi}, {Miyazaki}, \&
  {Uehara}}]{tsuboi2015}
{Tsuboi}, M., {Miyazaki}, A., \& {Uehara}, K. 2015, \pasj, 67, 109

\bibitem[{{van der Tak} \& {van Dishoeck}(2000)}]{vandertak2000}
{van der Tak}, F.~F.~S. \& {van Dishoeck}, E.~F. 2000, \aap, 358, L79

\bibitem[{{van Dishoeck}(2014)}]{vandishoeck2014}
{van Dishoeck}, E.~F. 2014, Faraday Discussions, 168, 9

\bibitem[{{Vastel} {et~al.}(2018){Vastel}, {Qu{\'e}nard}, {Le Gal}, {Wakelam},
  {Andrianasolo}, {Caselli}, {Vidal}, {Ceccarelli}, {Lefloch}, \&
  {Bachiller}}]{vastel2018}
{Vastel}, C., {Qu{\'e}nard}, D., {Le Gal}, R., {et~al.} 2018, \mnras, 478, 5514

\bibitem[{{Viti} {et~al.}(2001){Viti}, {Caselli}, {Hartquist}, \&
  {Williams}}]{viti2001}
{Viti}, S., {Caselli}, P., {Hartquist}, T.~W., \& {Williams}, D.~A. 2001, \aap,
  370, 1017

\bibitem[{{Viti} {et~al.}(2004){Viti}, {Collings}, {Dever}, {McCoustra}, \&
  {Williams}}]{viti2004}
{Viti}, S., {Collings}, M.~P., {Dever}, J.~W., {McCoustra}, M. R.~S., \&
  {Williams}, D.~A. 2004, \mnras, 354, 1141

\bibitem[{{Viti} {et~al.}(2002){Viti}, {Natarajan}, \& {Williams}}]{viti2002}
{Viti}, S., {Natarajan}, S., \& {Williams}, D.~A. 2002, \mnras, 336, 797

\bibitem[{{Wakelam} {et~al.}(2024){Wakelam}, {Gratier}, {Loison}, {Hickson},
  {Penguen}, \& {Mechineau}}]{kida2024}
{Wakelam}, V., {Gratier}, P., {Loison}, J.~C., {et~al.} 2024, \aap, 689, A63

\bibitem[{{Walker} {et~al.}(2021){Walker}, {Longmore}, {Bally}, {Ginsburg},
  {Kruijssen}, {Zhang}, {Henshaw}, {Lu}, {Alves}, {Barnes}, {Battersby},
  {Beuther}, {Contreras}, {G{\'o}mez}, {Ho}, {Jackson}, {Kauffmann}, {Mills},
  \& {Pillai}}]{walker2021}
{Walker}, D.~L., {Longmore}, S.~N., {Bally}, J., {et~al.} 2021, \mnras, 503, 77

\bibitem[{{Xu} {et~al.}(2025){Xu}, {Lu}, {Wang}, {Liu}, {Ginsburg}, {Liu},
  {Zhang}, {Budaiev}, {Tang}, {Schilke}, {Zhang}, {Jiao}, {Jiao}, {Zheng},
  {Jones}, {Kruijssen}, {Battersby}, {Walker}, {Mills}, {Kauffmann},
  {Longmore}, \& {Pillai}}]{xu2025}
{Xu}, F., {Lu}, X., {Wang}, K., {et~al.} 2025, \aap, 697, A164

\bibitem[{{Yang} {et~al.}(2025){Yang}, {Lu}, {Zhang}, {Liu}, {Ginsburg}, {Liu},
  {Cheng}, {Feng}, {Liu}, {Zhang}, {Mills}, {Walker}, {Inutsuka}, {Battersby},
  {Longmore}, {Tang}, {Kauffmann}, {Gu}, {Li}, {Luo}, {Kruijssen}, {Pillai},
  {Qiao}, {Qiu}, \& {Shen}}]{yang2025}
{Yang}, K., {Lu}, X., {Zhang}, Y., {et~al.} 2025, \aap, 694, A86

\bibitem[{{Zeng} {et~al.}(2018){Zeng}, {Jim{\'e}nez-Serra}, {Rivilla},
  {Mart{\'\i}n}, {Mart{\'\i}n-Pintado}, {Requena-Torres},
  {Armijos-Abenda{\~n}o}, {Riquelme}, \& {Aladro}}]{zeng2018}
{Zeng}, S., {Jim{\'e}nez-Serra}, I., {Rivilla}, V.~M., {et~al.} 2018, \mnras,
  478, 2962

\bibitem[{{Zeng} {et~al.}(2023){Zeng}, {Rivilla}, {Jim{\'e}nez-Serra}, {Colzi},
  {Mart{\'\i}n-Pintado}, {Tercero}, {de Vicente}, {Mart{\'\i}n}, \&
  {Requena-Torres}}]{zeng2023}
{Zeng}, S., {Rivilla}, V.~M., {Jim{\'e}nez-Serra}, I., {et~al.} 2023, \mnras,
  523, 1448

\bibitem[{{Zeng} {et~al.}(2020){Zeng}, {Zhang}, {Jim{\'e}nez-Serra}, {Tercero},
  {Lu}, {Mart{\'\i}n-Pintado}, {de Vicente}, {Rivilla}, \& {Li}}]{zeng2020}
{Zeng}, S., {Zhang}, Q., {Jim{\'e}nez-Serra}, I., {et~al.} 2020, \mnras, 497,
  4896

\end{thebibliography}
